\documentclass[draftcls, journal, onecolumn]{IEEEtran}

\usepackage[english]{babel}
\usepackage[latin1]{inputenc}
\usepackage{amssymb,amstext, amsfonts}
\usepackage[cmex10]{amsmath}
\usepackage[ruled,vlined,titlenotnumbered]{algorithm2e}
\usepackage{verbatim} 
\usepackage{float}
\usepackage[normalem]{ulem}
\usepackage{color}

\usepackage{enumerate}

\usepackage{comment}
\excludecomment{hide}

\usepackage{cite}

\ifCLASSINFOpdf
   \usepackage[pdftex]{graphicx}
   \DeclareGraphicsExtensions{.pdf,.jpeg,.png}
\else
   \usepackage[dvips]{graphicx}
   \DeclareGraphicsExtensions{.eps}
\fi
\usepackage[tight,footnotesize]{subfigure}

\usepackage{tikz}

\newtheorem{theorem}{Theorem}

\usepackage{mymaths}

\newcommand{\nota}[1]{{\slshape\color{blue}[#1]}}
\renewcommand{\nota}[1]{}

\renewcommand{\B}[1]{\boldsymbol{#1}}
\newcommand{\tr}[1]{\mathrm{tr} \left( #1 \right)}

\DeclareFontFamily{OT1}{pzc}{}
\DeclareFontShape{OT1}{pzc}{m}{it}{<-> s * [1.10] pzcmi7t}{}
\DeclareMathAlphabet{\mathpzc}{OT1}{pzc}{m}{it}

\begin{document}

%


\title{\huge Compressive Classification of a Mixture of Gaussians:\\ Analysis, Designs and Geometrical Interpretation} 

%
%
%

\author{Hugo Reboredo,~\IEEEmembership{Student Member,~IEEE,}
        Francesco Renna,~\IEEEmembership{Member,~IEEE,}
	Robert Calderbank,~\IEEEmembership{Fellow,~IEEE,}
        and~Miguel R. D. Rodrigues,~\IEEEmembership{Member,~IEEE}
        \thanks{\footnotesize This paper was presented in part at the 2013 IEEE International Symposium on Information Theory and the 2013 IEEE Global Conference on Signal and Information Processing. The work of H. Reboredo was supported by Funda\c{c}\~{a}o para a Ci\^{e}ncia e Tecnologia, Portugal, through the doctoral grant SFRH/BD/81543/2011. The work of M. R. D. Rodrigues was supported by the EPSRC through the research grant EP/K503459/1. This work was also supported by the Royal Society International Exchanges Scheme IE120996. \vspace{0.2cm}}

\thanks{\footnotesize H. Reboredo and F. Renna are with the Instituto de Telecomunica\c{c}\~{o}es and the Departamento
de Ci\^{e}ncia de Computadores da Faculdade de Ci\^{e}ncias da Universidade do Porto, Portugal 
({\sf email: \{hugoreboredo, frarenna\}@dcc.fc.up.pt}).}
\thanks{\footnotesize R. Calderbank is with the Department of Electrical and Computer Engineering, Duke University, NC, USA
({\sf email: robert.calderbank@duke.edu}).}
\thanks{\footnotesize M. R. D. Rodrigues is with the Department of Electronic and Electrical Engineering, University College London, United Kingdom 
({\sf email: m.rodrigues@ucl.ac.uk}).}
}

\maketitle

\begin{abstract}

This paper derives fundamental limits on the performance of compressive classification when the source is a mixture of Gaussians. It provides an asymptotic analysis of a Bhattacharya based upper bound on the misclassification probability for the optimal Maximum-A-Posteriori (MAP) classifier that depends on quantities that are dual to the concepts of diversity-order and coding gain in multi-antenna communications. The diversity-order of the measurement system determines the rate at which the probability of misclassification decays with signal-to-noise ratio (SNR) in the low-noise regime. The counterpart of coding gain is the measurement gain which determines the power offset of the probability of misclassification in the low-noise regime. These two quantities make it possible to quantify differences in misclassification probability between random measurement and (diversity-order) optimized measurement. Results are presented for two-class classification problems first with zero-mean Gaussians then with nonzero-mean Gaussians, and finally for multiple-class Gaussian classification problems. The behavior of misclassification probability is revealed to be intimately related to certain fundamental geometric quantities determined by the measurement system, the source and their interplay. Numerical results, representative of compressive classification of a mixture of Gaussians, demonstrate alignment of the actual misclassification probability with the Bhattacharya based upper bound. The connection between the misclassification performance and the alignment between source and measurement geometry may be used to guide the design of dictionaries for compressive classification.

\end{abstract}




%
\IEEEpeerreviewmaketitle

\section{Introduction}
%
%
%
%

\subsection{The Compressive Classification Problem}

Compressive sensing (CS) is an emerging paradigm that offers the means to simultaneously sense and compress a signal without any loss of information~\cite{Candes06a,Candes06b,Candes06c,Donoho06b}. The sensing process is based on the projection of the signal of interest onto a set of vectors, which are typically constituted randomly~\cite{Baraniuk08,Candes05,Candes06a,Candes06b,Candes06c,Donoho06b}, and the recovery process is based on the resolution of an inverse problem. The result that has captured the imagination of the signal and information processing community is that it is possible to perfectly reconstruct an $n$-dimensional $s$-sparse signal (sparse in some orthonormal dictionary or frame) with overwhelming probability with only $\mathcal{O}\left(s\log\left(n/s\right)\right)$ linear random measurements or projections~\cite{Baraniuk08,Candes06a,Donoho06b} using tractable $\ell_1$ minimization methods~\cite{Candes06b} or iterative methods, like greedy matching pursuit~\cite{Mallat93,Chen98,Tropp10}. 
As such, compressive sensing has been proposed for a myriad of applications ranging from signal, image and video compression and processing, to communications and to medicine~\cite{Romberg08,Healy08,Lustig08,Haupt08}.

The focus of compressive sensing has been primarily on exact or near-exact signal reconstruction from a set of linear signal measurements. However, it is also natural to leverage the paradigm to perform other relevant information processing tasks, such as detection, classification and estimation of certain parameters, from the set of compressive measurements. One could in fact argue that the paradigm is a better fit to decision support tasks such as signal detection, signal classification or pattern recognition rather than signal reconstruction, since it may be easier to discriminate between signal classes than reconstruct an entire signal using only partial information about the source signal. Compressive information processing, recently proposed by Davenport \emph{et al.}~\cite{Davenport10}, thus advocates the resolution of various information processing tasks directly in the compressive measurement domain rather than the original possibly high-dimensional signal domain, which would entail resorting to full-scale signal reconstruction. In particular, Davenport \emph{et al.}~\cite{Davenport10} argue that the use of the conventional random compressive measurement strategies that are agnostic to the exact form of the original signal -- so applicable to a large class of signals -- is key to develop very efficient and flexible compressive sensing hardware that can be used for the acquisition and processing of a large variety of signals.

This paper aims to study in detail the performance of a particular compressive information processing task: the classification of (possibly high-dimensional) signals from a set of compressive linear and noisy measurements. This problem is fundamental to the broad fields of signal and image processing~\cite{Ashok08,Baheti08,Baheti09}, computer vision~\cite{Mairal08,Hitomi11} and machine learning~\cite{Wright09,Calderbank09,Chen12}, and pre-processing often relies on dimension reduction to increase the speed and reliability of classification as well as reduce the complexity and cost of data processing and computation. The question then becomes that of understanding how compressive measurements affect the classification performance as a function of the model parameters.

\subsection{Prior Work on Compressive Classification}

Compressive classification appears in the machine learning literature as feature extraction or supervised dimensionality reduction. 
For example, linear dimensionality reduction methods based on geometrical characterizations of the source have been developed, with linear discriminant analysis (LDA)~\cite{duda00} and principal component analysis (PCA)~\cite{duda00} just depending on second order statistics. In particular, LDA, which is one of the most well-known supervised dimensionality reduction methods~\cite{Fisher36}, addresses simultaneously the between-class scattering and the within-class scattering of the projected data. It has been proven that under mild conditions this method is Bayes optimal~\cite{Hamsici08}. However, this method has two disadvantages: i) the dimensionality of the projected space can only be less than the number of data classes, which greatly restricts its applicability; and ii) LDA only uses first and second order statistics of the data, ignoring the higher-order information. Linear dimensionality reduction methods based on higher-order statistics of the data have therefore also been developed~\cite{William12,Chen12,Erdogmus04,Hild06,Kaski03,Liu12,Nenadic07,Tao09,Torkkola01,Torkkola03,Wright09}. In particular, an information-theoretic supervised dimensionality reduction inspired approach, which uses the mutual information between the data class labels and the data projections~\cite{Chen12,William12} or approximations of the mutual information such as quadratic mutual information (with quadratic R\'enyi entropy) ~\cite{Torkkola01,Torkkola03,Hild06} as a criterion to linearly reduce dimensionality, have been shown to lead to state-of-the-art classification results. The rationale for using mutual information relates to the fact that the Bayes classification error is bounded by the mutual information (based on a Shannon entropy measure) between the data labels and the data projections~\cite{Nenadic07}. 
In addition, nonlinear (supervised) dimensionality reduction methods have become popular recently~\cite{Tenenbaum00}.

In turn,~\cite{Calderbank09} and~\cite{CalderbankJafarpour12} consider compressed learning, i.e. learning directly in the compressed domain rather than in the original data domain. Of particular relevance, bearing witness to the value of the compressive classification paradigm, it is shown that the linear kernel support vector machine (SVM) classifier (trained and working) in the measurement domain, with high probability, has true accuracy close to the accuracy of the best linear threshold classifier (trained and working) in the data domain. It is also shown that for a family of well-known compressed sensing matrices, compressed learning is universal, in the sense that learning and classification in the measurement domain works provided that the data are sparse in some, even unknown, basis -- therefore, compressed sensing matrices also provide the desired properties of good linear dimensionality reduction matrices. That is, compressed sensing can be used as an efficient transformation from the data domain to the measurement domain (when the data admits a sparse representation even in an unknown basis) that preserves the learnability (and separability) of the data while bypassing the computational cost and curse of dimensionality.

Compressive classification also appears in the compressive information processing literature in view of recent advances in compressive sensing~\cite{Davenport10,Wright09,Duarte07,Davenport07,Duarte06,Haupt06,Haupt07b,Wim12,Krishnamurthy10a,Krishnamurthy10b}. References~\cite{Duarte07} and~\cite{Davenport07} explore random compressive measurements to perform manifold-based image classification. References~\cite{Duarte06},~\cite{Haupt06},~\cite{Haupt07b} and~\cite{Wim12} study the performance of compressive detection and compressive classification. 
Reference \cite{Davenport10} considers various compressive information processing problems, including compressive detection and compressive classification. 
References \cite{Krishnamurthy10a} and~\cite{Krishnamurthy10b} consider the problem of detection of spectral targets based on noisy incoherent projections. Reference~\cite{Wright09} notes that a small number of random measurements captures sufficient information to allow robust face recognition. The common thread in this line of research relates to the demonstration that the detection and classification problems can be solved directly in the measurement domain, without requiring the transformation of the data from the compressive to the original data domain, i.e. without requiring the reconstruction of the data.

Other works associated with compressive classification that have arisen in the computational imaging literature, and developed under the rubric of task-specific sensing, include~\cite{Neifeld07,Ashok08,Baheti08,Baheti09,Ke10,Duarte-Carvajalino11,Duarte-Carvajalino13}. In particular, task-specific sensing, which advocates that the sensing procedure has to be matched to the task-specific nature of the sensing application, has been shown to lead to substantial gains in performance over compressive sensing in applications such as localization~\cite{Neifeld07}, target detection~\cite{Ashok08}, (face) recognition~\cite{Baheti08,Baheti09}, and reconstruction~\cite{Ke10}.

Another instance of compressive classification appears under the guise of sparse support recovery (also known as model selection) associated with compressive sensing problems \cite{Aeron2010,Akcakaya2009,Fletcher09,Rad11,Reeves11,Reeves12,Reeves13,Tang10,Tulino13,Wang10,Wainwright09a,Wainwright09b,Zhao06}.  
These works provide necessary and sufficient (high-dimensional) scalings on the triplet $(n,s,m)$, where $n$ relates to the signal dimension, $s$ relates to the signal sparsity and $m$ is the number of linear measurements, for successful or partially successful support recovery under various metrics, various decoders (optimal and sub-optimal decoders), various measurement matrices, and various sparsity regimes (e.g. the linear sparsity regime and the sub-linear sparsity regime). 
References \cite{Reeves11,Reeves12} and~\cite{Reeves13} provide a more refined analysis associated with the tradeoff between the number of measurements and the number of detection errors in the high-dimensional setting, where the sparsity rate (i.e. the fraction of nonzero entries) and the per-sample signal-to-noise ratio (SNR) are finite constants, independent of the signal dimension.

\subsection{A Characterization of the Incremental Value of Measurement}

This paper provides a finer grained characterization of the performance of a compressive classification system. It goes beyond the phase-transition inspired high-dimensional characterizations that describe as a function of the source signal and system parameters whether or not one can reliably recover or partially recover the sparse signal support (i.e. presence or absence of a misclassification probability floor)~\cite{Aeron2010,Akcakaya2009,Fletcher09,Rad11,Reeves11,Reeves12,Reeves13,Tang10,Tulino13,Wang10,Wainwright09a,Wainwright09b,Zhao06}. It quantifies the information that can be learned from incremental measurement as a function of SNR through the concepts of diversity-order and measurement gain.

\begin{enumerate}[a)] 
\item The diversity-order determines the decay (in a $\log\left(\mathrm{SNR}\right)$ scale) of the misclassification probability at low-noise levels (high-SNR levels);

\item The measurement gain determines the horizontal offset (in a $\log\left(\mathrm{SNR}\right)$ scale) of the misclassification probability at low-noise levels (high-SNR levels), i.e. the measurement gain distinguishes further characterizations that exhibit identical diversity-order.
\end{enumerate}

These metrics, which determine the power of measurement to discriminate classes, will be naturally described in terms of the geometry of the source and the measurement system, and this description will reveal how measurement leads to separation of classes. We suggest that these metrics be used as proxies when designing classification systems.

\emph{Enter wireless communications.} The performance characterizations we propose for the compressive classification problem are parallel to those adopted in the wireless communications field~\cite{Tarokh98,Tarokh99,Tse05,ZheTse03}: diversity-order and coding gain in wireless communications problems are the counterparts of the diversity-order and measurement gain in the compressive classification problem. 
The reason for this correspondence is the signal model; the assumption that measurement noise is Gaussian and that the distribution of the source under each class (hypothesis) is multivariate Gaussian with a certain known mean and a certain known (possibly rank-deficient) covariance matrix. It is the source model that leads to a fundamental duality between compressive classification and non-coherent wireless communication with multiple antennas [66]-[70]. This duality means that fundamental limits derived in the wireless domain can be transferred to the classification domain and vice versa.\footnote{These dualities also lead to characterizations that disclose fundamental tradeoffs between the performance of a compressive classification problem (via the diversity-order) and the number of classes in the compressive classification problem (a so-called discrimination gain). Such a diversity-discrimination tradeoff~\cite{Nokleby13}, which offers a more illuminating view of the performance of the compressive classification, is reminiscent of the diversity-multiplexing tradeoff  in multiple-antenna wireless communications~\cite{Zheng02b,ZheTse03}.} There are two other notable features associated with this model:

\begin{enumerate}[a)] 

\item The source model adopted for the compressive classification problem also relates to well-known models adopted in compressive sensing, most notably the Gaussian mixture model (GMM). 
The GMM~\cite{Chen10,Yu11,Yu12,Duarte-Carvajalino13}, which is typically used in conjunction with the Bayesian compressed sensing formalism~\cite{Ji08}, relates to various well-known structured models in the literature including union of sub-spaces~\cite{Blumensath09,Stojnic09,Eldar09,Eldar10}, wavelet trees~\cite{Blumensath09,Baraniuk10} or manifolds~\cite{Baraniuk09,Chen10}, that aim to capture additional signal structure beyond primitive sparsity in order to yield further gains. For example, a (low-rank) GMM can be seen as a Bayesian counterpart of the union of subspaces model. In fact, a signal drawn from a (low-rank) GMM lies in a union of subspaces, where each subspace corresponds to the image of each class conditioned covariance matrix in the model.\footnote{More generally, a signal drawn from a GMM model lies in a union of affine spaces rather than linear subspaces, where each affine space is associated with the mean and covariance of each class in the GMM model.} A low-rank GMM can also be seen as an approximation to a compact manifold. Compact manifolds can be covered by a finite collection of topological disks that can be represented by high-probability ellipsoids living on the principal hyperplanes corresponding to the different components of a low-rank GMM~\cite{Chen10}. However, one of the key advantages of adopting a GMM, {\it in lieu} of these other structured models, is associated with the fact that reconstruction of a signal drawn from a GMM from compressive linear measurements in Gaussian noise can be very effectively performed via a closed-form inversion formula~\cite{Chen10}.

\item This source model also leads to state-of-the-art results in various compressive classification scenarios such as character and digit recognition as well as image classification~\cite{Chen12,Torkkola01,Torkkola03}. Other successful instances -- beyond classification -- associated with the use of GMMs include problems in image processing, such as interpolation, zooming, deblurring~\cite{Yu11,Yu12,Duarte-Carvajalino13}, and problems in dictionary learning~\cite{Chen10}.

\end{enumerate}

\subsection{Contributions of the Article}

The main contributions include:
\begin{enumerate}
\item  Characterization of the behavior of an upper bound to the probability of misclassification for linear random Gaussian measurements contaminated by white Gaussian noise. The characterization unveils how the error floor, the diversity-order and the measurement gain behave as a function of the properties of the source (i.e. the number of classes and the means and covariances of the class conditioned multivariate Gaussian distributions);

\item Characterization of the behavior of an upper bound to the probability of misclassification for linear designed measurements contaminated by Gaussian noise. In particular, we construct projection designs that maximize the diversity-order 
subject to a certain measurement budget as a function of the properties of the source;

\item Extension of the performance characterizations from the low-noise to the high-noise regime. Such an extension showcases the key operational differences between the two regimes;

\item Connection of the performance behavior of the compressive classification problem, with random or designed measurements, to certain geometrical quantities associated with the measurement system and associated with the source, in order to provide additional insight.

\end{enumerate}

These contributions differ from other related contributions in the prior work on compressive classification (including the literature on sparse support recovery) in various aspects:

\begin{enumerate}

\item Prior work on performance measures  has focused on memoryless source models (e.g. the Bernoulli-Gaussian model or the models akin to the Bernoulli-Gaussian model associated with sparse support recovery problems)~\cite{Aeron2010,Akcakaya2009,Fletcher09,Rad11,Reeves11,Reeves12,Reeves13,Tang10,Tulino13,Wang10,Wainwright09a,Wainwright09b,Wim12,Wu12,Zhao06}; in contrast, we define the behavior of the probability of misclassification for source models that inherently incorporate memory;

\item Prior analysis of performance measures is typically asymptotic in various of the problem dimensions (e.g. the signal dimension, the signal sparsity, the number of measurements and certain signal parameters for certain scalings)~\cite{Aeron2010,Akcakaya2009,Fletcher09,Rad11,Reeves11,Reeves12,Reeves13,Tang10,Tulino13,Wang10,Wainwright09a,Wainwright09b,Wu12,Zhao06}; in contrast, we define the behavior of the probability of misclassification non-asymptotically in various of the problem dimensions and asymptotically only in the signal-to-noise ratio;

\item Prior work has concentrated on identifying phase transitions, whereas we concentrate on characterizations that articulate about the presence of absence of an error floor, the diversity-order and the measurement gain.

\end{enumerate}

Some elements of our approach are present in the characterizations presented in~\cite{Wim12,Davenport10,Haupt06,Haupt07b} but their focus is not on the incremental value of measurement to classification. In particular, the additional refinement offered by our approach can also be appreciated by specializing our setting, where the source signal lives in a union of sub-spaces, to the traditional sparse recovery setting, where the source signals are taken to be sparse in the canonical basis. For example, let us consider a $n$-dimensional {GMM} source formed by $\sum_{i=1}^k\binom{n}{i}$ equiprobable classes, where each Gaussian class is represented by a zero-mean vector and a diagonal covariance matrix with rank less than or equal to $k$, whose nonzero elements correspond to one out of the $\sum_{i=1}^k\binom{n}{i}$ possible supports of cardinality less than or equal to $k$ in $\mathbb{R}^n$. The results in~\cite{Wainwright09b,Wang10} show that it is possible to recover the sparse signal support pattern (classify the signal sub-space) with $\mathcal{O}(k \log n)$ measurements\footnote{Note that these results hold in particular when the minimum amplitude of the nonzero entries in the input vector is assumed to decrease as $\mathcal{O}(1/\sqrt{k})$ and for a fixed value of the noise power $\sigma^2$.}. On the other hand, our results show that it is possible to classify the signal sub-space with exactly $k$ random measurements when the noise power tends to zero. In addition, our results also lead to a sharper characterization of the behavior of the misclassification probability: it decays with at least a slope of $1/4$ on a log(SNR) scale.

\subsection{Structure of the Article}
This article is structured as follows: Section~\ref{PS} introduces the compressive classification problem and the associated system, signal and noise models. Sections~\ref{two_class} and~\ref{mul_class} study the performance of compressive classification for the two-class and the multiple-class scenarios, respectively, with random measurements in the regime of low noise. In turn, Section~\ref{sec:design} studies the performance of two- and multiple-class compressive classification problems with designed measurements also in the low-noise regime. The purpose of Section~\ref{high_noise} is to extend the analysis from the low-noise to the high-noise regime. A comprehensive set of numerical results that sheds further light on the performance of compressive classification problems is provided in Section~\ref{num_res}. Concluding remarks are made in Section~\ref{conclusions}. The technical proofs of the main results are delegated to the Appendices.

\subsection{Notation}
The article adopts the following notation: boldface upper-case letters denote matrices $\left(\mathbf{X}\right)$, boldface lower-case letters denote column vectors $\left(\mathbf{x}\right)$ and italics denote scalars $\left(x\right)$; the context defines whether the quantities are deterministic or random.  $\mathbf{I}_{N}$ represents the $N \times N$ identity matrix, $\B{0}_{M \times N}$ represents the $M \times N$ zero matrix (the subscripts that refer to the dimensions of such matrices will be droppep when evident from the context) and $\mathrm{diag} \left(a_1,a_2,\ldots,a_N\right)$ represents an $N \times N$ diagonal matrix with diagonal elements $a_1, a_2,\ldots,a_N$. 
The operators $\left(\cdot\right)^T$ , $\mathrm{rank}\left(\cdot\right)$, $\mathrm{det}\left(\cdot\right)$, $\mathrm{pdet}\left(\cdot\right)$ and $\mathrm{tr}\left(\cdot\right)$ represent the transpose operator, the rank operator, the determinant operator, the pseudo-determinant operator and the trace operator, respectively. $\mathrm{Null}\left(\cdot\right)$ and $\mathrm{im}\left(\cdot\right)$ denote the null space and the (column) image of a matrix, respectively, and $\mathrm{dim}\left(\cdot\right)$ denotes the dimension of a linear subspace. $\mathcal{E}\left\{\cdot\right\}$ represents the expectation operator. The multi-variate Gaussian distribution with mean $\B{\mu}$ and covariance matrix $\B{\Sigma}$ is denoted by $\mathcal{N} \left(\B{\mu},\B{\Sigma}\right)$. $\log\left(\cdot\right)$  denotes the natural logarithm. The article also uses the big $\mathcal{O}$ notation where $g\left(x\right) = \mathcal{O}\left(f\left(x\right)\right)$ if $\displaystyle \lim_{x \to \infty} \frac{g\left(x\right)}{f\left(x\right)} = c$, where $c$ is a constant; and the little $o$ notation where $g\left(x\right) = o\left(f\left(x\right)\right)$ if $\displaystyle \lim_{x \to \infty} \frac{g\left(x\right)}{f\left(x\right)} = 0$.

\section{The Compressive Classification Problem}
\label{PS}
We consider the standard measurement model given by:
\begin{IEEEeqnarray}{c}
\label{s_model}
{\bf y} =  {\bf \Phi} {\bf x} + {\bf n},
\end{IEEEeqnarray}
where ${\bf y} \in \mathbb{R}^M$ represents the measurement vector, $\mathbf{x} \in \mathbb{R}^N$ represents the source vector, ${\bf \Phi} \in \mathbb{R}^{M \times N}$ represents the measurement matrix or kernel\footnote{We refer to ${\bf \Phi}$ as the measurement, sensing or projection matrix/kernel interchangeably throughout the paper.} and $\mathbf{n} \sim \mathcal{N} \left({\bf 0},\sigma^2 \cdot {\bf I}\right) \in \mathbb{R}^M $ represents standard white Gaussian noise. We consider both random measurement kernel designs, where the entries of ${\bf \Phi}$ are drawn i.i.d. from a zero-mean, fixed-variance, Gaussian distribution, which is common in the CS literature~\cite{Candes06a,Donoho06b} as well as optimized kernel designs that aim to improve the classification performance.

We suppose that the source is described by a Gaussian Mixture Model (GMM); the source signal is drawn from one of $L$ classes, the \emph{a priori} probability of class $i$ is $P_i$ and the distribution of the source conditioned on class $i$ is Gaussian with mean $\B{\mu}_i \in \mathbb{R}^N$ and (possibly rank-deficient) covariance matrix ${\bf \Sigma}_i \in  \mathbb{R}^{N \times N}$. We should point out that we use a low-rank modelling approach even though natural signals (e.g. patches extracted from natural images) are not always low-rank but rather ``approximately'' low-rank \cite{Chen10}. The justification for the use of such low-rank modeling approach is two-fold: first, a low-rank representation is often a very good approximation to real scenarios, particularly as the eigenvalues of the class conditioned covariances often decay rapidly; second, it is then standard practice to account for the mismatch between the low-rank and the ``approximately'' low-rank model by adding extra noise in the measurement model in (\ref{s_model}) (see~\cite{RecJournal}).

Our objective is then to produce an estimate of the true signal class given the measurement vector. The Maximum-A-Posteriori (MAP) classifier, which minimizes the probability of misclassification~\cite{duda00}, produces the estimate given by:
\begin{IEEEeqnarray}{LcR}
\hat{C} = {\arg} {\max_{C \in \{1, \cdots, L\}} P\left(C \mid {\bf y}\right)} & = &  \arg \max_{C \in \{1, \cdots, L\}} p\left({\bf y} \mid C \right) P_{C},
\label{class}
\end{IEEEeqnarray}
where $P(C \mid \mathbf{y})$ represents the \textit{a posteriori} probability of class $C$ given the measurement vector $\mathbf{y}$  and $p(\mathbf{y} \mid C)$ represents the probability density function of the measurement vector $\mathbf{y}$ given the class $C$.

Our performance analysis concentrates both on the low-noise regime, where $\sigma^2 \to 0$, and on the high-noise regime, where $\sigma^2 \to \infty$. The performance analysis -- in line with the standard practice in multiple-antenna communications systems~\cite{Tarokh98,Tse05}  --  also concentrates on the asymptotic behavior of an upper bound to the probability of misclassification of the MAP classifier ${P}_{err}^{UB}$ , rather than the exact probability of misclassification of such classifier $P_{err}$. In the high-noise regime, we use standard Taylor series expansions to describe the asymptotic behavior of the probability of misclassification.

In contrast, in the low-noise regime, we use two measures that describe the low-noise asymptotics of the upper bound to the probability of misclassification. In particular, we define the (lower bound to the) diversity-order of the measurement model in~(\ref{s_model}) as:
\begin{equation}
	d = \lim_{\sigma^2 \to 0} \frac{\log{P}_{err}^{UB}(\sigma^2)}{\log{\sigma^2}},
\label{diversity}	
\end{equation}
that determines how (the upper bound to) the misclassification probability decays (in the $\log{\sigma^2}$ scale) at low-noise levels~\cite{ZheTse03,Paul03}. We also define the (lower bound to the) measurement gain of the measurement model in~(\ref{s_model}) as:
\begin{IEEEeqnarray}{LcR}
{g_m} =  \lim_{\sigma^2 \rightarrow 0}  \sigma^2 \cdot \frac{1}{ \sqrt[d]{{P}_{err}^{UB}(\sigma^2)}  } ,
\label{gm}
\end{IEEEeqnarray}
that determines the power offset of (the upper bound to) the misclassification error probability at low-noise levels: note that the measurement gain refines the asymptotic description of the upper bound to the misclassification probability, by distinguishing further characterizations that exhibit identical diversity-order.

The characterization of the performance measures in (\ref{diversity}) and (\ref{gm}) will be expressed via quantities that relate to the geometry of the measurement model, namely, the rank and the pseudo-determinant of certain matrices. In particular, we define the behavior of (\ref{diversity}) and (\ref{gm}) in two ways. The geometric interplay of the linear transformation with the class covariance matrices is described by the quantities::
\begin{equation}
r_i = \mathrm{rank} (\mathbf{\Phi} \mathbf{\Sigma}_i \mathbf{\Phi}^T)
\end{equation}
and
\begin{equation}
v_i = \mathrm{pdet} (\mathbf{\Phi} \mathbf{\Sigma}_i \mathbf{\Phi}^T) 
\end{equation}
which measure the dimension of the sub-space spanned by the linear transformation of the signals in class $i$ and the volume of the parallelepiped in $\mathbb{R}^M$ where those signals are mostly concentrated, respectively; and the quantities:
\begin{equation}
r_{ij} = \mathrm{rank} (\mathbf{\Phi} (\mathbf{\Sigma}_i + \mathbf{\Sigma}_j) \mathbf{\Phi}^T)
\end{equation}
and
\begin{equation}
v_{ij} = \mathrm{pdet} (\mathbf{\Phi} (\mathbf{\Sigma}_i + \mathbf{\Sigma}_j) \mathbf{\Phi}^T)
\end{equation}
which measure the dimension of the direct sum of sub-spaces spanned by the linear transformation of the signals in classes $i$ or $j$ and the volume of the parallelepiped in $\mathbb{R}^M$ where signals in classes $i$ and $j$ are mostly concentrated, respectively. 

In addition, we also define the behavior of the measures in (\ref{diversity}) and (\ref{gm}) via the geometry of the original source signal, by using the quantities:
\begin{equation}
r_{\mathbf{\Sigma}_i} = \mathrm{rank}(\mathbf{\Sigma}_i)
\end{equation}
which relates to the dimension of the sub-space spanned by source signals in class $i$ and
\begin{equation}
r_{\mathbf{\Sigma}_{ij}} = \mathrm{rank}(\mathbf{\Sigma}_i +  \mathbf{\Sigma}_j)
\end{equation}
which relates to the dimension of the direct sum of sub-spaces spanned by source signals in classes $i$ or $j$. Finally, we will also be using the quantity:
\begin{equation}
\label{nodim}
NO_{Dim} = r_{\B{\Sigma}_{ij}} - \left[\left(r_{\B{\Sigma}_{i}} + r_{\B{\Sigma}_{j}}\right) - r_{\B{\Sigma}_{ij}}\right]
\vspace{-0.05cm}
\end{equation}
that relates to the difference between the dimension of the  sub-spaces spanned by source signals in classes $i$ or $j$ and the dimension of the intersection of such sub-spaces. This can also be interpreted as the number of non-overlapping dimensions between the sub-spaces spanned by the eigenvectors of the covariance matrices pertaining to the two classes.

It turns out - as shown in the sequel - that the asymptotics of the upper bound to the misclassification probability mimic closely the behavior of the true misclassification probability, attesting to the value of the approach.

\section{Two-Class Compressive Classification with Random Measurements\\ in the Low-Noise Regime}
\label{two_class}

We set $L=2$ and consider two-class compressive classification using random measurements. The misclassification probability associated to the MAP classifier can be written as
\begin{equation}
P_{err} = \int_{-\infty}^{+\infty}  \min \left(  P_{1} \cdot p(\mathbf{y}| C=1),   P_{2} \cdot p(\mathbf{y}| C=2) \right)  d \mathbf{y}.
\end{equation}
and, by exploiting the fact that $\min\left(a,b\right) \leq a^t b^{1-t}, ~~ 0 \leq t \leq 1, ~~ a,b>0$, an upper bound to the misclassification probability can also be immediately written as\cite{duda00}:
\begin{equation}
\label{P_err_Chern}
{P}_{err} \leq P_{1}^t P_{2}^{1-t}~ \int_{-\infty}^{+\infty}  p^t\left({\bf y} \mid C=1\right)  p^{1-t}\left({\bf y} \mid C=2\right) d{\bf y}, ~~ 0 \leq t \leq 1.
\end{equation}

It turns out that the Bhattacharyya bound~\cite{Bhat43}, which corresponds to a specialization of the bound in~\eqref{P_err_Chern} for $t=0.5$, admits the closed-form expression for class-conditioned Gaussian distributions given by: \footnote{Note that Bhattacharyya upper bound corresponds to a value of $t = 0.5$ in~\eqref{P_err_Chern} whereas the Chernoff bound corresponds to the value of t that minimizes~\eqref{P_err_Chern}.}
\begin{IEEEeqnarray}{rCl}
\label{P_err_Bhat}
{P}_{err}^{UB} & = & \sqrt{P_{1}P_{2}}~ \int_{-\infty}^{+\infty}  \sqrt{p\left({\bf y} \mid C=1\right)  p\left({\bf y} \mid C=2\right)} d{\bf y} 
 = \sqrt{P_{1}P_{2}}~ e^{-K_{12}},
\end{IEEEeqnarray}
where
\begin{IEEEeqnarray}{rCl}
\nonumber
{K_{ij}} & =  &   \frac{1}{8}{\left[{\bf \Phi}\left(\boldsymbol{\mu}_i - \boldsymbol{\mu}_j\right)\right]^T \left[\frac{{\bf \Phi}\left({\bf \Sigma}_i+{\bf \Sigma}_j\right){\bf \Phi}^T + 2 \sigma^2 \mathbf{I}}{2}\right]^{-1} \left[{\bf \Phi}\left(\boldsymbol{\mu}_i - \boldsymbol{\mu}_j\right)\right]  } \\ 
&& { + \frac{1}{2}\log\frac{\mathrm{det} \left(\frac{{\bf \Phi}\left({\bf \Sigma}_i+{\bf \Sigma}_j\right){\bf \Phi}^T + 2 \sigma^2 \mathbf{I}}{2}\right)}{\sqrt{\mathrm{det} \left({\bf \Phi}{\bf \Sigma}_i{\bf \Phi}^T + \sigma^2 \mathbf{I}\right)  \mathrm{det} \left({\bf \Phi}{\bf \Sigma}_j{\bf \Phi}^T + \sigma^2 \mathbf{I}\right)}} }.
\label{exp_Bhat}
\end{IEEEeqnarray}

This Bhattacharyya based upper bound on the probability of misclassification encapsulated in (\ref{P_err_Bhat}) and (\ref{exp_Bhat}) is the basis of our analysis. We treat the case where the classes are zero-mean, i.e. $\B{\mu}_1 = \B{\mu}_2 = \B{0}$, and the case where classes are nonzero-mean, i.e. $\B{\mu}_1 \neq \B{0}$ or $\B{\mu}_2 \neq \B{0}$, separately. The zero-mean case exhibits the main operational features associated with the compressive classification problem; the nonzero-mean case can also exhibit additional operational features, e.g. infinite diversity-order.

\subsection{Zero-Mean Classes}

The following Theorem offers a view of the asymptotic behavior of the (upper bound to the) probability of misclassification for the two-class compressive classification problem with zero-mean classes, by leveraging directly the geometry of the linear transformation of the source signal effected by the measurement kernel. Note that, since the matrices $\mathbf{\Phi} \mathbf{\Sigma}_1 \mathbf{\Phi}^T$ and $\mathbf{\Phi} \mathbf{\Sigma}_2 \mathbf{\Phi}^T$ are positive semidefinite, it is straightforward to show that $\frac{r_1 + r_2}{2} \leq r_{12}$; in addition, it is also possible to show that $\frac{r_1 + r_2}{2} = r_{12}$ if and only if $\mathrm{im} (\mathbf{\Phi} \mathbf{\Sigma}_1 \mathbf{\Phi}^T) = \mathrm{im} (\mathbf{\Phi} \mathbf{\Sigma}_2 \mathbf{\Phi}^T)$, so that the two sub-spaces associated with the two classes overlap completely~\cite[Lemma 2]{RecJournal}.

\vspace{0.30cm}
\begin{theorem}
\label{theorem1}
Consider the measurement model in \eqref{s_model} where ${\bf x} \sim \mathcal{N} ({\bf 0},{\bf \Sigma}_1)$ with probability $P_{1}$ and ${\bf x} \sim \mathcal{N} ({\bf 0},{\bf \Sigma}_2)$ with probability $P_{2} = 1 - P_{1}$. Then, in the regime of low noise where $\sigma^2 \to 0$, the upper bound to the probability of misclassification in \eqref{P_err_Bhat} behaves as:
\begin{itemize}
  \item	If $\frac{r_1+r_2}{2} = r_{12}$ then, ~$\displaystyle {P}_{err}^{UB} =  \mathcal{O}\left(1\right),~~\sigma^2 \to 0$;\\
\end{itemize}

\begin{itemize}
  \item If $\frac{r_1+r_2}{2} < r_{12}$ then, ~$\displaystyle {P}_{err}^{UB} = \left(\frac{g_m}{\sigma^2} \right) ^ {- d} + o \left(\left(\frac{1}{\sigma^2}\right)^{- d}\right),~~\sigma^2 \to 0$, where:
\end{itemize}

\begin{equation}
\label{diver}
	d = -\frac{1}{2}\left(\frac{r_1+r_2}{2}- r_{12}\right)
\end{equation}
and 
\begin{equation}
\label{measure}
{g_m} = 
{ \left[2^{\frac{r_{12}}{2}} \sqrt{P_{1}P_{2}}\left[\frac{v_{12}}{\sqrt{v_1 v_2}}\right]^{-\frac{1}{2}}\right]^{-\frac{1}{d}}.
}
\end{equation}
\end{theorem}

\begin{IEEEproof}
The proof is presented in Appendix \ref{app_A}.
\end{IEEEproof}
\vspace{0.5cm}

It is now instructive to probe further onto the characterizations embodied in Theorem \ref{theorem1} to infer some operational features associated with the two-class compressive classification problem. Such characterization admits a very simple interpretation:

\begin{itemize}

\item If $\frac{r_1 + r_2}{2} = r_{12}$, then the sub-spaces spanned by the linear transformation of the signals in classes $C=1$ and $C=2$ overlap completely -- the upper bound to the misclassification probability exhibits an error floor because it is not possible to distinguish the classes perfectly as the noise level approaches zero;

\item If $\frac{r_1 + r_2}{2} < r_{12}$, then the sub-spaces spanned by the linear transformation of the signals in classes $C=1$ and $C=2$ do not overlap completely -- the upper bound to the misclassification error probability (and the true error probability) does not exhibit an error floor as it is possible to distinguish the classes perfectly as the noise level approaches zero. The lower the degree of overlap, the higher the diversity-order -- this is measured via the interplay of the various ranks, $r_1$, $r_2$ and $r_{12}$ in terms of the difference between the dimensions of the sub-spaces corresponding to the linear transformation of the signals in classes $C=1$ and $C=2$ and the dimension of their intersection; in fact, it can be shown that:
\begin{IEEEeqnarray}{rCl}
\nonumber
2r_{12} -r_1 -r_2 &  = & \dim \mathrm{im}\left(\mathbf{\Phi} \mathbf{\Sigma}_1 \mathbf{\Phi}^T \right) - \dim\left( \mathrm{im}\left(\mathbf{\Phi} \mathbf{\Sigma}_1 \mathbf{\Phi}^T \right)   \cap  \mathrm{im}\left(\mathbf{\Phi} \mathbf{\Sigma}_2 \mathbf{\Phi}^T \right)  \right)   \\
&& +   \dim \mathrm{im}\left(\mathbf{\Phi} \mathbf{\Sigma}_2 \mathbf{\Phi}^T \right) - \dim\left( \mathrm{im}\left(\mathbf{\Phi} \mathbf{\Sigma}_1 \mathbf{\Phi}^T \right)   \cap  \mathrm{im}\left(\mathbf{\Phi} \mathbf{\Sigma}_2 \mathbf{\Phi}^T \right)  \right).
\label{diffdim}
\end{IEEEeqnarray} 

\end{itemize}

The following Theorem now describes the asymptotic behavior of the probability of misclassification for the two-class compressive classification problem with zero-mean classes, by leveraging directly the geometry of the source signals -- this has the advantage of showcasing how the number of measurements together with the properties of the source affect performance. The result uses the fact that $N \geq r_{{\bf \Sigma}_{12}} \geq \max\left(r_{{\bf \Sigma}_1},r_{{\bf \Sigma}_2}\right)$ and, with  probability 1, $r_1 = \min\left(M,r_{{\bf \Sigma}_1}\right)$, $r_2= \min\left(M,r_{{\bf \Sigma}_2}\right)$ and 
$r_{12} = \min\left(M,r_{{\bf \Sigma}_{12}}\right)$. The result also assumes, without loss of generality, that $r_{{\bf \Sigma}_1} \leq r_{{\bf \Sigma}_2}$. Once more, note that since the matrices $\mathbf{\Sigma}_1$ and $\mathbf{\Sigma}_2$ are positive semidefinite, it follows that $\frac{r_{{\bf \Sigma}_{1}} + r_{{\bf \Sigma}_{2}}}{2} \leq r_{{\bf \Sigma}_{12}}$, and that $\frac{r_{{\bf \Sigma}_{1}} + r_{{\bf \Sigma}_{2}}}{2} = r_{{\bf \Sigma}_{12}}$ if and only if $\mathrm{im} (\mathbf{\Sigma}_1) =\mathrm{im} (\mathbf{\Sigma}_2) $~\cite[Lemma 2]{RecJournal}.

\vspace{0.30cm}
\begin{theorem}
\label{theorem2}
Consider the measurement model in \eqref{s_model} where ${\bf x} \sim \mathcal{N} ({\bf 0},{\bf \Sigma}_1)$ with probability $P_{1}$ and ${\bf x} \sim \mathcal{N} ({\bf 0},{\bf \Sigma}_2)$ with probability $P_{2} = 1 - P_{1}$. Then, in the regime of low noise where $\sigma^2 \to 0$, the upper bound to the probability of misclassification in \eqref{P_err_Bhat} behaves as:
\begin{itemize}
\item If $\frac{r_{{\bf \Sigma}_1}+r_{{\bf \Sigma}_2}}{2} = r_{{\bf \Sigma}_{12}}$ then, ~$\displaystyle  {P}_{err}^{UB} =  \mathcal{O}\left(1\right),~~\sigma^2 \to 0$; \\

  \item  If $\frac{r_{{\bf \Sigma}_1}+r_{{\bf \Sigma}_2}}{2} < r_{{\bf \Sigma}_{12}}$ then, 

\textendash~~~when $M \leq r_{{\bf \Sigma}_1} \leq r_{{\bf \Sigma}_2} \leq r_{{\bf \Sigma}_{12}}$, ~$\displaystyle {P}_{err}^{UB} =  \mathcal{O}\left(1\right),~~\sigma^2 \to 0$;

\vspace{0.15cm}

\textendash~~~otherwise, ~$\displaystyle {P}_{err}^{UB} = \left(\frac{g_m}{\sigma^2} \right) ^ {- d} + o \left(\left(\frac{1}{\sigma^2}\right)^{- d}\right),~~\sigma^2 \to 0$
where the measurement gain is given by:
\begin{equation}
{g_m} = 
{ \left[2^{\frac{\min(M,r_{\mathbf{\Sigma}_{12}})}{2}}\sqrt{P_{1}P_{2}}\left[\frac{v_{12}}{\sqrt{v_1 v_2}}\right]^{-\frac{1}{2}}\right]^{-\frac{1}{d}}
}
\label{gm2}
\end{equation}
and the diversity-order is given by:
\begin{equation}
d = \threecases{-\frac{1}{2}\left(\frac{r_{{\bf \Sigma}_1} - M}{2}\right)}{if $r_{{\bf \Sigma}_1} < M \leq r_{{\bf \Sigma}_2} \leq r_{{\bf \Sigma}_{12}}$}{-\frac{1}{2}\left(\frac{r_{{\bf \Sigma}_1}+r_{{\bf \Sigma}_2}}{2}- M\right)}{if $r_{{\bf \Sigma}_1} \leq r_{{\bf \Sigma}_2} < M < r_{{\bf \Sigma}_{12}}$}{-\frac{1}{2}\left(\frac{r_{{\bf \Sigma}_1}+r_{{\bf \Sigma}_2}}{2}- r_{{\bf \Sigma}_{12}}\right)}{if $r_{{\bf \Sigma}_1} \leq r_{{\bf \Sigma}_2} \leq r_{{\bf \Sigma}_{12}} \leq M$}.
\end{equation}

\end{itemize}

\end{theorem}

\begin{IEEEproof}
The proof is presented in Appendix \ref{app_B}.
\end{IEEEproof}
\vspace{0.5cm}

Theorem~\ref{theorem2} provides insight into the interplay between the number of measurements and the source geometry. In particular:
\begin{itemize}

\item When the sub-spaces spanned by the signals in classes $C=1$ and $C=2$ overlap completely, i.e., $\frac{r_{{\bf \Sigma}_1}+r_{{\bf \Sigma}_2}}{2} = r_{{\bf \Sigma}_{12}}$, it is not possible to construct a random measurement kernel that will be able to distinguish between the signals from classes $C=1$ and $C=2$. In such scenario, the (upper bound to the) probability of misclassification will exhibit an error floor at low-noise levels.

\item When the sub-spaces spanned by the signals in classes $C=1$ and $C=2$ do not overlap completely, i.e., $\frac{r_{{\bf \Sigma}_1}+r_{{\bf \Sigma}_2}}{2} < r_{{\bf \Sigma}_{12}}$, the number of random measurements $M$ defines the behavior of the (upper bound to the) probability of misclassification as follows:

i) if  $M \leq r_{{\bf \Sigma}_{1}}$, the upper bound will exhibit an error floor at low-noise levels, because the random measurements will affect the signals in a way that the sub-spaces associated with the linearly transformed classes completely overlap -- that is, the random measurements do not provide the sufficient degrees of freedom to distinguish between the signals from classes $C=1$ and $C=2$. Note that in this case both the sub-spaces corresponding to the randomly projected classes occupy the entire space $\mathbb{R}^M$ thus they are completely overlapping;

ii) otherwise, if $M > r_{{\bf \Sigma}_{1}}$ the upper bound will not exhibit such an error floor at low-noise levels, since it is possible to randomly transform the original signals such that the corresponding sub-spaces do not overlap completely; in fact, it can be shown directly from~\eqref{diffdim} that the difference between the dimensions of these sub-spaces and that of their intersection is at least one;

iii) once again, note that the diversity-order is a function of the difference between the dimensions of the sub-spaces associated with the two classes and the dimension of their intersection, as given by \eqref{diver} and (\ref{diffdim}): by gradually increasing the number of measurements $M$ from 1 up to $r_{{\bf \Sigma}_{12}}$ it is possible to increase the diversity-order up to the maximum value $\frac{1}{4} NO_{Dim}$; however, increasing the number of measurements past $r_{{\bf \Sigma}_{12}}$ does not offer a higher diversity-order -- instead, it only translates into a higher measurement gain. In fact, as $M$ ranges from $1$ to $r_{{\bf \Sigma}_{12}}$ we can gradually unveil the ``degree of non-overlap" between the original sub-spaces because the number of non-overlapping dimensions in the projected domain approaches the number of non-overlapping dimensions in the original domain, achieving it when $M = r_{{\bf \Sigma}_{12}}$. In other terms, when $M=r_{{\bf \Sigma}_{12}}$, the performance of classification -- defined via the diversity-order -- in the measurement domain equals that in the data domain~\cite{CalderbankJafarpour12,Calderbank09}. In contrast, for  $M > r_{{\bf \Sigma}_{12}}$ projecting the original subspaces into an higher dimensional linear space does not increase the number of original non-overlapping dimensions. One then understands the role of measurement as a way to probe the differences between the classes, providing the degrees of freedom in order to distinguish between the signals from classes $C=1$ and $C=2$. 

\end{itemize}

On the other hand, the measurement gain is a function of the exact geometry of the classes in the Gaussian mixture model. It increases with the ratio of the product of the nonzero eigenvalues of  ${\bf \Phi}\left({\bf \Sigma}_1+{\bf \Sigma}_2\right){\bf \Phi}^T$ to the product of the nonzero singular values of ${\bf \Phi}{\bf \Sigma}_1{\bf \Phi}^T$ and ${\bf \Phi}{\bf \Sigma}_2{\bf \Phi}^T$.

\subsection{Nonzero-Mean Classes}

The following Theorem now generalizes the description of the asymptotic behavior of the probability of misclassification from the zero-mean to the nonzero-mean, two-class compressive classification problem.

\vspace{0.30cm}

\begin{theorem}
\label{theo:nonzero}
Consider the measurement model in (\ref{s_model}) where $\mathbf{x} \sim \mathcal{N} (\B{\mu}_1,\mathbf{\Sigma}_1)$ with probability $P_{1}$ and $\mathbf{x} \sim \mathcal{N} (\B{\mu}_2,\mathbf{\Sigma}_2)$ with probability $P_{2} = 1 - P_{1}$. If 
\begin{equation}
\B{\Phi} (\B{\mu}_1 - \B{\mu}_2) \notin \mathrm{im}( \mathbf{\Phi} (\mathbf{\Sigma}_1 + \mathbf{\Sigma}_2) \mathbf{\Phi}^{T}),
  \label{eq:im}
\end{equation}
then the upper bound to the probability of misclassification in (\ref{P_err_Bhat}) decays exponentially with $1/\sigma^2$ as $\sigma^2 \rightarrow 0$; otherwise,
\begin{equation}
								{P}_{err}^{UB} = \left(\frac{a \cdot g_m}{\sigma^2} \right) ^ {- d} + o \left(\left(\frac{1}{\sigma^2}\right)^{- d}\right),~~\sigma^2 \to 0
								\end{equation}
where $a$ is a finite constant which depends only on the first term in~(\ref{exp_Bhat}), with $a = 1$ for $\B{\mu}_1 = \B{\mu}_2$ and $a >1 $ for $\B{\mu}_1 \neq \B{\mu}_2$, whereas $g_m$ and $d$ are as in Theorems \ref{theorem1} and \ref{theorem2}.
\end{theorem}

\begin{IEEEproof}
The proof is presented in Appendix \ref{app_C}.
\end{IEEEproof}
\vspace{0.5cm}

The characterization embodied in Theorem~{\ref{theo:nonzero}} illustrates that the asymptotic behavior of the upper bound of the error probability for classes with nonzero-mean can be radically different from that for classes with zero-mean. The differences in behavior trace back to the fact that $M >r_{\mathbf{\Sigma}_{12}}$ represents a necessary condition for condition (\ref{eq:im}) to hold (see Appendix~\ref{app_C}). In the nonzero-mean case, choosing $M >r_{\mathbf{\Sigma}_{12}}$ (provided that condition (\ref{eq:im}) also holds) leads to a diversity-order $d = \infty$; in contrast,  in the zero-mean case choosing $M >r_{\mathbf{\Sigma}_{12}}$ does not affect the diversity-order and only affects the measurement gain. In contrast, letting $M  \leq r_{\mathbf{\Sigma}_{12}}$ induces the same diversity-order both for nonzero-mean and zero-mean classes; the presence of the nonzero-mean here may only impact the measurement gain since $a \geq 1$.

In other terms, we can argue that we will not achieve infinite diversity-order if the projection (according to $\B{\Phi}$) of the difference $\B{\mu}_1- \B{\mu}_2$ lies in the direct sum of the projected sub-spaces corresponding to the two classes. In this case, as expected, the fact that the two classes have a nonzero-mean does not change the diversity-order as the mean vectors do not provide a way to separate the projected sub-spaces corresponding to the two classes. On the other hand, when the projection of the difference $\B{\mu}_1- \B{\mu}_2$ does not lie in the direct sum of the projected sub-spaces corresponding to the two classes, the upper bound on the misclassification probability decays exponentially with $1/\sigma^2$ when $\sigma^2\to 0$. Geometrically, this result associated with infinite diversity reflects the fact that, when embedded in a higher dimensional space ($\mathbb{R}^M$ in our cases), the affine spaces spanned by the linear transformation of the signals in the two classes do not intersect.

\section {Multiple-Class Compressive Classification with Random Measurements\\ in the Low-Noise Regime}
\label{mul_class}

We now consider the characterization of the performance of a multiple-class compressive classification problem, where $L \geq 3$, with random measurements. The generalization of the two-class results to the multiple-class case is possible by using the union bound in conjunction with the two-class Bhattacharyya bound that leads to the upper bound to the probability of misclassification given by~\cite{Wim12}:
\begin{IEEEeqnarray}{LcR}
{{P}_{err}^{UB} }= \sum_{i=1}^L \sum_{\substack{
   j = 1 \\
   j \neq i
  }}^L P_{i} ~ e^{-K_{ij}},
  \label{multiclass}
\end{IEEEeqnarray}
where $K_{ij}$ is also given by \eqref{exp_Bhat}.

The fact that the form of the upper bound in (\ref{multiclass}) is akin to the form of the upper bound in (\ref{P_err_Bhat}), involving in addition only the various pairwise misclassification terms that capture the interaction between the different classes, leads to the immediate generalization of the results encapsulated in the previous Theorems. In fact, we can easily conclude from~\eqref{diversity} and~\eqref{multiclass} that the diversity-order for the multiple-class classification problem is given by
\begin{equation}
d = \min_{\substack{
   i,j \\
   j \neq i
  }} d(i,j),
\label{diversity_mult_rand}
\end{equation}
where ${d} \left( i,j\right)$ represents the diversity-order of a given pair of classes $C=i$ and $C=j$ in the two-class classification problem\footnote{Note that the diversity-order $d(i,j)$ does not depend on the exact value of the a priori probability of the classes $P_{i}, P_{j}$, provided that $P_{i}, P_{j}>0$.}. In addition, we can also conclude from (\ref{gm}) and (\ref{multiclass}) that the measurement gain for the multiple-class classification problem is given by
\begin{equation}
g_m = \left[   \sum_{   (i,j) \in \mathcal{S}_d}  P_i 2^{r_{ij}} \left[     \frac{v_{ij}}{   \sqrt{v_i v_j}  }   \right]^{-\frac{1}{2}}  \right]^{-\frac{1}{d}}
\label{gm_mult}
\end{equation}
where $\mathcal{S}_d$ is the set of pairs of indexes corresponding to pairs of classes with minimum diversity-order, that is,
$\mathcal{S}_d = \{ (i,j) : i\neq j,  d(i,j) =d \}
$.

In particular, we can argue that the upper bound to the misclassification probability will exhibit an error floor if at least one of the pairwise misclassification probabilities also exhibits an error floor. Conversely, the upper bound to the misclassification probability (and the true misclassification probability) will tend to zero as $\sigma^2$ tends to zero if all the pairwise upper bounds to the misclassification probabilities also tend to zero. We can also understand how the performance is affected by the geometry of the problem via~\eqref{diversity_mult_rand}, (\ref{gm_mult}) together with Theorems~\ref{theorem1},~\ref{theorem2} and~\ref{theo:nonzero}.

\section{Measurements Design in the Low-Noise Regime}
\label{sec:design}
It is also of interest to investigate how one can further improve performance by using designed measurements in \emph{lieu} of the conventional random ones in the low-noise regime. In particular, we investigate this question by posing a measurement design problem where the objective is to maximize the diversity-order subject to a given measurement budget, i.e.
\begin{equation}
\label{max_d_1}
\max_{\B{\Phi}} d\left(\B{\Phi}\right),
\end{equation}
subject to:
\begin{equation}
\label{max_d_2}
\mathrm{rank} \left(\B{\Phi}\right) \leq M.
\end{equation}
where we express explicitly the diversity-order in terms of the measurement matrix. \footnote{In view of the fact that we use the diversity-order and the measurement gain as a proxy to measure performance in the low-noise regime, we could also conceive measurement design problems where one would first define the set of kernels that maximize the diversity-order subject to the measurement budget and one would then define -- out of the diversity-order maximizing designs - the kernel that maximizes the measurement gain. One could also impose other additional constraints, such as an average power constraint. Our designs also respect this constraint.}

In the sequel, and in line with the previous analytical procedure, we will consider measurement designs for a two-class scenario followed by measurement designs for the multiple-class case. For the two-class problem, it is possible to solve the optimization problem in \eqref{max_d_1}--\eqref{max_d_2}; for multiple-class problems, it does not seem possible to conceive a closed-form solution to such optimization problem -- therefore, we only put forth an, in general suboptimal, algorithm which is inspired by the optimal solution of the two-class scenario: this algorithm attempts at maximizing the diversity-order while satisfying the measurement budget constraint.

\subsection{Two-Class Case}
We first consider kernel designs for two-class compressive classification problems.

\subsubsection{Zero-Mean Classes}
The following Theorem defines the kernel design that solves the optimization problem in \eqref{max_d_1} and~\eqref{max_d_2} for the compressive classification of two zero-mean classes.
\begin{theorem}
\label{theorem4}
Consider the measurement model in \eqref{s_model} where ${\bf x} \sim \mathcal{N} ({\bf 0},{\bf \Sigma}_1)$ with probability $P_{1}$ and ${\bf x} \sim \mathcal{N} ({\bf 0},{\bf \Sigma}_2)$ with probability $P_{2} = 1 - P_{1}$. Assume that the measurement budget is such that $M \geq NO_{Dim}$. Then, the maximum possible diversity-order is given by:
\begin{equation}
\label{dmax}
d_{max} = \frac{1}{4}NO_{Dim}
\end{equation}
which is achieved by a measurement matrix design that obeys the following necessary and sufficient condition:
\begin{equation}
\label{nscond}
2r_{12} - r_1 - r_2  = NO_{Dim}.
\end{equation}
A measurement matrix $\B{\Phi}$ that achieves the maximum diversity-order is
\begin{equation}\label{dmax_mat}
\B{\Phi} = \left[\mathbf{v}_1, \mathbf{v}_2,	\ldots, \mathbf{v}_{n_{\B{\Sigma}_{1}}},\mathbf{w}_1,	\mathbf{w}_2,\ldots,\mathbf{w}_{n_{\B{\Sigma}_{2}}}\right]^T,
\end{equation}
where the sets of vectors $\left[\B{u}_1,\ldots,\B{u}_{n_{12}}\right],~\left[\B{u}_1,\ldots,\B{u}_{n_{12}},\B{v}_1,\ldots,\B{v}_{n_{\B{\Sigma}_{1}}}\right]$, $\left[\B{u}_1,\ldots,\B{u}_{n_{12}},\B{w}_1,\ldots,\B{w}_{n_{\B{\Sigma}_{2}}}\right]$, $\B{u}_i, \B{v}_i, \B{w}_i  \in \mathbb{R}^N$, constitute an orthonormal basis of the linear spaces $\mathrm{Null}\left(\B{\Sigma}_{1}\right)  \bigcap \mathrm{Null}\left(\B{\Sigma}_{2}\right)$, $\mathrm{Null}\left(\B{\Sigma}_{1}\right)$ and $\mathrm{Null}\left(\B{\Sigma}_{2}\right)$, respectively, and $n_{12}=N-r_{\mathbf{\Sigma}_{12}}$, $n_{\B{\Sigma}_{1}} = N-n_{12}-r_{\mathbf{\Sigma}_{1}}$ and  $n_{\B{\Sigma}_{2}} = N-n_{12}-r_{\mathbf{\Sigma}_{2}}$. 

Assume now that the measurement budget is such that $M < NO_{Dim}$. Then, the maximum possible diversity-order is given by:
\begin{equation}
\label{dmax_2}
d = \frac{1}{4}M
\end{equation}
which is achieved, if and only if, the measurement matrix design is such that $r_{12} = M$ and $r_1 + r_2 = r_{12}$. A measurement matrix $\B{\Phi}$ that achieves such a diversity-order can be obtained from the measurement matrix $\B{\Phi}$ in \eqref{dmax_mat}, by choosing arbitrarily only $M$ out of the $n_{\mathbf{\Sigma}_1}+ n_{\mathbf{\Sigma}_2}$ row vectors.

\end{theorem}

\vspace{0.5cm}

\begin{IEEEproof}
The proof is presented in Appendix \ref{app_D}.
\end{IEEEproof}
\vspace{0.5cm}

This Theorem encapsulates key operational features associated with optimized measurements:

\begin{itemize}

\item For a sufficient measurement budget, the Theorem reveals that the maximum diversity-order is directly related to the number of non-overlapping dimensions between the two classes. The measurements design that achieves such maximum diversity-order in~\eqref{dmax} is then required to satisfy $2r_{12} - r_1 - r_2 = NO_{Dim}$ that implies -- as shown in Appendix \ref{app_D} -- that i) it measures all the non-overlapping dimensions, i.e. $M \geq NO_{Dim}$ and ii) it measures all the dimensions in each class that are not contained in the intersection of the corresponding sub-spaces, i.e. $r_1 \geq r_{\B{\Sigma}_{12}} - r_{\B{\Sigma}_{2}}$ and $r_2 \geq r_{\B{\Sigma}_{12}} - r_{\B{\Sigma}_{1}}$. In contrast, for an insufficient measurement budget, the Theorem reveals that maximum diversity-order is related instead to the number of available measurements. The measurement designs that achieve this maximal diversity-order in~\eqref{dmax_2} now only probe a limited number of the non-overlapping dimensions. Therefore we can argue that optimal measurement employs features associated with linear subspaces that are contained in the spaces spanned by the individual classes but not contained in their intersection.

\item In particular, it follows from Theorems \ref{theorem2} and \ref{theorem4} that a designed kernel can offer marked improvements over a random one in the low-noise regime. These include:

i) the ability to achieve perfect separation of the projected signals with a single measurement -- with a random measurement kernel, according to Theorem~\ref{theorem2}, we require $M > r_{\B{\Sigma}_{1}}$ in order to drive the (upper bound to the) probability of misclassification to zero as the noise level approaches zero;

ii) the ability to achieve the maximum diversity-order with a lower number of measurements -- with a random measurement kernel, according to Theorem~\ref{theorem2}, we require $M= r_{\B{\Sigma}_{12}}$ in order to extract the maximum diversity-order, but with a designed measurement kernel we only require $M = NO_{Dim} \leq r_{\B{\Sigma}_{12}}$;

\end{itemize}

\subsubsection{Nonzero-Mean Classes}

The following Theorem now defines the kernel design that solves the optimization problem in~\eqref{max_d_1} and~\eqref{max_d_2} for the compressive classification of two nonzero-mean classes.

\begin{theorem}
\label{theorem5}
Consider the measurement model in (\ref{s_model}) where $\mathbf{x} \sim \mathcal{N} (\B{\mu}_1,\mathbf{\Sigma}_1)$ with probability $P_{1}$ and $\mathbf{x} \sim \mathcal{N} (\B{\mu}_2,\mathbf{\Sigma}_2)$ with probability $P_{2} = 1 - P_{1}$. Assume that:
\begin{equation}
\label{nonzero_design}
  (\B{\mu}_1 - \B{\mu}_2) \notin  \mathrm{im}\left(\mathbf{\Sigma}_1 + \mathbf{\Sigma}_2\right).
\end{equation}

Then, the maximum diversity-order is $d = \infty$ and a matrix design that achieves such a diversity-order is:
\begin{equation}
\B{\Phi} = \begin{bmatrix}
	\B{\phi}^T     
     \end{bmatrix},
\label{phi_nonzero}
\end{equation}
where $\B{\phi}$ can be any vector that belongs to  $\mathrm{Null}\left(\mathbf{\Sigma}_1 + \mathbf{\Sigma}_2\right)$.

Assume now that:
\begin{equation}
  (\B{\mu}_1 - \B{\mu}_2) \in  \mathrm{im}\left(\mathbf{\Sigma}_1 + \mathbf{\Sigma}_2\right).
\end{equation}

Then, the maximum diversity-order is given by \eqref{dmax} or \eqref{dmax_2}, depending on the number of available measurements and a matrix design that achieves such a diversity-order is given by Theorem \ref{theorem4}.
\end{theorem}
\vspace{0.50cm}

\begin{IEEEproof}
The proof is presented in Appendix \ref{app_E}.
\end{IEEEproof}
\vspace{0.5cm}

The most important feature associated with this Theorem relates to the fact that, under certain conditions associated with the geometry of the classification problem, it is possible to attain a diversity-order  $d = \infty$, or exponential decay of the error probability, by taking a single measurement. This feature, which is also unique to nonzero-mean classes, also bears witness to the advantage of kernel design: recall that random measurement kernel requires $M>r_{\B{\Sigma}_{12}}$ in order to achieve a diversity-order  $d = \infty$ (see Theorem~\ref{theo:nonzero}). Note also that the existence of a $\B{\phi} \in \mathrm{Null}\left(\mathbf{\Sigma}_1 + \mathbf{\Sigma}_2\right)$, which is used to construct the measurement kernel in~\eqref{phi_nonzero}, is guaranteed by the condition in (\ref{nonzero_design}).

\IncMargin{1em}
\begin{algorithm}[!tbp]
\NoCaptionOfAlgo
\SetKwFor{For}{for}{}{}
\SetKwFor{While}{while}{}{}
\SetKwIF{If}{ElseIf}{Else}{if}{}{else if}{-- else}{}

\SetKwData{Left}{left}\SetKwData{This}{this}\SetKwData{Up}{up}
\SetKwFunction{Union}{Union}\SetKwFunction{FindCompress}{FindCompress}
\SetKwInOut{Input}{Input}\SetKwInOut{Output}{Output}
\Input{Number of classes $L$, input covariance matrices ${\bf \Sigma}_i$, $i=1,\cdots,L$, number of measurements $M$.}
\Output{Measurement matrix $\B{\Phi}$.}
\BlankLine

\nlset{Step 1} {\bf--} Determine $\displaystyle \left(i^*,j^*\right) = \arg \min_{\substack{i,j \\i \neq j}} ~~NO_{Dim}\left(i,j\right)$, where $NO_{Dim}\left(i,j\right)$ is as expressed in (\ref{nodim});

\nlset{Step 2} {\bf--}  \For{$i = 1,\ldots,L-1$}{

	{\bf--}  \For{$j = i+1,\ldots,L$}{

		{{\bf--}  Construct $\mathbf{\Phi}_{ij} \in \mathbb{R}^{NO_{Dim}(i^*,j^*) \times N}$ according to Theorem~\ref{theorem4} applied to classes $i$ and $j$;

		}
		
	}
}

\nlset{Step 3} {\bf --}    $\B{\Phi}  \leftarrow \left[	\B{\Phi}_{12}^T,  \B{\Phi}_{13}^T, \ldots,  \mathbf{\Phi}_{(L-1)L}^T   \right]^T$;

\nlset{Step 4} {\bf --} Compute the rank  $r_{\B{\Phi}} = \mathrm{rank} (\B{\Phi})$:

{\bf--}  \eIf{$r_{\B{\Phi}} \leq M$}{
			{\bf--}  Return the matrix obtained by selecting the nonzero rows of the row echelon form of $\B{\Phi}$;
		}
		{{\bf--} $\mathbf{\Phi}_{ij}  \leftarrow \left[  \mathbf{I}_{M_{ij}-1},  \mathbf{0}_{(M_{ij}-1) \times 1} \right] \B{\Phi}_{ij}  $, where $M_{ij}$ denotes the number of rows of $\B{\Phi}_{ij} $; 
		
		{\bf--} Go to Step 3;
}
	
\caption{\centerline{{\bfseries Table I:}} Algorithmic approach to maximize the diversity-order subject to a measurement budget in a multiple-class classification problem with zero-mean classes.}\label{alg1}
\end{algorithm}\DecMargin{1em}

\subsection{Multiple-Class Case}
We now consider kernel designs for multiple-class compressive classification problems. In particular, we propose, possibly suboptimal, algorithmic approaches, inspired by the two-class case designs in Theorems \ref{theorem4} and \ref{theorem5}, that attempt to achieve the maximum diversity-order while satisfying the constraint on the number of measurements. We leverage the fact that the diversity-order associated with a multiple-class classification problem corresponds  to the lowest of the diversity-orders of the pairwise classification problems (see \eqref{diversity_mult_rand}).

\subsubsection{Zero-Mean Classes}\label{multi_des}

Table I puts forth an algorithmic approach to design projections for multiple-class classification problems with zero-mean classes. The algorithm consists of four main steps. In the first step, we identify the pair of classes associated with the minimum value of the diversity-order (for the two-class compressive classification problem). In the second step
we construct a set of measurement matrices $\B{\Phi}_{ij}, \forall i,j$ as in Theorem \ref{theorem4} for all possible pairs of classes, consisting of a number of measurements necessary to achieve the minimum value of the diversity-order determined in step 1. In the third step we build a measurement matrix by concatenating all the measurements designed for the different pairs of classes. Then, in the fourth step we compute the rank of such matrix. If the rank is less than or equal to the number of available measurements, then the algorithm returns the matrix obtained by picking the nonzero rows of the row echelon form of matrix $\B{\Phi}$. Otherwise, for all pair of classes, we erase the last measurement from the corresponding matrix $\B{\Phi}_{ij}$ and iterate again from step 3 until the rank of $\B{\Phi}$ satisfies the measurement constraint.

Note that the construction in Algorithm~\ref{alg1} satisfies the measurement budget constraint, but it does not necessarily lead to the global optimum solution of the problem in \eqref{max_d_1}--\eqref{max_d_2}, as different choices of measurements in step 2 and step 4 might lead to different values of the diversity-order.\footnote{Note that, in step 2, we are arbitrarily choosing $NO_{Dim}(i^*,j^*)$ linearly independent measurements from a linear space of dimension $NO_{Dim}(i,j)$. Moreover, in step 4 we could decide to delete any row of $\B{\Phi}_{ij}$ instead of the last one.} However, observe that the matrix constructed in step 2 achieves the maximum diversity-order, and each iteration of step 4 decreases the achieved diversity-order of a factor at most equal to $1/4$. Finally, we underline the fact that Algorithm \ref{alg1} might eventually output an empty measurement matrix. This can happen if, after $NO_{Dim}(i^*,j^*)-1$ iterations of step 4, the single measurements corresponding to the $L(L-1)/2$ pairs of classes still span a linear space of dimension greater than $M$. In order to avoid that the algorithm outputs such an empty matrix, we could propose to delete measurements only from a single randomly selected matrix $\B{\Phi}_{ij}$ or some randomly selected subset of matrices $\B{\Phi}_{ij}$ {\it in lieu} of a single measurement from all such matrices: this procedure would never result in the output of a null matrix, but could result in an error floor in the upper bound of the misclassification probability. Despite this possible issues, numerical results show that matrices designed according to this algorithm appear to lead to very good performance.

\subsubsection{Nonzero-Mean Classes}

The algorithmic approach to design projections for multiple-class classification problems with nonzero-mean classes is shown in Table II.

The algorithm also consists of four steps. In step 1, we check if $(\B{\mu}_i - \B{\mu}_j) \notin  \mathrm{im}\left(\mathbf{\Sigma}_i + \mathbf{\Sigma}_j\right)$ holds for every pair of classes. In this case, we can construct the measurement kernel by picking vectors from the null spaces $\mathrm{Null}\left(\mathbf{\Sigma}_i + \mathbf{\Sigma}_j\right)$ for every pair of classes, thus constituting a measurement matrix $\B{\Phi}$. In step 3 we check if the rank of such matrix satisfies the measurement constraint. If this is the case, the algorithm returns the matrix obtained by picking the nonzero rows of the row echelon form of matrix $\B{\Phi}$. Otherwise, or if there is a pair of classes for which $(\B{\mu}_i - \B{\mu}_j) \in  \mathrm{im}\left(\mathbf{\Sigma}_i + \mathbf{\Sigma}_j\right)$, then, it is not possible to achieve infinite diversity-order with $M$ measurements extracted from $\B{\Phi}$ and we construct the measurement kernel according to Algorithm \ref{alg1}.

\IncMargin{1em}
\begin{algorithm}[!tbp]
\NoCaptionOfAlgo
\SetKwFor{For}{for}{}{}
\SetKwFor{While}{while}{}{}
\SetKwIF{If}{ElseIf}{Else}{if}{}{else if}{-- else}{}

\SetKwData{Left}{left}\SetKwData{This}{this}\SetKwData{Up}{up}
\SetKwFunction{Union}{Union}\SetKwFunction{FindCompress}{FindCompress}
\SetKwInOut{Input}{Input}\SetKwInOut{Output}{Output}
\Input{Number of classes $L$, input covariance matrices ${\bf \Sigma}_i$ and mean vectors $\B{\mu}_i$, $i=1,\cdots,L$, number of measurements $M$.}
\Output{Measurement matrix $\B{\Phi}$.}
\BlankLine

\nlset{Step 1} {\bf--} \For{$i = 1,\ldots,L-1$}{

	{\bf--} \For{$j = i+1,\ldots,L$}{

		{\bf--} \eIf{$(\B{\mu}_i - \B{\mu}_j) \notin  \mathrm{im}\left(\mathbf{\Sigma}_i + \mathbf{\Sigma}_j\right)$}{
			{\bf--} Choose $\B{\phi}_{ij} \in \mathrm{Null}\left(\mathbf{\Sigma}_i + \mathbf{\Sigma}_j\right)$ as in Theorem~\ref{theorem5};
		}	
		{ {\bf--} Go to Step 4;
		}

	}
}

\nlset{Step 2} {\bf --}    $\B{\Phi}  \leftarrow \left[	\B{\phi}_{12},  \B{\phi}_{13}, \ldots,  \mathbf{\phi}_{(L-1)L}   \right]^T$;

\nlset{Step 3} {\bf --} Compute the rank  $r_{\B{\Phi}} = \mathrm{rank} (\B{\Phi})$:

{\bf--}  \eIf{$r_{\B{\Phi}} \leq M$}{
			{\bf--}  Return the matrix obtained by selecting the nonzero rows of the row echelon form of $\B{\Phi}$;
		}
		{
			{\bf--} Go to Step 4;
}

\nlset{Step 4} {\bf--} Construct $\B{\Phi}$ according to Algorithm \ref{alg1};

\caption{\centerline{{\bfseries Table II:}} Algorithmic approach to maximize the diversity-order subject to a measurement budget in the multiple-class classification problem with nonzero-mean classes.}\label{alg2}
\end{algorithm}\DecMargin{1em}

\section{High-Noise Regime}
\label{high_noise}
It is also of interest to briefly contrast the behavior of the upper bound to the misclassification probability in the low-noise regime ($\sigma^2 \to 0$) to the high-noise regime ($\sigma^2 \to \infty$). For example, for a two-class compressive classification problem where $\mathbf{x} \sim \mathcal{N} (\B{\mu}_1,\mathbf{\Sigma}_1)$ with probability $P_{1}$ and $\mathbf{x} \sim \mathcal{N} (\B{\mu}_2,\mathbf{\Sigma}_2)$ with probability $P_{2} = 1 - P_{1}$, this can be done by expanding the Bhattacharyya upper bound to the misclassification probability, in (\ref{P_err_Bhat}) and (\ref{exp_Bhat}), as follows\footnote{The expansion and ensuing discussion also generalize immediately to multiple-class problems via the usual union bound arguments.}:
\begin{equation}
{P}_{err}^{UB}\left(\frac{1}{\sigma^2}\right)  = {P}_{err}^{UB}\left(0\right) + \frac{d~P_{err}^{UB} \left(\frac{1}{\sigma^2}\right)} {d~\frac{1}{\sigma^2}} \Big|_{\frac{1}{\sigma^2} = 0} \frac{1}{\sigma^2} + \frac{d^2~P_{err}^{UB} \left(\frac{1}{\sigma^2}\right)} {d~\left(\frac{1}{\sigma^2}\right)^2} \Big|_{\frac{1}{\sigma^2} = 0} \frac{\left(\frac{1}{\sigma^2}\right)^2}{2} + o\left(\left(\frac{1}{\sigma^2}\right)^2\right).
\label{exp_highnoise}
\end{equation}

Then, for zero-mean classes, the expansion of the upper bound to the probability of misclassification reduces to:
\begin{equation}
{P}_{err}^{UB}\left(\frac{1}{\sigma^2}\right) = \sqrt{P_{1}P_{2}}~\left[1 + \frac{1}{4} ~ A ~ \left(\frac{1}{\sigma^2}\right)^2\right] + o \left( \left(\frac{1}{\sigma^2}\right)^2 \right), 
\label{high_zm}
\end{equation}
where:
\begin{IEEEeqnarray}{rCl}
\label{lowsnrA}
A & = & \mathrm{tr} \left[\left(\frac{{\bf \Phi}\left({\bf \Sigma}_1+{\bf \Sigma}_2\right){\bf \Phi}^T}{2}\right)^2\right] - \frac{1}{2} \mathrm{tr} \left[\left({{\bf \Phi}{\bf \Sigma}_1{\bf \Phi}^T}\right)^2\right] - \frac{1}{2} \mathrm{tr} \left[\left({{\bf \Phi}{\bf \Sigma}_2{\bf \Phi}^T}\right)^2\right] \nonumber \\
& &  + \mathrm{tr} \left[{{\bf \Phi}{\bf \Sigma}_1{\bf \Phi}^T}\right]\mathrm{tr} \left[{{\bf \Phi}{\bf \Sigma}_2{\bf \Phi}^T}\right] - \mathrm{tr}^2 \left[\frac{{\bf \Phi}\left({\bf \Sigma}_1+{\bf \Sigma}_2\right){\bf \Phi}^T}{2}\right],
\end{IEEEeqnarray}
and for nonzero-mean classes, it reduces to:
\begin{IEEEeqnarray}{rcl}
{P}_{err}^{UB} \left(  \frac{1}{\sigma^2} \right) & = & \sqrt{P_{1}P_{2}}~\left[1 - \frac{1}{8} \left\|{\bf \Phi}\left(\boldsymbol{\mu}_1 - \boldsymbol{\mu}_2\right)\right\|^2 \frac{1}{\sigma^2}   \right] + o\left( \frac{1}{\sigma^2} \right).
\label{high_nzm}
\end{IEEEeqnarray}

Note that the behavior of the upper bound to the misclassification probability for zero-mean classes is fundamentally different from that for nonzero-mean classes. In particular, the first-order term in~\eqref{exp_highnoise} is always equal to zero for zero-mean classes and is nonzero for nonzero-mean classes, implying that the upper bound to the misclassification probability decays faster in the later case as $1/\sigma^2 \to 0$. In addition, the behavior of the high-noise expansions is also fundamentally different from that of the low-noise ones. In the low-noise case, we can approximate the upper bound to the misclassification probability by a line defined via its slope (the diversity-order) and its offset (the measurement gain) on a log-log scale. Of course, its slope can occasionally be infinite as per Theorem \ref{theo:nonzero}. In contrast, in the high-noise case we approximate the upper bound to the misclassification probability with a line (for nonzero-mean classes) or with a parabola (for zero-mean classes) on a linear, rather than logarithmic, scale.

We can also derive further insight by studying the behavior of the average, with respect to the measurement matrix, of the upper bounds to the misclassification probability by assuming that the measurement matrix is such that its elements are drawn i.i.d. from a zero-mean unit-variance Gaussian distribution. In particular, it is simple to establish that the average of the upper bound to the misclassification probability behaves as follows:
\begin{IEEEeqnarray}{rcl}
\bar{P}_{err}^{UB} \left(\frac{1}{\sigma^2}\right) = \mathcal{E}\left\{{P}_{err}^{UB}\left(\frac{1}{\sigma^2}\right)\right\} = \sqrt{P_{1}P_{2}}~\left[1 + \frac{1}{4} ~ \mathcal{E}\left\{A\right\} ~ \left(\frac{1}{\sigma^2}\right)^2\right] + o \left( \left(\frac{1}{\sigma^2}\right)^2 \right),
\end{IEEEeqnarray}
for zero-mean classes and:
\begin{IEEEeqnarray}{rcl}
\bar{P}_{err}^{UB} \left(\frac{1}{\sigma^2}\right) = \mathcal{E}\left\{{P}_{err}^{UB}\left(\frac{1}{\sigma^2}\right)\right\} = \sqrt{P_{1}P_{2}}~\left[1 - \frac{1}{8} ~ \mathcal{E}\left\{\left\|{\bf \Phi}\left(\boldsymbol{\mu}_1 - \boldsymbol{\mu}_2\right)\right\|^2\right\} \frac{1}{\sigma^2}   \right] + o\left( \frac{1}{\sigma^2} \right),
\end{IEEEeqnarray}
for nonzero-mean classes. Via random matrix theory, we can further calculate:

\begin{IEEEeqnarray}{rcl}
\mathcal{E}\left\{\mathrm{tr} \left[{{\bf \Phi}{\bf \Sigma}_1{\bf \Phi}^T}\right]\mathrm{tr} \left[{{\bf \Phi}{\bf \Sigma}_2{\bf \Phi}^T}\right]\right\} & = & M\left(\mathrm{tr}\left[{\bf \Sigma}_1\right] \mathrm{tr}\left[{\bf \Sigma}_2\right] + 2 \mathrm{tr}\left[{\bf \Sigma}_1{\bf \Sigma}_2\right] \right) 
+ M\left(M-1\right) \mathrm{tr}\left[{\bf \Sigma}_1\right] \mathrm{tr}\left[{\bf \Sigma}_2\right] \label{trace_first}
\end{IEEEeqnarray}
\begin{IEEEeqnarray}{rcl}
\mathcal{E}\left\{\mathrm{tr}^2 \left[\frac{{\bf \Phi}\left({\bf \Sigma}_1+{\bf \Sigma}_2\right){\bf \Phi}^T}{2}\right]\right\} & = & M\left\{\mathrm{tr}^2\left[\frac{\left({\bf \Sigma}_1+{\bf \Sigma}_2\right)}{2}\right]+ 2 \mathrm{tr}\left[\left(\frac{\left({\bf \Sigma}_1+{\bf \Sigma}_2\right)}{2}\right)^2\right] \right\} \nonumber \\ 
& + & M\left(M-1\right)  \mathrm{tr}^2\left[\frac{\left({\bf \Sigma}_1+{\bf \Sigma}_2\right)}{2}\right] \IEEEeqnarraynumspace
\end{IEEEeqnarray}
\begin{IEEEeqnarray}{rcl}
\mathcal{E}\left\{\mathrm{tr}\left[\left({\bf \Phi}{\bf \Sigma}_1{\bf \Phi}^T\right)^2\right]\right\} & = & M \left[\sum_{j=1 }^N m_4 {\lambda_{{\bf \Sigma}_1}}_j^2 + \sum_{\substack{j,k=1 \\ j \neq k}}^N m_2 {\lambda_{{\bf \Sigma}_1}}_j {\lambda_{{\bf \Sigma}_1}}_k\right] 
+ M\left(M-1\right) \sum_{j=1 }^N m_2 {\lambda_{{\bf \Sigma}_1}}_j^2
\end{IEEEeqnarray}
\begin{IEEEeqnarray}{rcl}
\mathcal{E}\left\{\mathrm{tr}\left[\left({\bf \Phi}{\bf \Sigma}_2{\bf \Phi}^T\right)^2\right]\right\} & = & M \left[\sum_{j=1 }^N m_4 {\lambda_{{\bf \Sigma}_2}}_j^2 + \sum_{\substack{j,k=1 \\ j \neq k}}^N m_2 {\lambda_{{\bf \Sigma}_2}}_j {\lambda_{{\bf \Sigma}_2}}_k\right] 
+ M\left(M-1\right) \sum_{j=1 }^N m_2 {\lambda_{{\bf \Sigma}_2}}_j^2
\end{IEEEeqnarray}
\begin{IEEEeqnarray}{rcl}
\mathcal{E}\left\{\mathrm{tr}\left[\left(\frac{{\bf \Phi}\left({\bf \Sigma}_1+{\bf \Sigma}_2\right){\bf \Phi}^T}{2}\right)^2\right]\right\} & = & M \left[\sum_{j=1 }^N m_4 {\lambda_{{\bf \Sigma}_{12}}}_j^2 + \sum_{\substack{j,k=1 \\ j \neq k}}^N m_2 {\lambda_{{\bf \Sigma}_{12}}}_j {\lambda_{{\bf \Sigma}_{12}}}_k\right] 
+ M\left(M-1\right) \sum_{j=1 }^N m_2 {\lambda_{{\bf \Sigma}_{12}}}_j^2 \IEEEeqnarraynumspace
\end{IEEEeqnarray}
\begin {equation}
\mathcal{E}\left\{ \|{\bf \Phi}\left(\boldsymbol{\mu}_1 - \boldsymbol{\mu}_2\right)   \|^2\right\} = M \|\boldsymbol{\mu}_1 - \boldsymbol{\mu}_2  \|^2
\label{trace_last}
\end{equation}
where ${\lambda_{{\bf \Sigma}_1}}_j$, $j=1, \ldots, N$ are the eigenvalues of the matrix ${{\bf \Sigma}_1}$,  ${\lambda_{{\bf \Sigma}_2}}_j$, $j=1, \ldots, N$ are the eigenvalues of the matrix ${{\bf \Sigma}_2}$, ${\lambda_{{\bf \Sigma}_{12}}}_j$, $j=1, \ldots, N$ are the eigenvalues of the matrix $\frac{\left({\bf \Sigma}_1+{\bf \Sigma}_2\right)}{2}$, and $m_2$ and $m_4$ are the second and fourth-order moments, respectively, of a zero-mean and 
unit variance Gaussian random variable.

Therefore, one observes that in the nonzero-mean case the behavior of the average value of the upper bound to the misclassification probability depends only on the number of measurements and the means of the classes -- moreover, the higher the number of measurements the lower the average upper bound; in contrast, in the zero-mean case the average value of the upper bound to the probability of the misclassification depends in a more intricate manner on the number of measurements and the source covariances via the eigenvalues of the matrices ${{\bf \Sigma}_1}$, ${{\bf \Sigma}_2}$ and $\frac{\left({\bf \Sigma}_1+{\bf \Sigma}_2\right)}{2}$.

It is more difficult to establish how the upper bound to the misclassification probability, in the high-noise regime, behaves in the presence of optimized measurements (though equations \eqref{high_zm} and \eqref{high_nzm} offer a means to carry out numerical optimizations). However, we can easily see that for a compressive classification problem with nonzero-mean classes the design which minimizes the first-order expansion of the upper bound to the probability of misclassification in the high-noise regime is obtained by aligning the measurements with the vector $\left(\boldsymbol{\mu}_1 - \boldsymbol{\mu}_2\right)$, i.e.:
\begin{equation}
{\bf \Phi} = \alpha\left(\boldsymbol{\mu}_1 - \boldsymbol{\mu}_2\right)^T,
\end{equation}
where the scalar $\alpha \neq 0$ determines the norm of the measurement vector ${\bf \Phi}$.

\section{Numerical Results}
\label{num_res}

We now present a series of results that illustrate the main operational features associated with the compressive classification of a mixture of Gaussians. In particular, we also compare the behavior of the upper bound to the probability of misclassification to the behavior of the true probability of misclassification, in order to determine whether the previous theoretical results are aligned with real ones.

\subsection{Random Measurements}\label{result_rand}

We first consider a compressive classification problem with two zero-mean classes. The two classes are such that $\B{\mu}_1 = \B{\mu}_2=0$, ${\bf \Sigma}_1 = {\bf U} \mathrm{diag}\left(1,1,0,0,0,0\right) {\bf U}^T$ and ${\bf \Sigma}_2 = {\bf U} \mathrm{diag}\left(0,1,1,1,0,0\right) {\bf U}^T$, where ${\bf U}$ is a randomly generated unitary matrix. Therefore, $r_{{\bf \Sigma}_1} = \mathrm{rank}\left({\bf \Sigma}_1\right) = 2$, $r_{{\bf \Sigma}_2} = \mathrm{rank}\left({\bf \Sigma}_2\right) = 3$,  $r_{{\bf \Sigma}_{12}} = \mathrm{rank}\left({\bf \Sigma}_1 + {\bf \Sigma}_2\right) = 4$ and $N=6$. The sensing matrix is such that:
\begin{equation}
{\bf \Phi} = \frac{M}{\mathrm{tr}\left({\bf \Phi}^{\prime}{{\bf \Phi}^{\prime}}^T\right)} {\bf \Phi}^{\prime},
\end{equation}
where ${\bf \Phi}^{\prime}$ has i.i.d. zero-mean unit-variance Gaussian entries. Note that this scenario is such that the sub-spaces corresponding to the two classes do not overlap completely and the number of non-overlapping dimensions in \eqref{nodim} is equal to 3. Therefore, accordingly to Theorems \ref{theorem1} and \ref{theorem2} it is possible to drive the upper bound to the misclassification probability to zero at low-noise levels.

\begin{figure}[thbp]
   \centerline{\subfigure[]{\includegraphics[width=3.2in, height=2.5in]{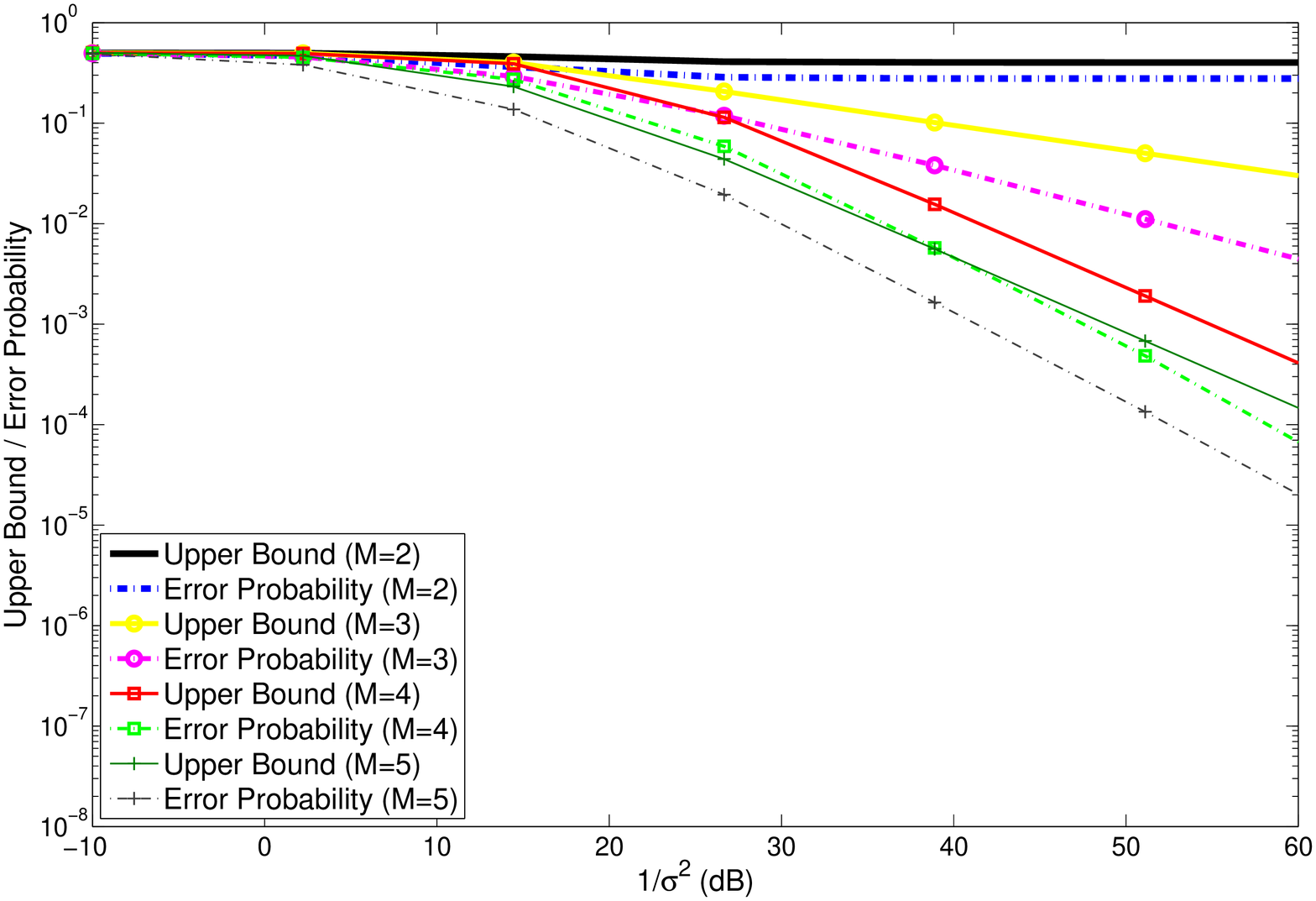}
       \label{fig:two_class_1}}
     \hfil
     \subfigure[]{\includegraphics[width=3.2in, height=2.5in]{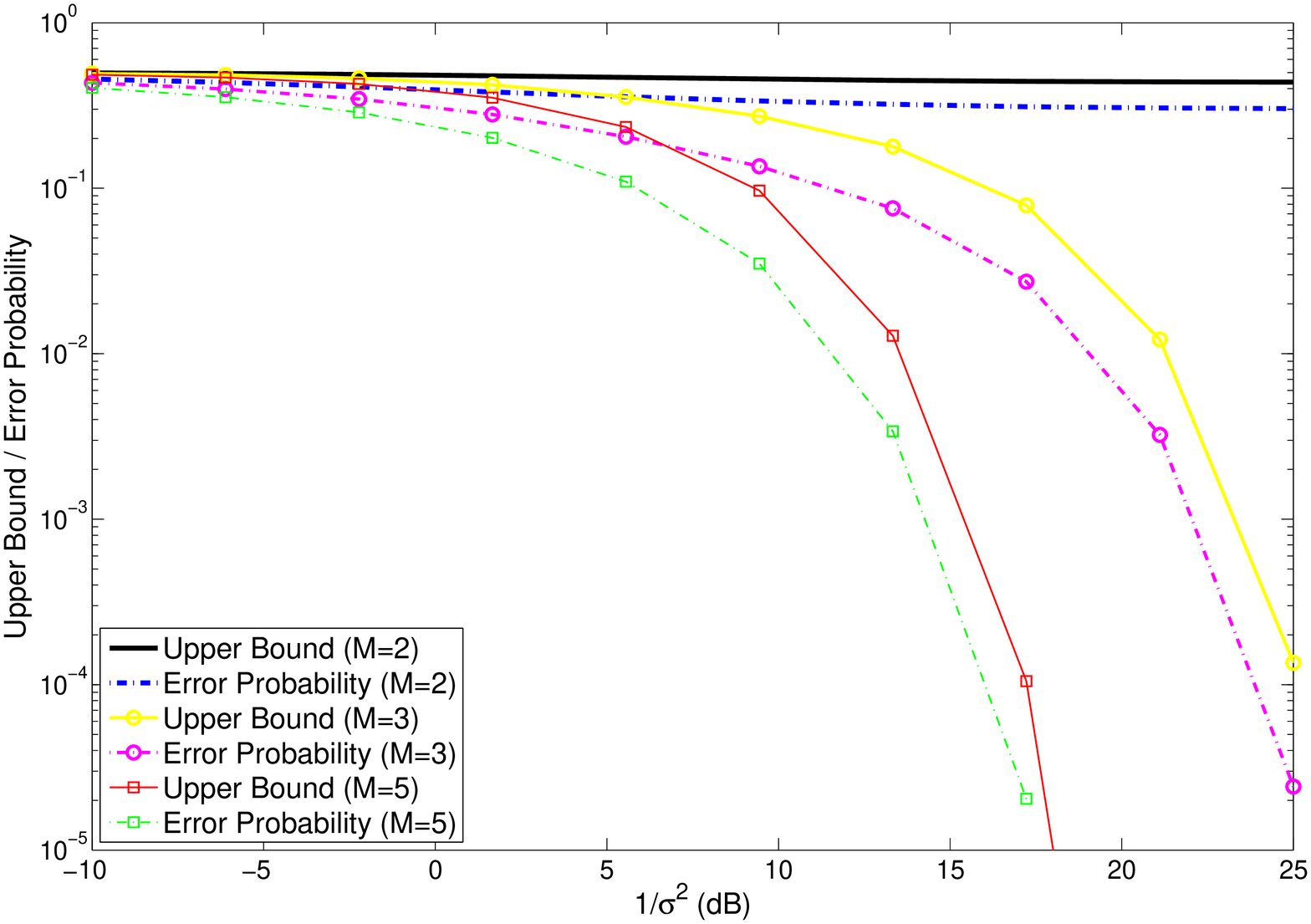}
       \label{fig:mu_nonzero}}}
\vspace{-0.4cm}
      \centerline{\subfigure[]{\includegraphics[width=3.2in, height=2.5in]{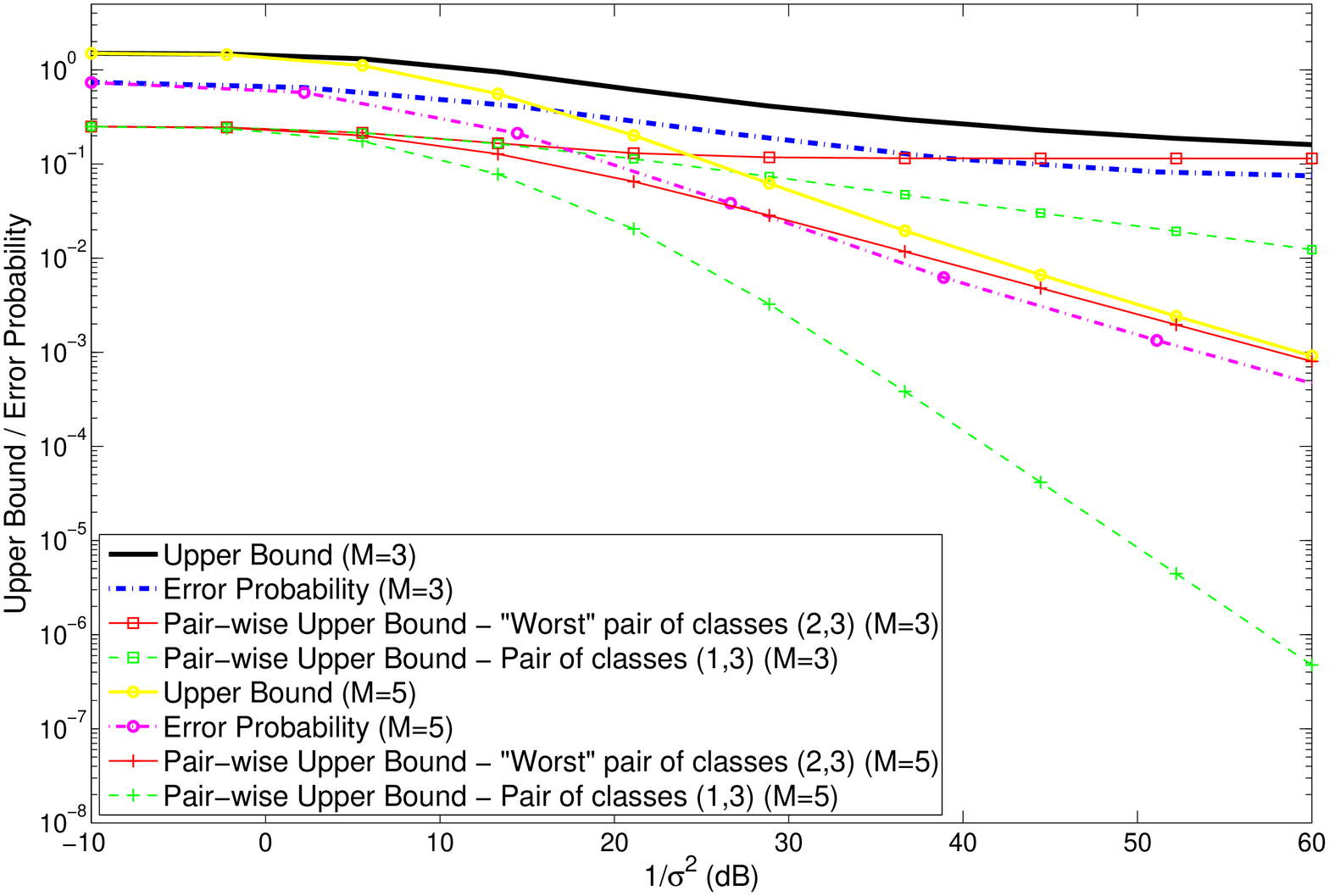}
       \label{fig:mul_class}}}
   \caption{Upper bound to error probability and true error probability \emph{vs.} $1/\sigma^2$ (in dB) for: (a) the two zero-mean classes; (b) the two nonzero-mean classes; (c) multiple-classes.}
   \label{fig:Fig_1}
 \end{figure}

Figure \ref{fig:two_class_1} shows that for $M = 2$ the upper bound exhibits an error floor and for $M > 2$ the upper bound tends to zero as the noise level also tends to zero, in accordance with Theorem \ref{theorem2}. Figure \ref{fig:two_class_1}  also shows that the increase of $M$ from 2 to 3 or 4 results in the increase in the diversity-order but  $M > r_{{\bf \Sigma}_{12}}$ does not result in further increases in diversity but only in measurement gain -- this is also consistent with Theorem \ref{theorem2}. We can also observe that the upper bound to the probability of misclassification is able to capture the behavior of the true probability of misclassification: it captures the presence or absence of an error floor and (except for $M = 2$) it also captures closely the diversity-order and increases in measurement gain in the true error probability.

We now consider a compressive classification problem with two nonzero-mean classes. Here, the two classes are such that $\B{\mu}_1 \neq \B{\mu}_2$ and ${\bf \Sigma}_1 = {\bf \Sigma}_2 = {\bf U} \mathrm{diag}\left(1,1,0,0,0,0\right) {\bf U}^T$, where ${\bf U}$ is also a randomly generated unitary matrix. Then, $r_{{\bf \Sigma}_1} = \mathrm{rank}\left({\bf \Sigma}_1\right) = 2$, $r_{{\bf \Sigma}_2} = \mathrm{rank}\left({\bf \Sigma}_2\right) = 2$, $r_{{\bf \Sigma}_{12}} = \mathrm{rank}\left({\bf \Sigma}_1 + {\bf \Sigma}_2\right) = 2$ and $N=6$. The sensing matrix is also generated randomly with i.i.d. zero-mean unit-variance Gaussian entries and normalized as in the previous scenario. Note now that this scenario is such that the affine spaces corresponding to the two classes are parallel, differing only by a translation determined by the vector $\B{\mu}_1 - \B{\mu}_2$.

Figure \ref{fig:mu_nonzero} -- in line with Theorem~\ref{theo:nonzero} -- shows that when $M \leq r_{{\bf \Sigma}_{12}} = 2$ the upper bound exhibits an error floor and when $M > r_{{\bf \Sigma}_{12}}$ the upper bounds to the misclassification probability tends to zero exponentially as the noise level also tends to zero. Once again, we can observe that the behavior of the upper bound to the probability of misclassification is consistent with the behavior of the true probability of misclassification.

We now turn the attention to a multi-class scenario where  the number of classes is $L= 4$, $\B{\mu}_1 = \B{\mu}_2 = \B{\mu}_3 = \B{\mu}_4 = 0$, ${\bf \Sigma}_1 = {\bf U} \mathrm{diag}\left(1,1,0,0,0,0\right) {\bf U}^T$, ${\bf \Sigma}_2 = {\bf U} \mathrm{diag}\left(0,1,1,1,0,0\right) {\bf U}^T$, ${\bf \Sigma}_3 = {\bf U} \mathrm{diag}\left(0,0,1,1,1,0\right) {\bf U}^T$ and ${\bf \Sigma}_4 = {\bf U} \mathrm{diag}\left(0,0,0,0,1,1\right) {\bf U}^T$ where ${\bf U}$ is a randomly generated unitary matrix. Now, $r_{{\bf \Sigma}_1} = \mathrm{rank}\left({\bf \Sigma}_1\right) = 2$, $r_{{\bf \Sigma}_2} = \mathrm{rank}\left({\bf \Sigma}_2\right) = 3$, $r_{{\bf \Sigma}_3} = \mathrm{rank}\left({\bf \Sigma}_3\right) = 3$, $r_{{\bf \Sigma}_4} = \mathrm{rank}\left({\bf \Sigma}_4\right) = 2$, $r_{{\bf \Sigma}_{12}} = \mathrm{rank}\left({\bf \Sigma}_1 + {\bf \Sigma}_2\right) = 4$, $r_{{\bf \Sigma}_{13}} = \mathrm{rank}\left({\bf \Sigma}_1 + {\bf \Sigma}_3\right) = 5$, $r_{{\bf \Sigma}_{14}} = \mathrm{rank}\left({\bf \Sigma}_1 + {\bf \Sigma}_4\right) = 4$, $r_{{\bf \Sigma}_{23}} = \mathrm{rank}\left({\bf \Sigma}_2 + {\bf \Sigma}_3\right) = 4$, $r_{{\bf \Sigma}_{24}} = \mathrm{rank}\left({\bf \Sigma}_2 + {\bf \Sigma}_4\right) = 5$, $r_{{\bf \Sigma}_{34}} = \mathrm{rank}\left({\bf \Sigma}_3 + {\bf \Sigma}_4\right) = 4$ and $N=6$. The sensing matrix is also generated randomly with i.i.d. zero-mean unit-variance Gaussian entries and normalized as previously noted. The pair of classes corresponding to the minimum pairwise diversity-order is $(2,3)$, offering a diversity-order that cannot exceed $-\frac{1}{2}\left(\frac{r_{{\bf \Sigma}_2}+r_{{\bf \Sigma}_3}}{2}- r_{{\bf \Sigma}_{23}}\right) = \frac{1}{2}$; all other pairs of classes offer a higher diversity-order.

Figure \ref{fig:mul_class} confirms that the behavior of the upper bound to the misclassification probability is indeed dominated by the behavior of the pairwise upper bound associated with classes $(2,3)$. The Figure also confirms that the upper bound to the misclassification probability approximates well  the true misclassification probability: in particular, it is able to predict the presence of absence of the error floor as well as the diversity-order of the true error probability.

\subsection{Designed Measurements}\label{result_design}

We now compare the performance of designed measurements to the performance of random measurements for two-class compressive classification problems with zero-mean and nonzero-mean classes and for zero-mean, multiple-class compressive classification problems.

\begin{figure*}[!tbp]
   \centerline{\subfigure[]{\includegraphics[width=3.2in, height=2.5in]{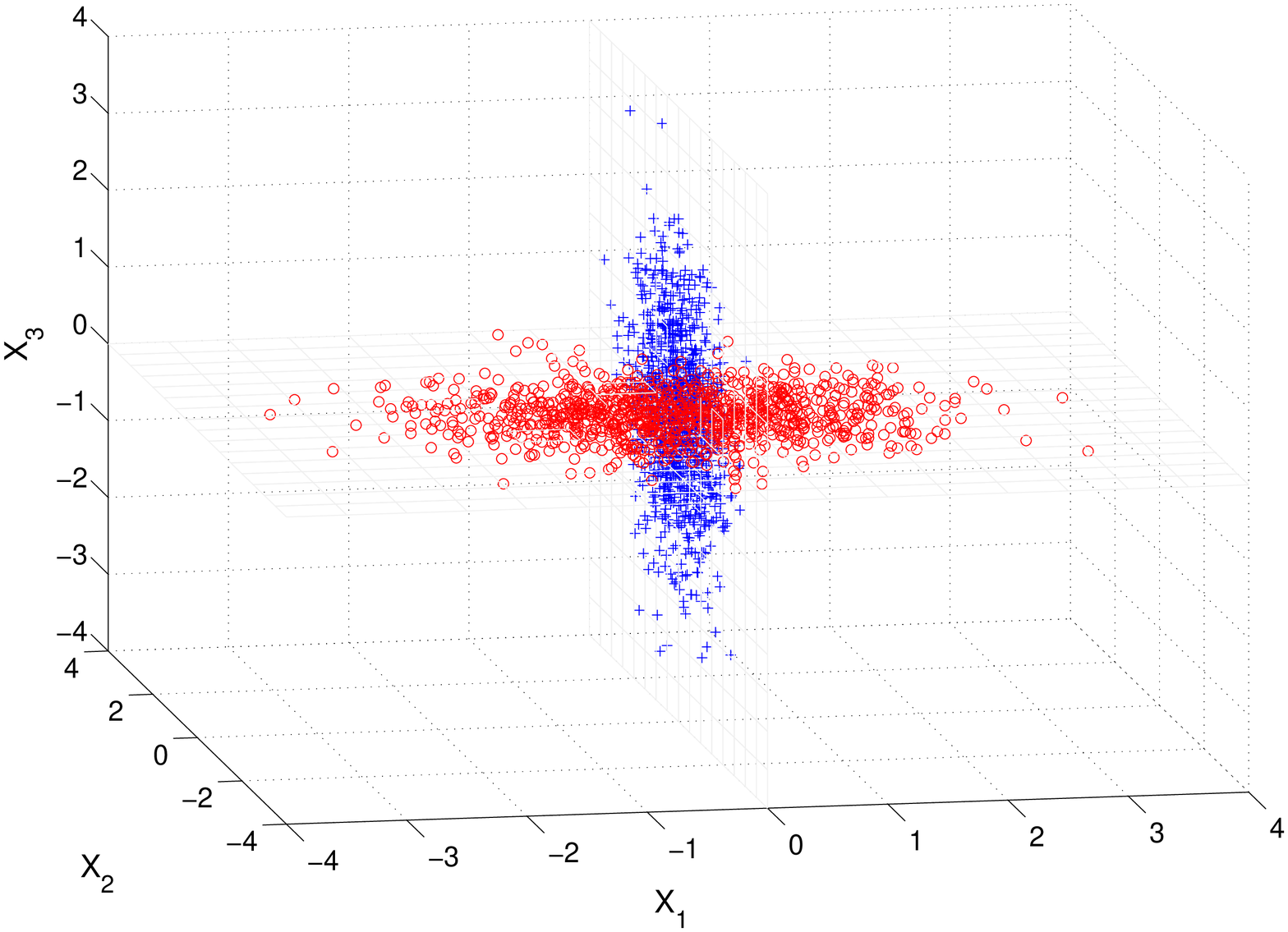}
       \label{fig:classes}}
     \hfil
     \subfigure[]{\includegraphics[width=3.5in, height=2.5in]{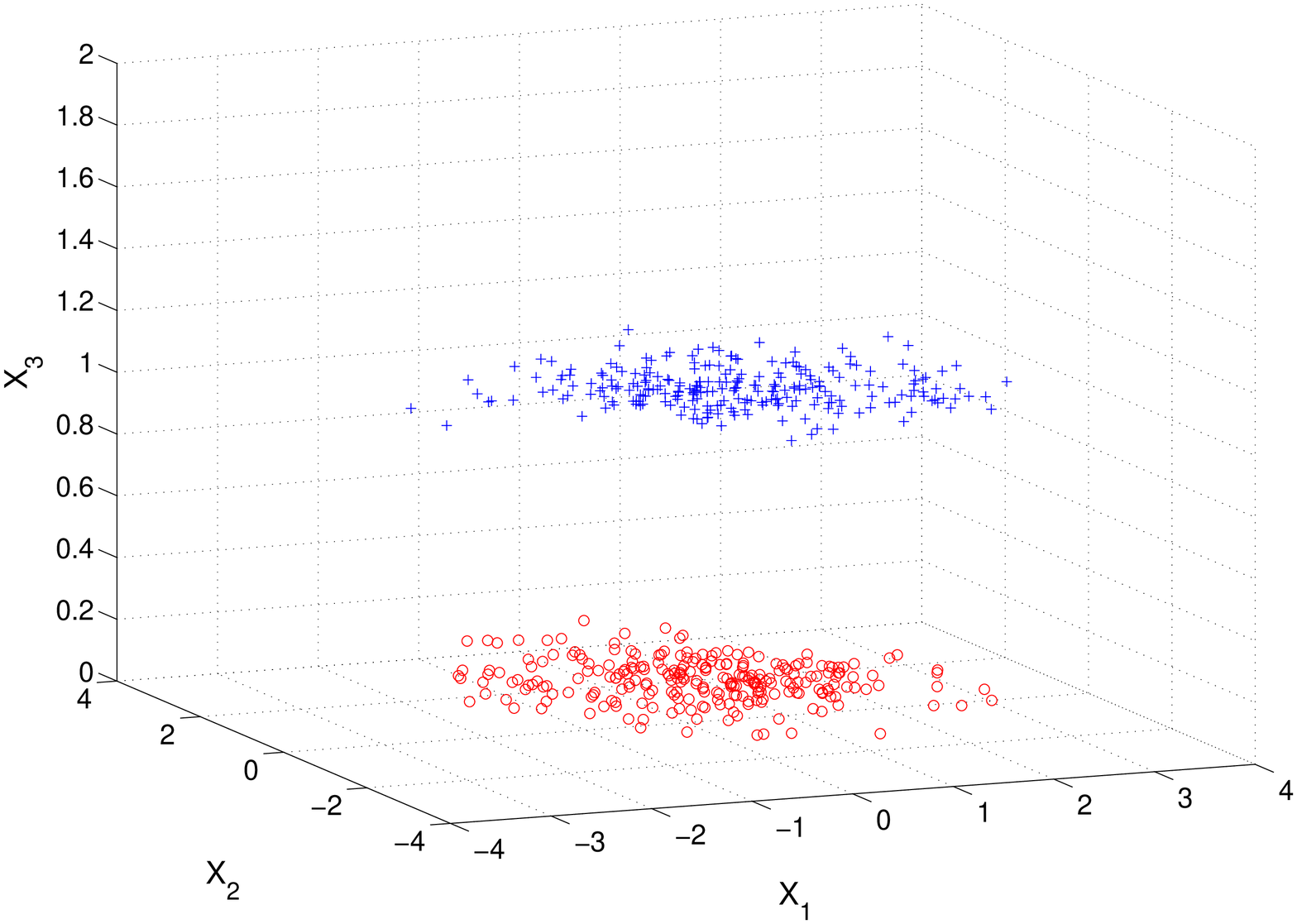}
       \label{fig:classes_nonzero}}}
   \caption{Spatial representation of realizations of the source signals from classes $1$ (in red circles) and $2$ (in blue crosses) for a) the zero-mean two-class problem; and b) the nonzero-mean two-class problem.}
   \label{fig:Fig_2}
 \end{figure*}

For zero-mean two-class problems, we let $\B{\mu}_1 = \B{\mu}_2 = 0$ and for simplicity (and without loss of generality) $\B{\Sigma}_1 = \mathrm{diag}\left(1,1,0\right)$ and $\B{\Sigma}_2 = \mathrm{diag}\left(0,1,1\right)$.  The realizations of the signals in class 1 live in the $x_1$--$x_2$ plane whereas the realizations of the signals in class 2 live in the $x_2$--$x_3$ plane; in addition, the dimension of the intersection of the sub-spaces associated with the signals in classes 1 and 2 is equal to one and the number of non-overlapping dimensions defined in \eqref{nodim} is equal to two. Realizations of the source are depicted in Figure \ref{fig:classes}.

The measurement matrix for this zero-mean two-class problem is constructed by taking the first $M$ rows of the matrix:
\begin{equation}
\mathbf{\Phi}_0 = \left[\begin{array}{cccc}
1&0&0\\ 0&0&1\\ 0&1&0
\end{array} \right],
\end{equation}
according to the desired number of measurements $M$. Note that the first two rows of this matrix, which are consistent with the optimal design in Theorem 4, enable us to achieve the maximum diversity-order $d_{max} = \frac{1}{4}NO_{Dim} = \frac{1}{2}$ and the third row only provides for additional measurement gain.

For nonzero-mean two-class problems, we let $\B{\mu}_1 = \left[0.328, 0.264, 0.114\right]^T$, $\B{\mu}_2 = \left[1,1,1\right]^T$ and once again for simplicity $\B{\Sigma}_1 = \B{\Sigma}_2 = \mathrm{diag}\left(1,1,0\right)$. Note that the realizations of the signals in classes 1 and 2 are in two parallel affine spaces that differ by a translation corresponding to the vector $\B{\mu}_1 - \B{\mu}_2$. Note also that the geometry is such that the condition in \eqref{nonzero_design} in Theorem \ref{theorem5} is satisfied. Realizations of the source are depicted in Figure \ref{fig:classes_nonzero}.

The measurement matrix for this nonzero-mean two-class problem is now constructed by taking the first $M$ rows of the matrix:
\begin{equation}
\mathbf{\Phi}_0  =\left[\begin{array}{ccc}
0&0&1\\ 1&0&0\\ 0&1&0
\end{array} \right],
\end{equation}
according to the desired number of measurements. Note now that the first row of this matrix is in $\mathrm{Null}\left(\mathbf{\Sigma}_1 + \mathbf{\Sigma}_2\right)$, which in accordance with Theorem \ref{theorem5}, provides for infinite diversity-order.

\begin{figure*}[!tbp]
   \centerline{\subfigure[]{\includegraphics[width=3.2in, height=2.5in]{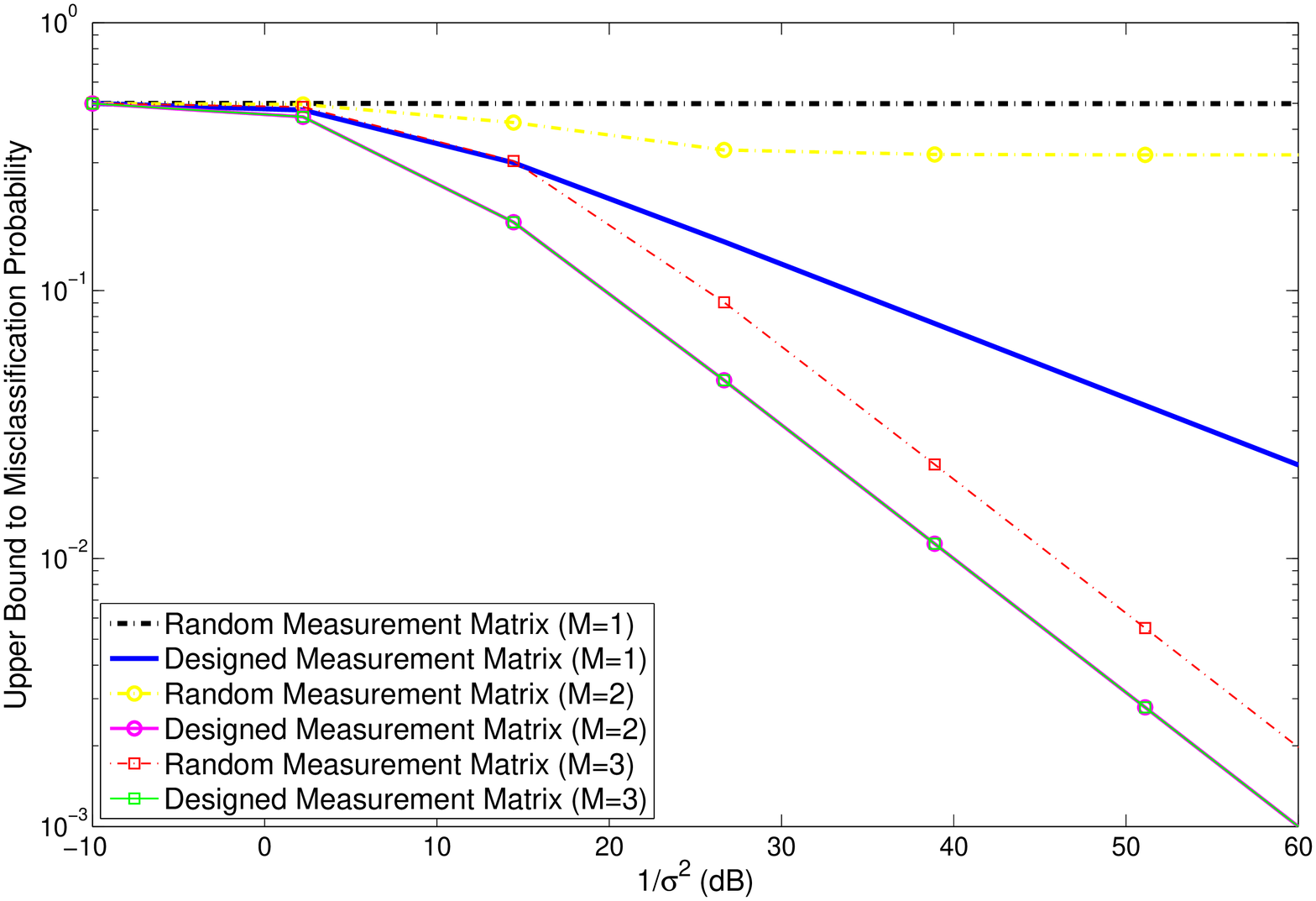}
       \label{fig:pe_comp_upper}}
     \hfil
     \subfigure[]{\includegraphics[width=3.2in, height=2.5in]{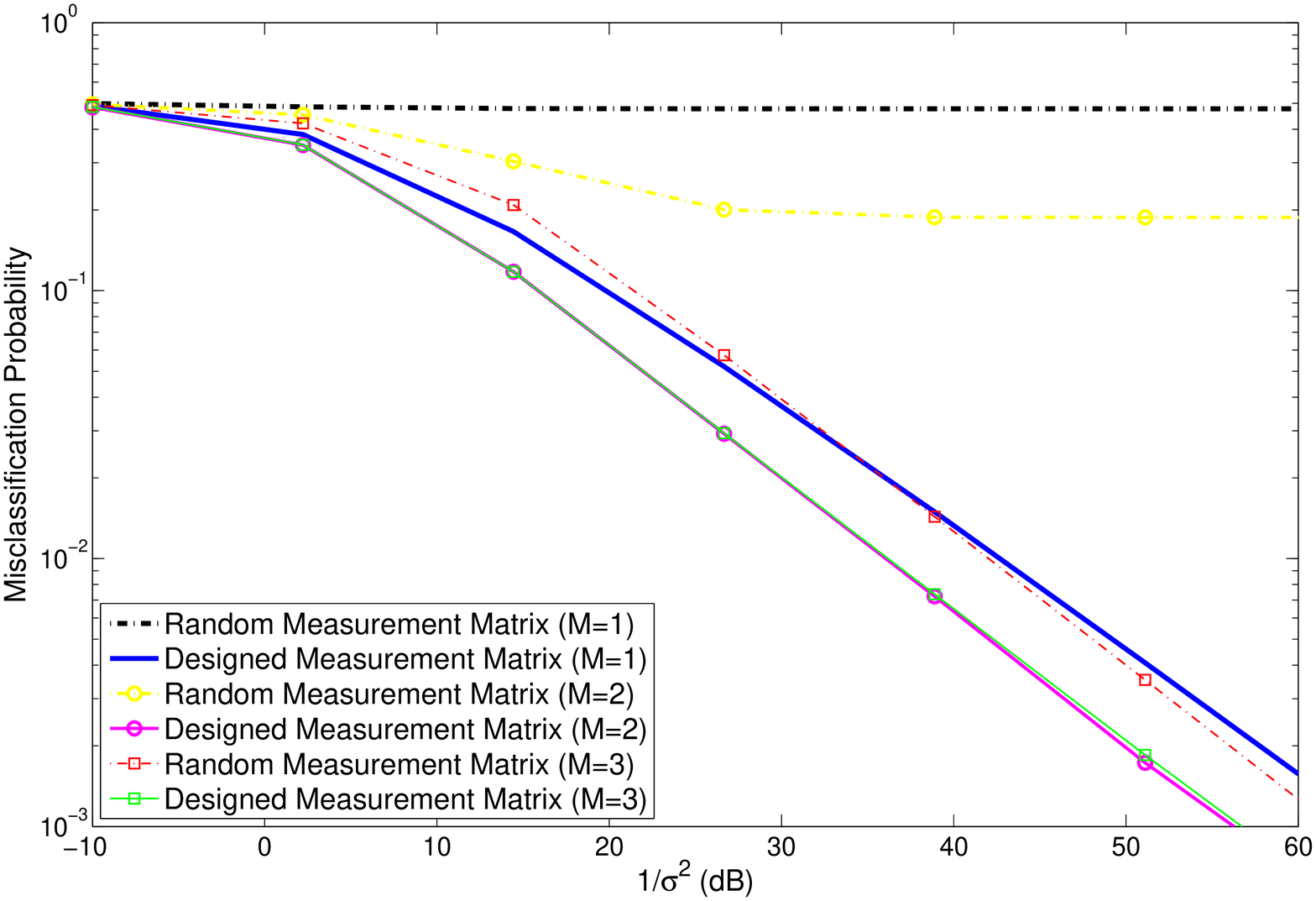}
       \label{fig:pe_comp_real}}}
\vspace{-0.20cm}
   \caption {Upper bound to the probability of misclassification (a) and true probability of misclassification (b) vs. $1/\sigma^2$ (in dB) for random and designed measurements (two zero-mean classes).}
   \label{fig:pe_comp}
\vspace{-0.50cm}
 \end{figure*}

\begin{figure*}[!tbp]
   \centerline{\subfigure[]{\includegraphics[width=3.2in, height=2.5in]{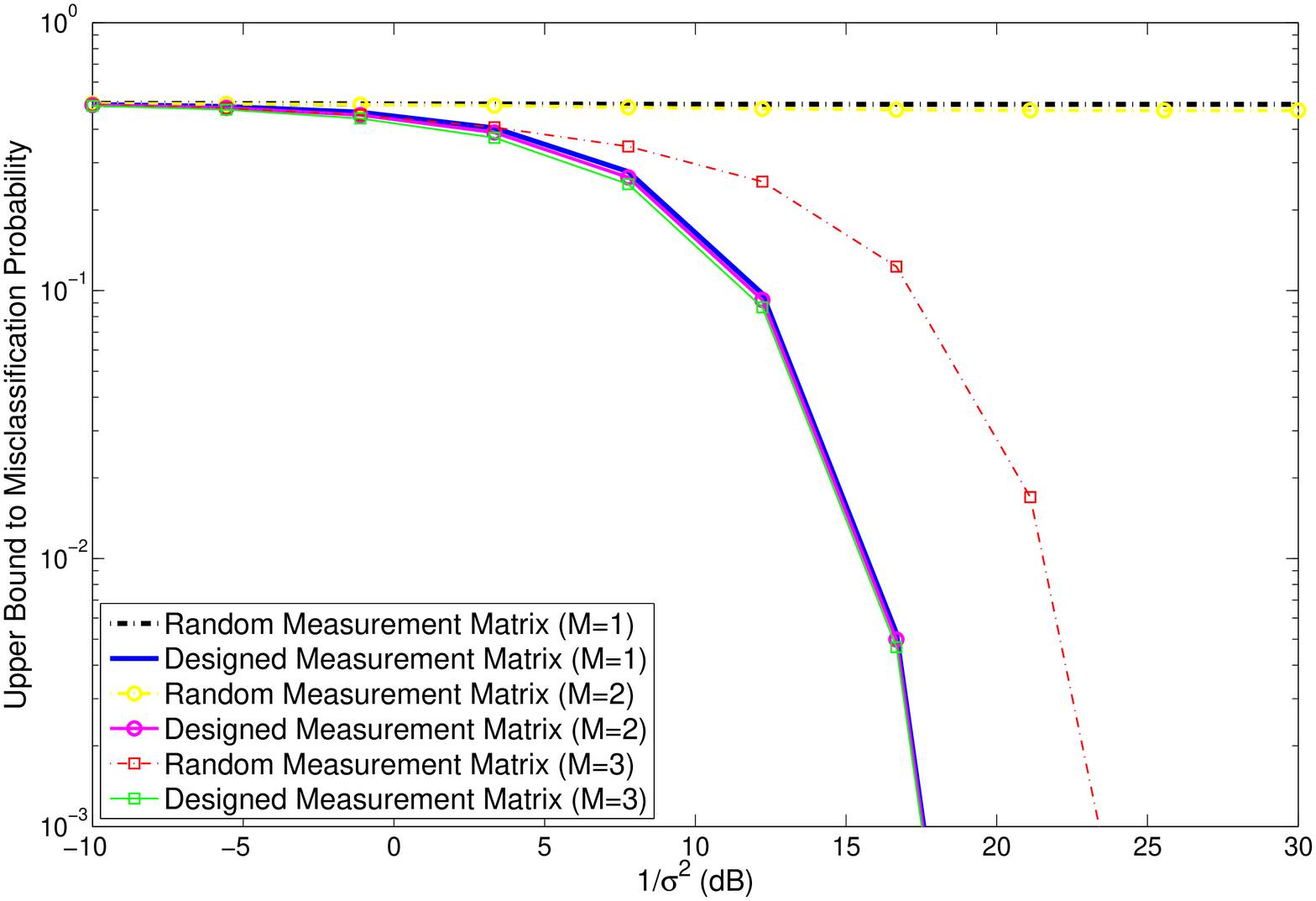}
       \label{fig:pe_comp_nonzero_upper}}
     \hfil
     \subfigure[]{\includegraphics[width=3.2in, height=2.5in]{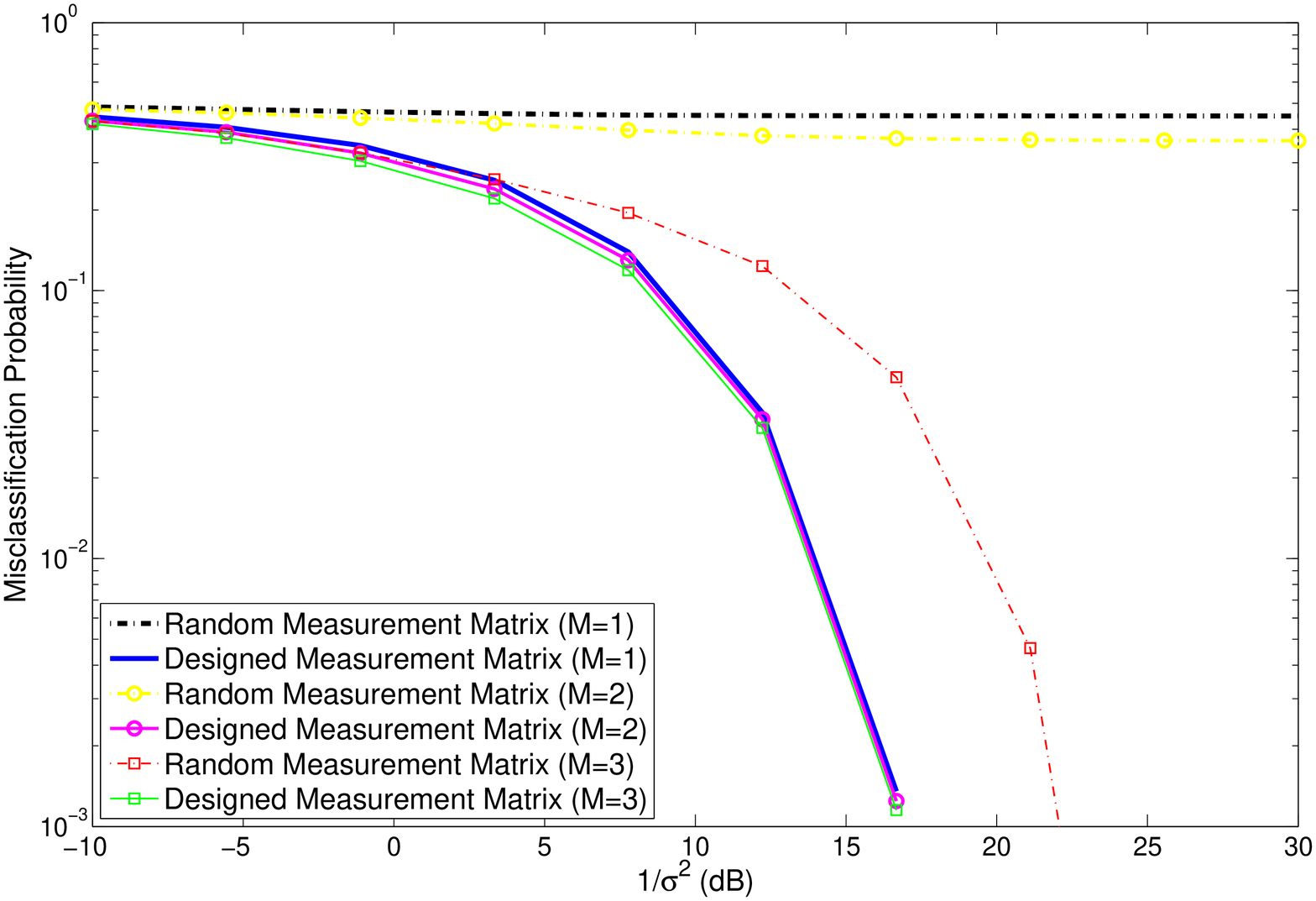}
       \label{fig:pe_comp_nonzero_real}}}
\vspace{-0.20cm}
   \caption {Upper bound to the probability of misclassification (a) and true probability of misclassification (b) vs. $1/\sigma^2$ (in dB) for random and designed measurements (two nonzero-mean classes).}
   \label{fig:pe_comp_nonzero}
\vspace{-0.50cm}
 \end{figure*}

\begin{figure*}[tbp]
   \centerline{\subfigure[Random Measurements, $M=1$]{\includegraphics[width=2.13in, height=1.8in]{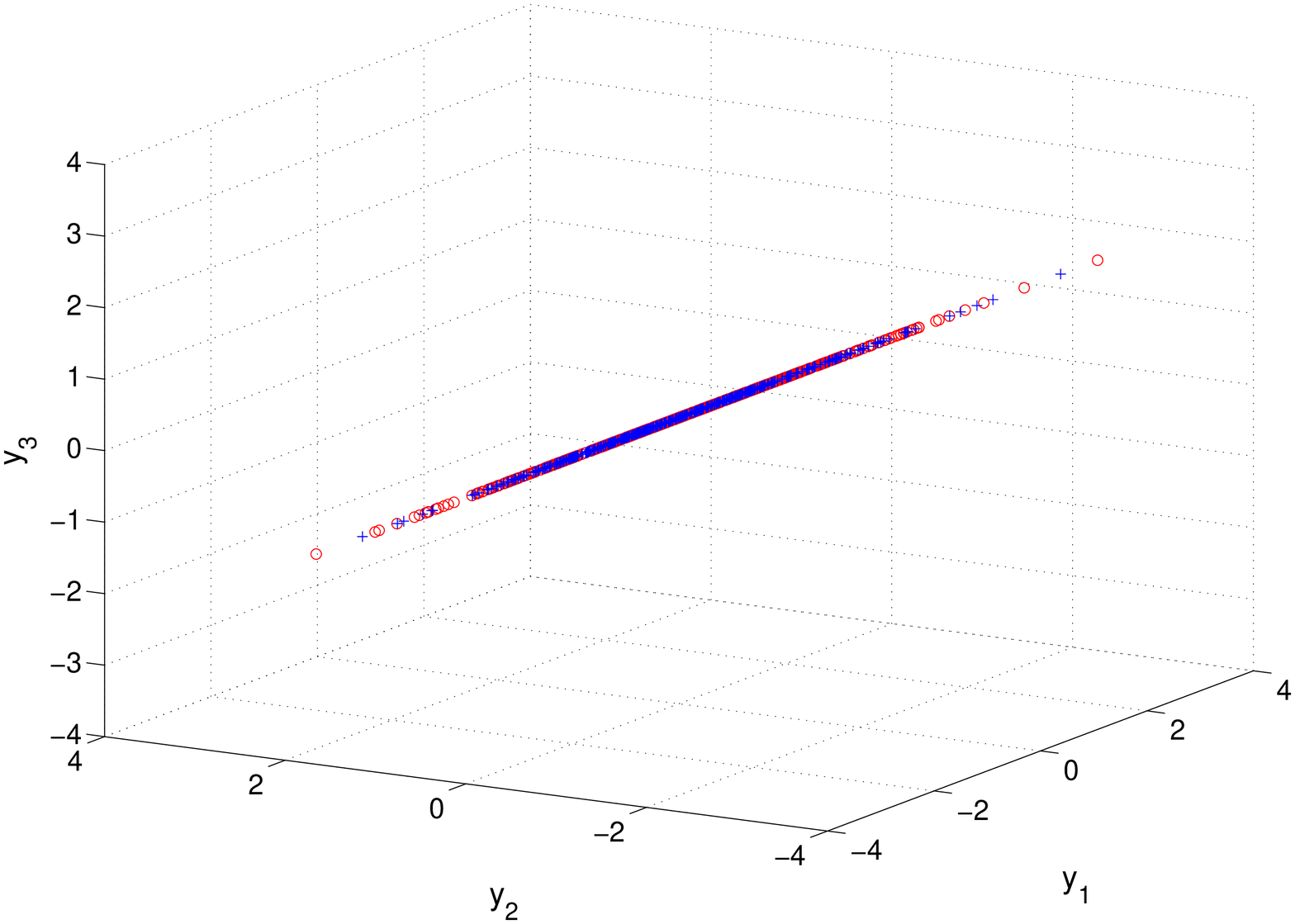}
       \label{fig:proj_r_1}}
     \hfil
     \subfigure[Random Measurements, $M=2$]{\includegraphics[width=2.13in, height=1.8in]{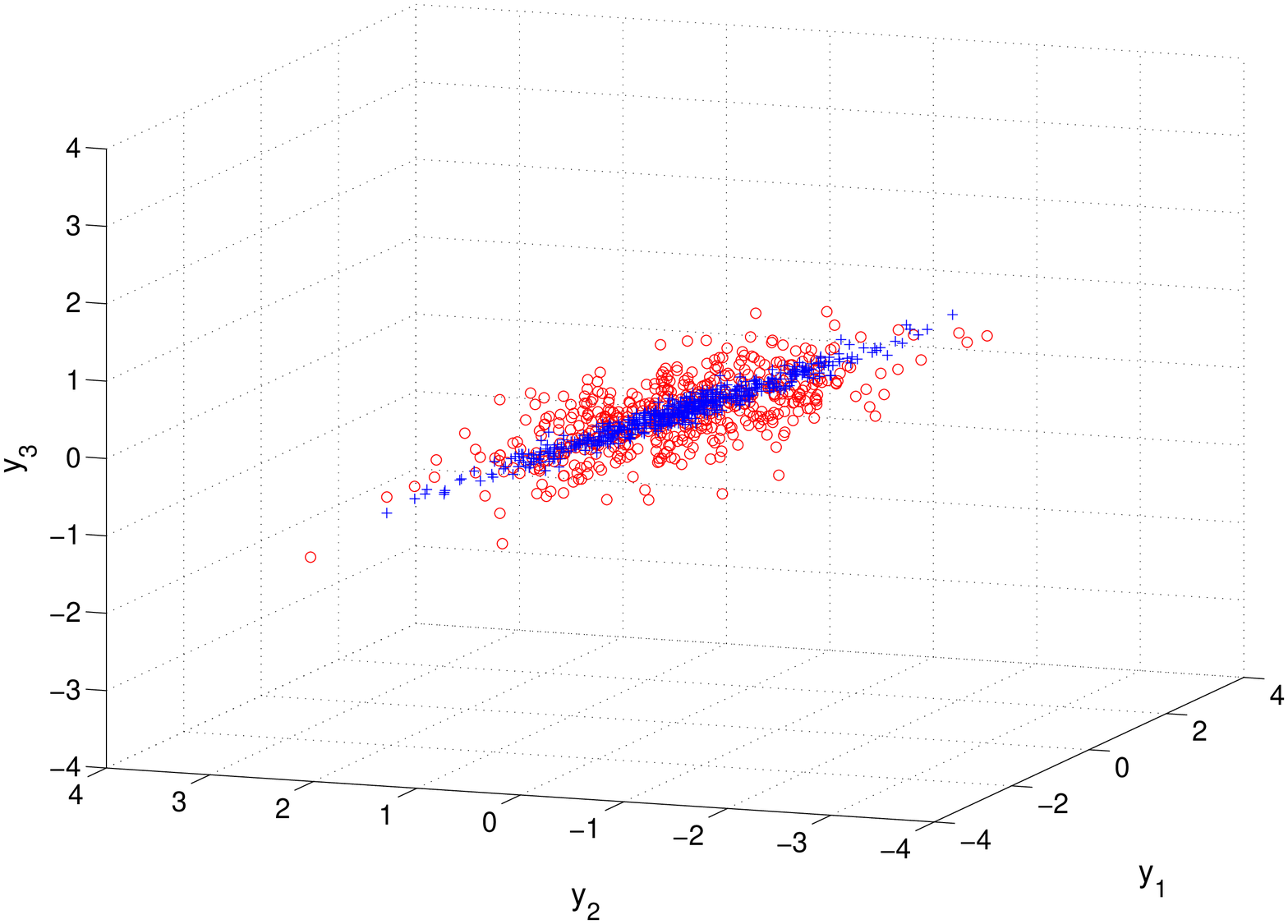}
       \label{fig:proj_r_2}}
\hfil
     \subfigure[Random Measurements, $M=3$]{\includegraphics[width=2.13in, height=1.8in]{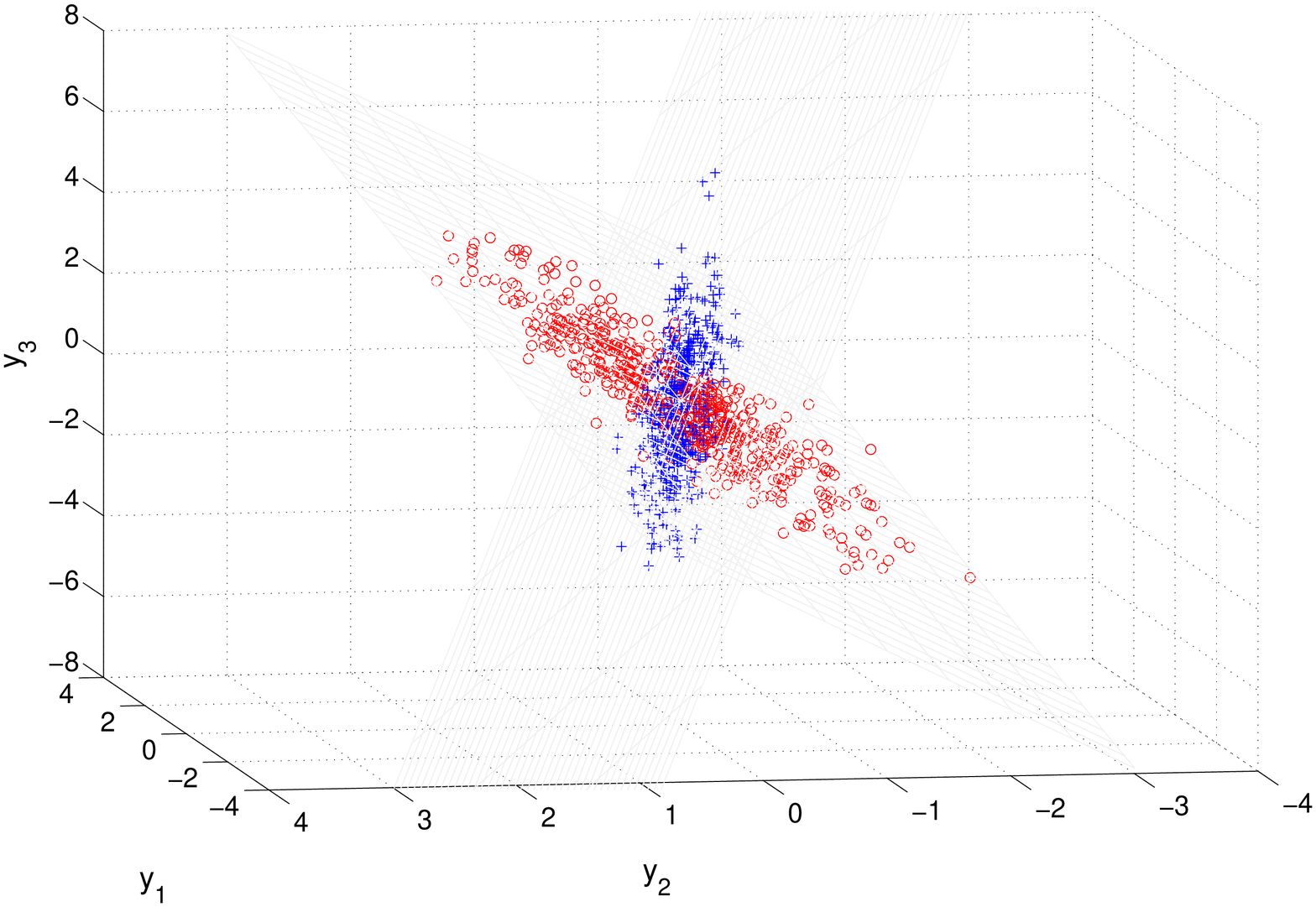}
       \label{fig:proj_r_3}}}
\vspace{-0.2cm}
   \centerline{\subfigure[Designed Measurements, $M=1$]{\includegraphics[width=2.13in, height=1.8in]{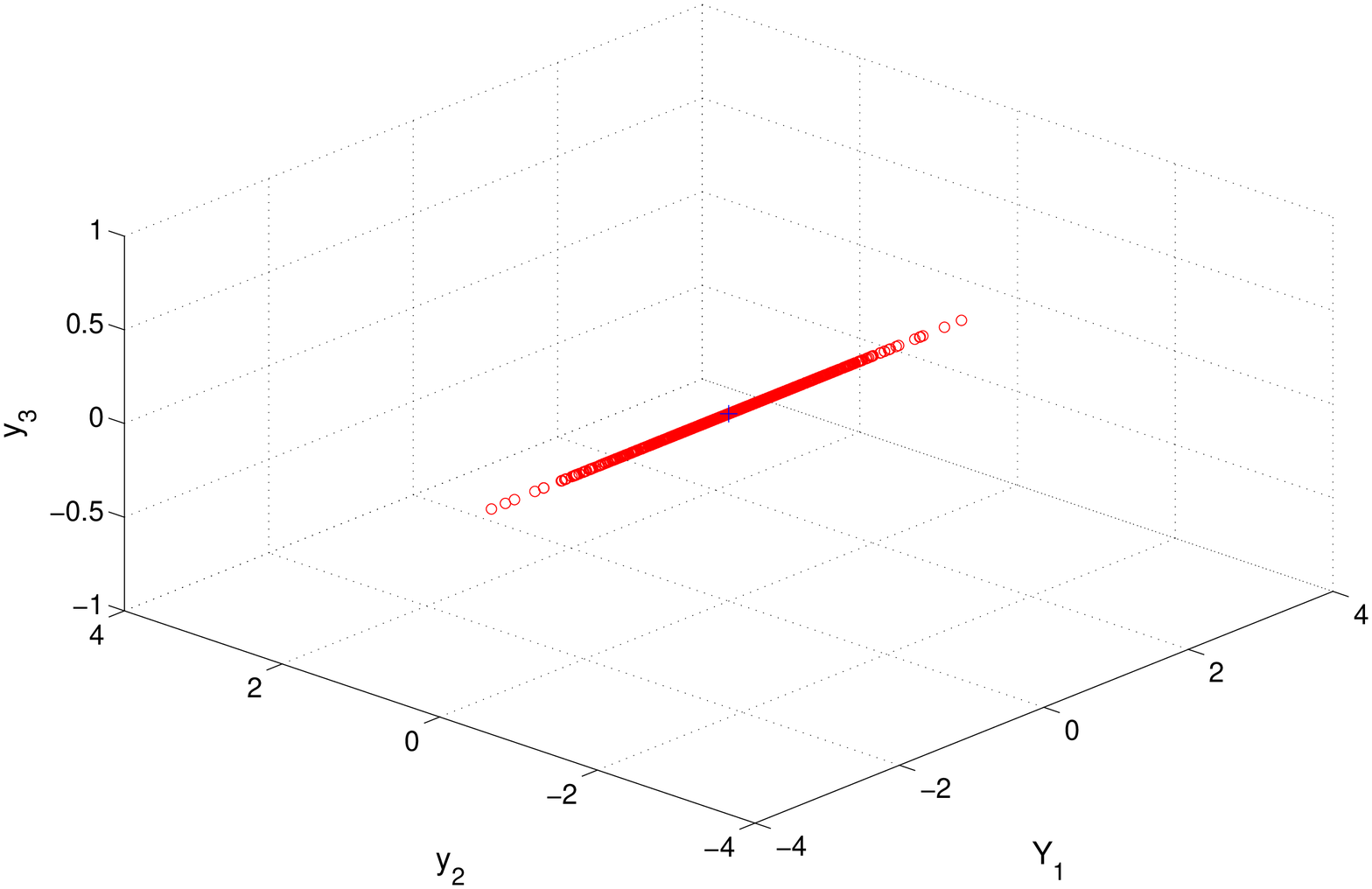}
       \label{fig:proj_d_1}}
     \hfil
     \subfigure[Designed Measurements, $M=2$]{\includegraphics[width=2.13in, height=1.8in]{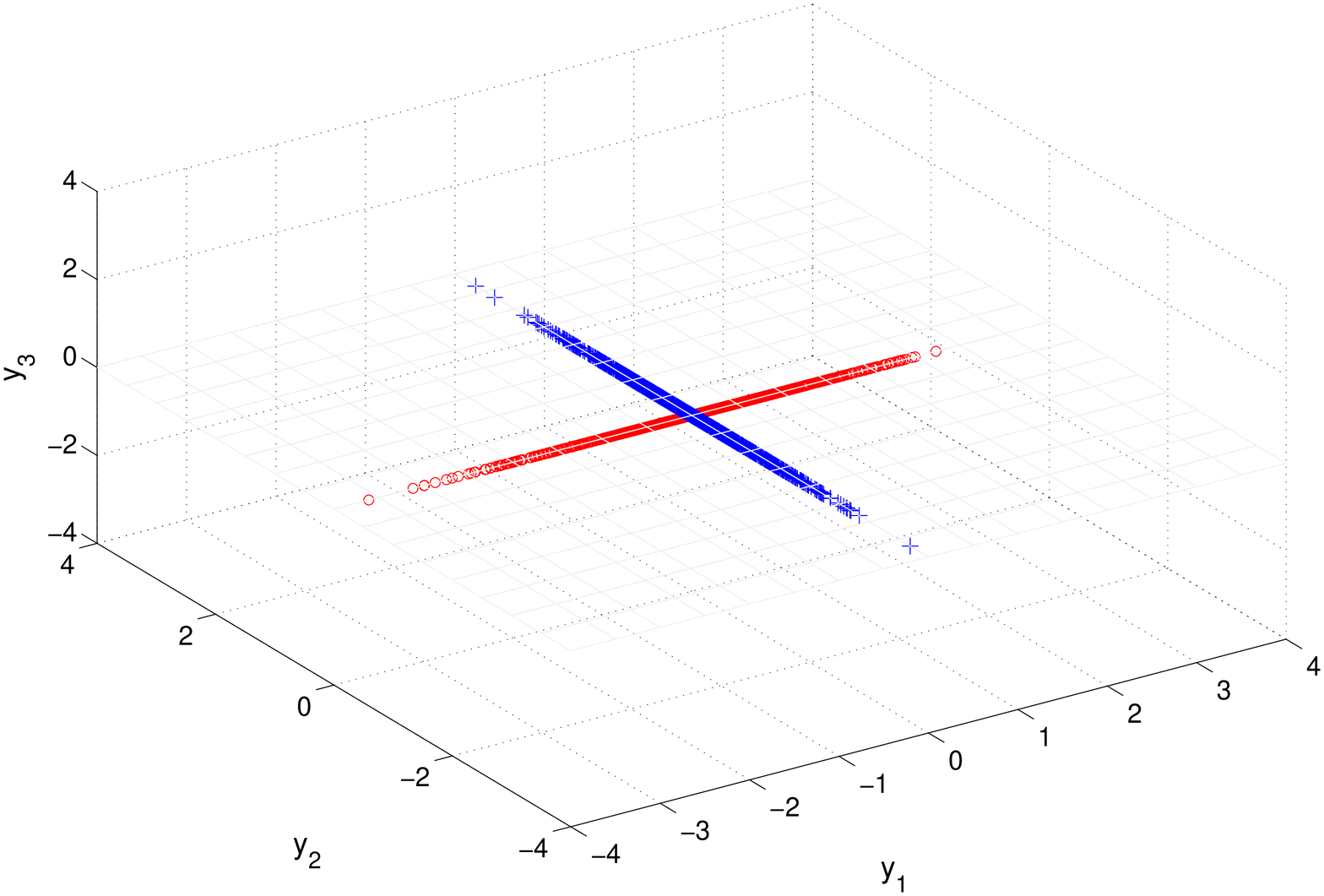}
       \label{fig:proj_d_2}}
\hfil
     \subfigure[Designed Measurements, $M=3$]{\includegraphics[width=2.13in, height=1.8in]{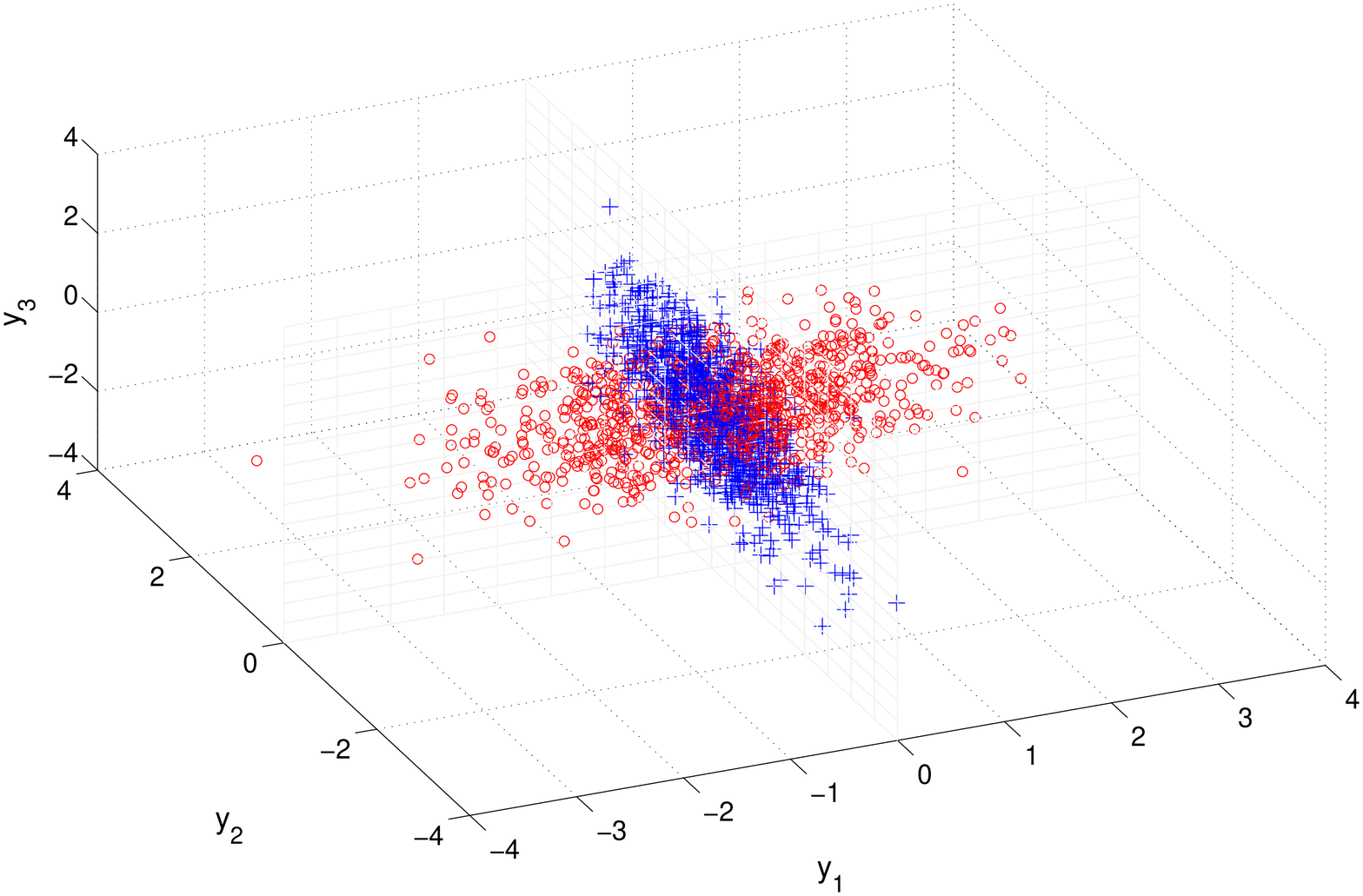}
       \label{fig:proj_d_3}}}
\vspace{-0.2cm}
   \caption{Spatial representation of realizations of noiseless projected source signals from classes $1$ (in red circles) and $2$ (in blue crosses) for zero-mean classes: (a) Random Measurements, $M=1$; (b) Random Measurements, $M=2$; (c) Random Measurements, $M=3$; (d) Designed Measurements, $M=1$; (e) Designed Measurements, $M=2$; (f) Designed Measurements, $M=3$.}
   \label{fig:Fig_3}
 \end{figure*}

\begin{figure*}[htbp]
   \centerline{\subfigure[Random Measurements, $M=1$]{\includegraphics[width=2.13in, height=1.8in]{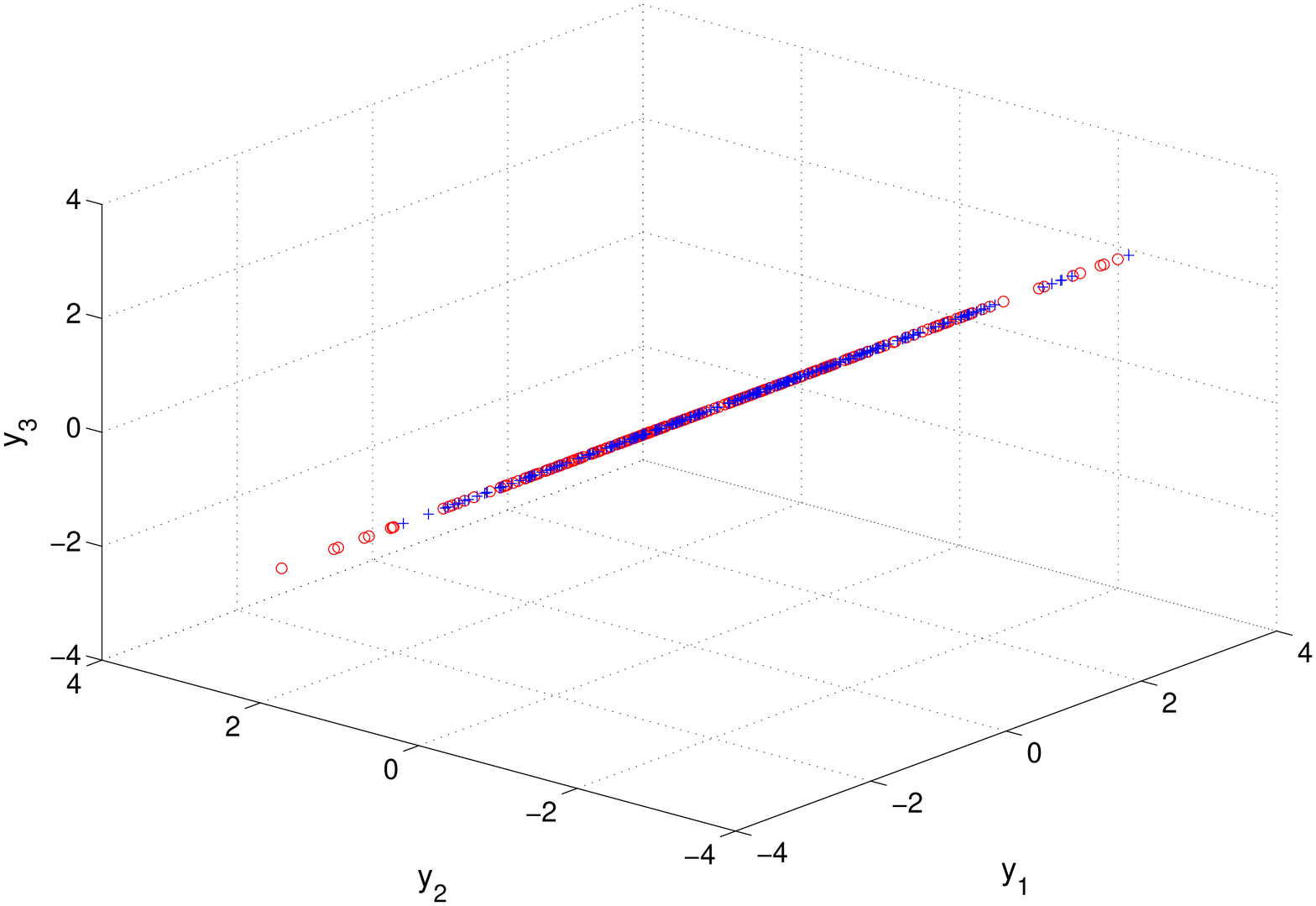}
       \label{fig:proj_r_1_nonzero}}
     \hfil
     \subfigure[Random Measurements, $M=2$]{\includegraphics[width=2.13in, height=1.8in]{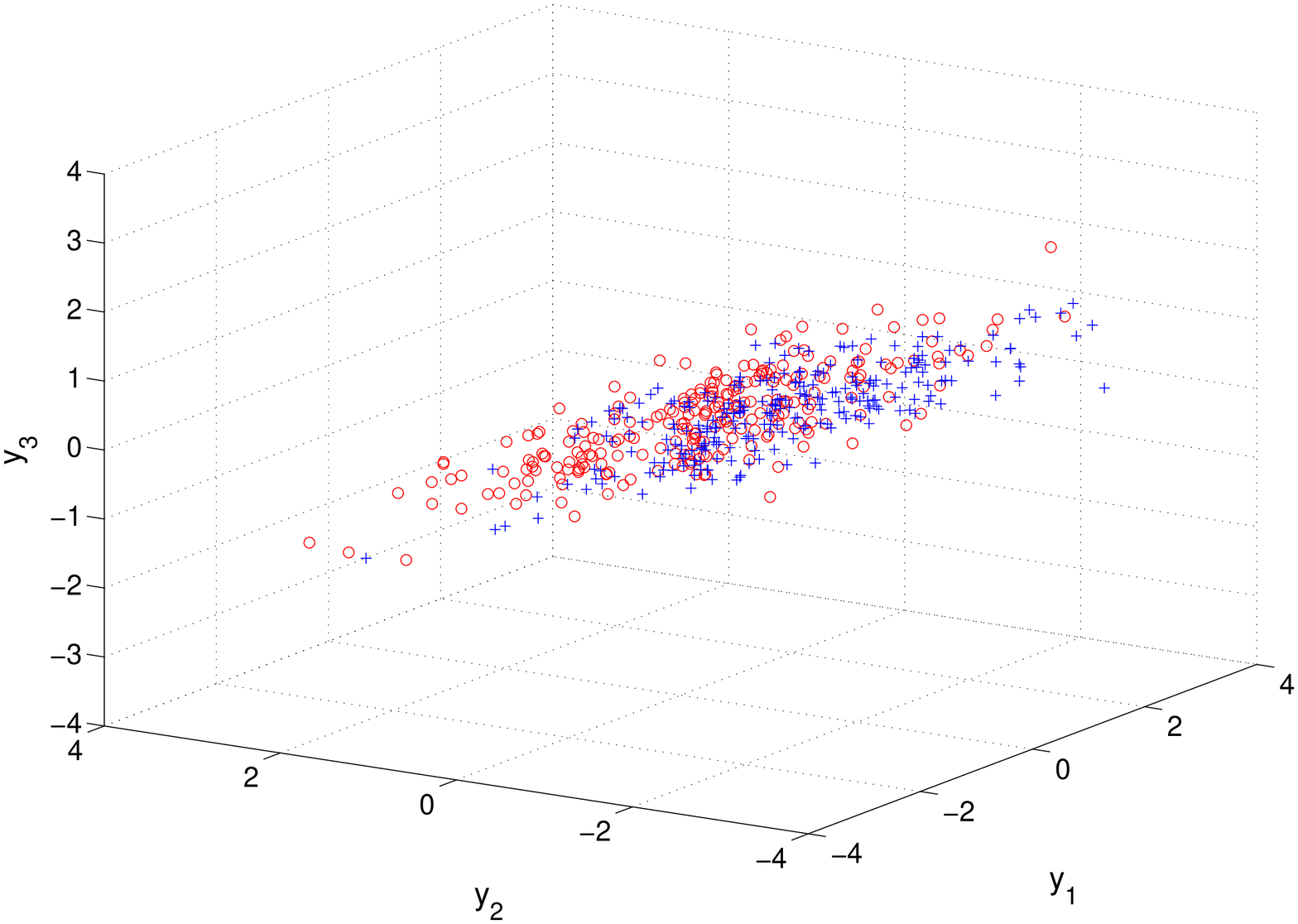}
       \label{fig:proj_r_2_nonzero}}
\hfil
     \subfigure[Random Measurements, $M=3$]{\includegraphics[width=2.13in, height=1.8in]{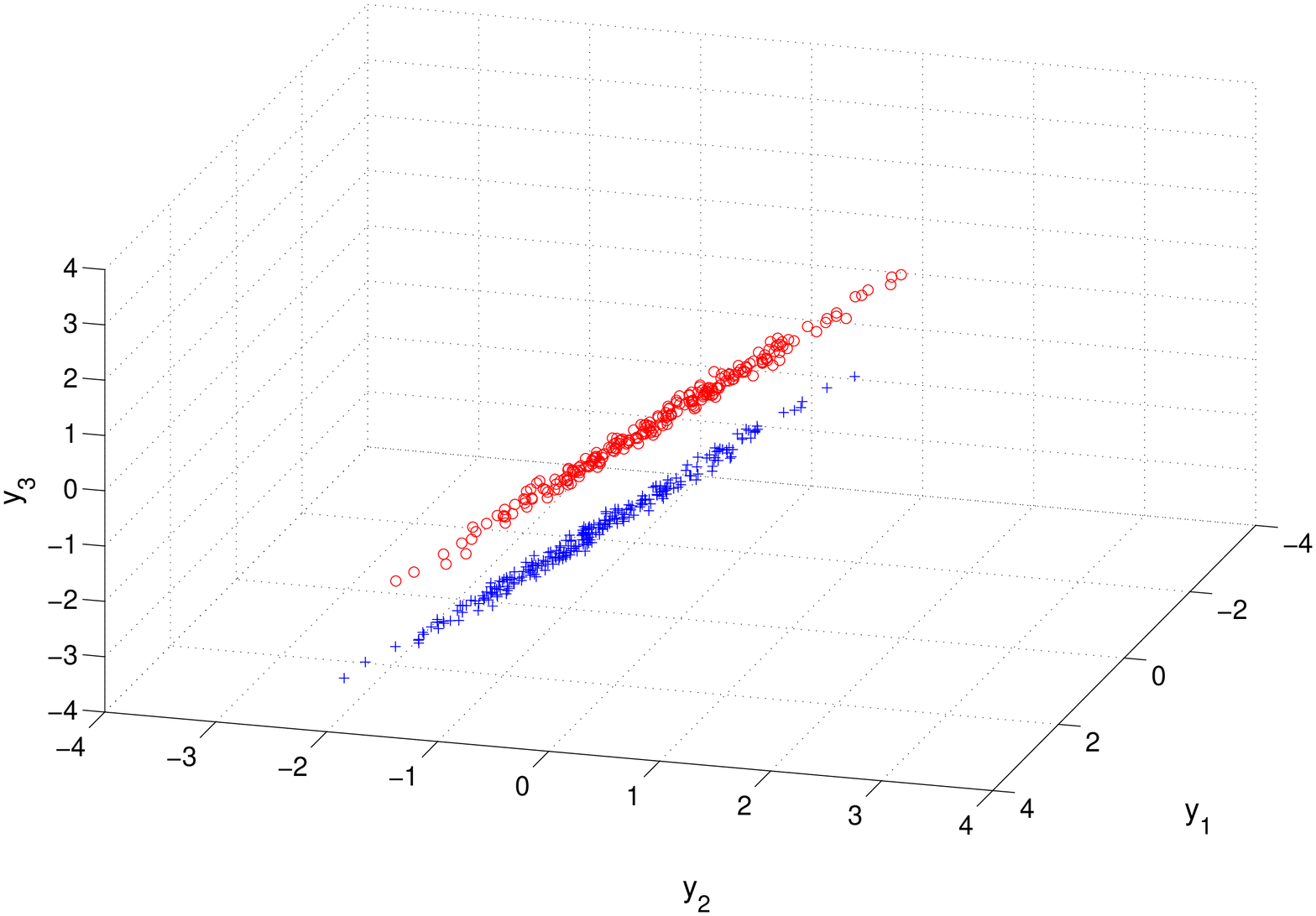}
       \label{fig:proj_r_3_nonzero}}}
\vspace{-0.2cm}
   \centerline{\subfigure[Designed Measurements, $M=1$]{\includegraphics[width=2.13in, height=1.8in]{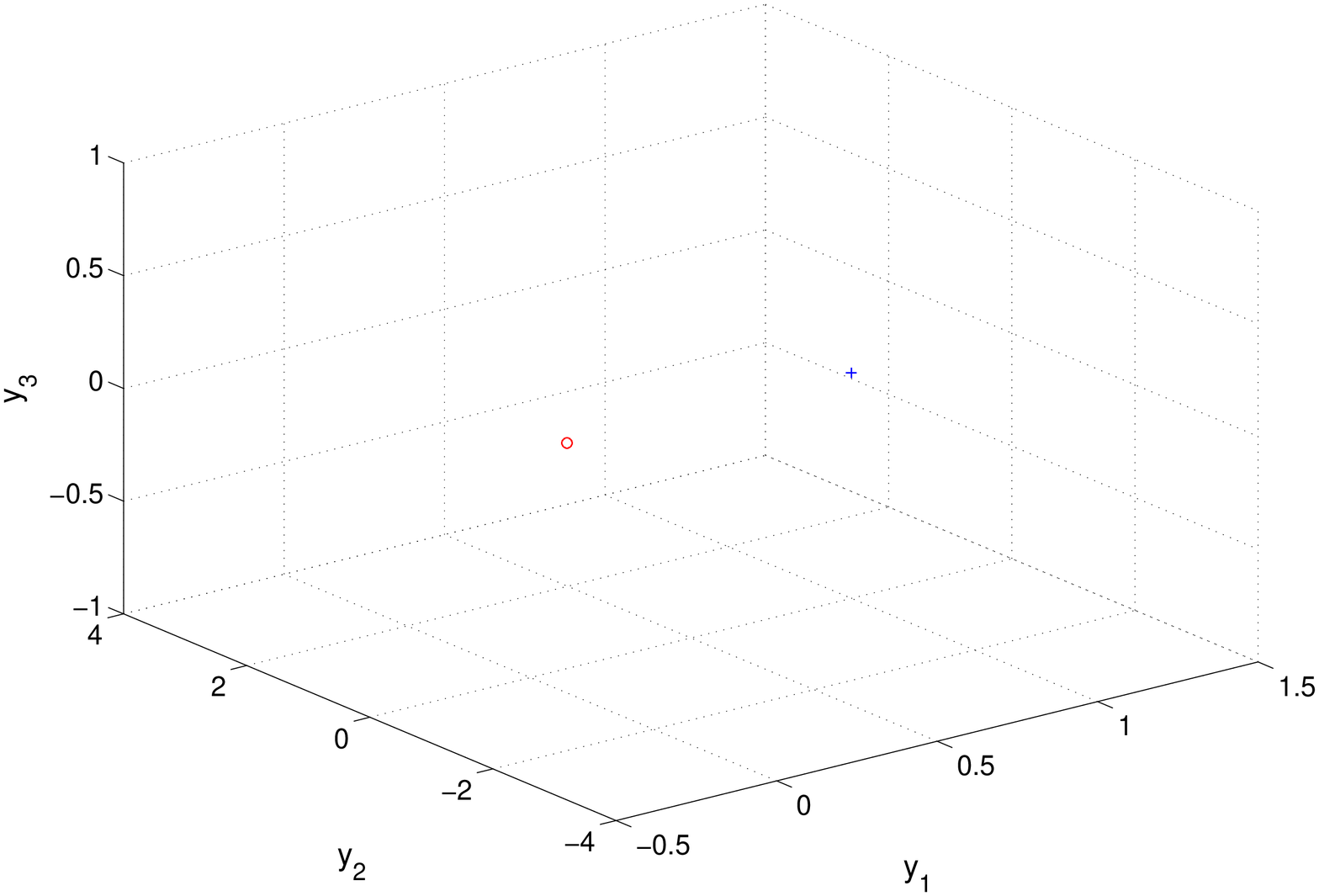}
       \label{fig:proj_d_1_nonzero}}
     \hfil
     \subfigure[Designed Measurements, $M=2$]{\includegraphics[width=2.13in, height=1.8in]{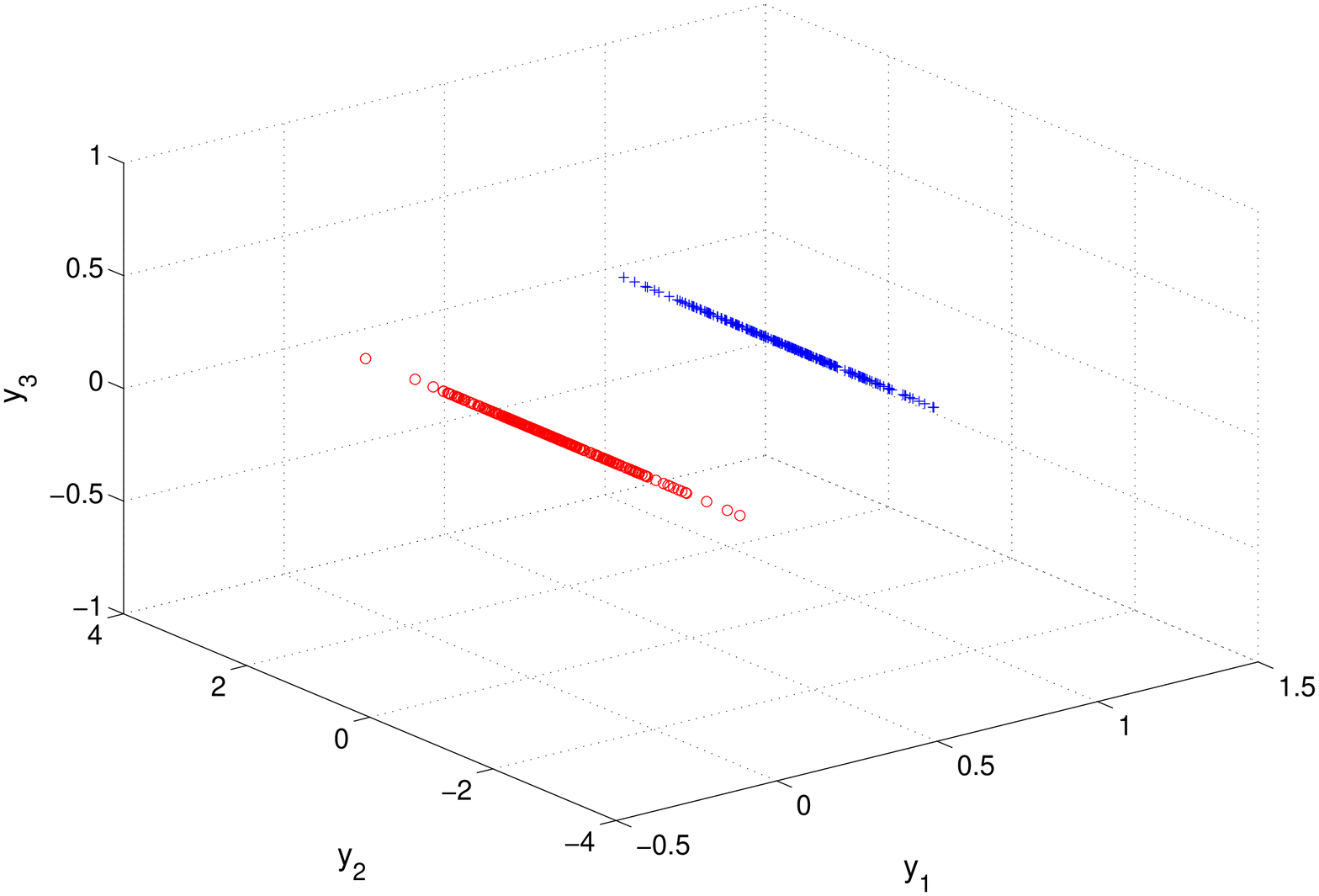}
       \label{fig:proj_d_2_nonzero}}
\hfil
     \subfigure[Designed Measurements, $M=3$]{\includegraphics[width=2.13in, height=1.8in]{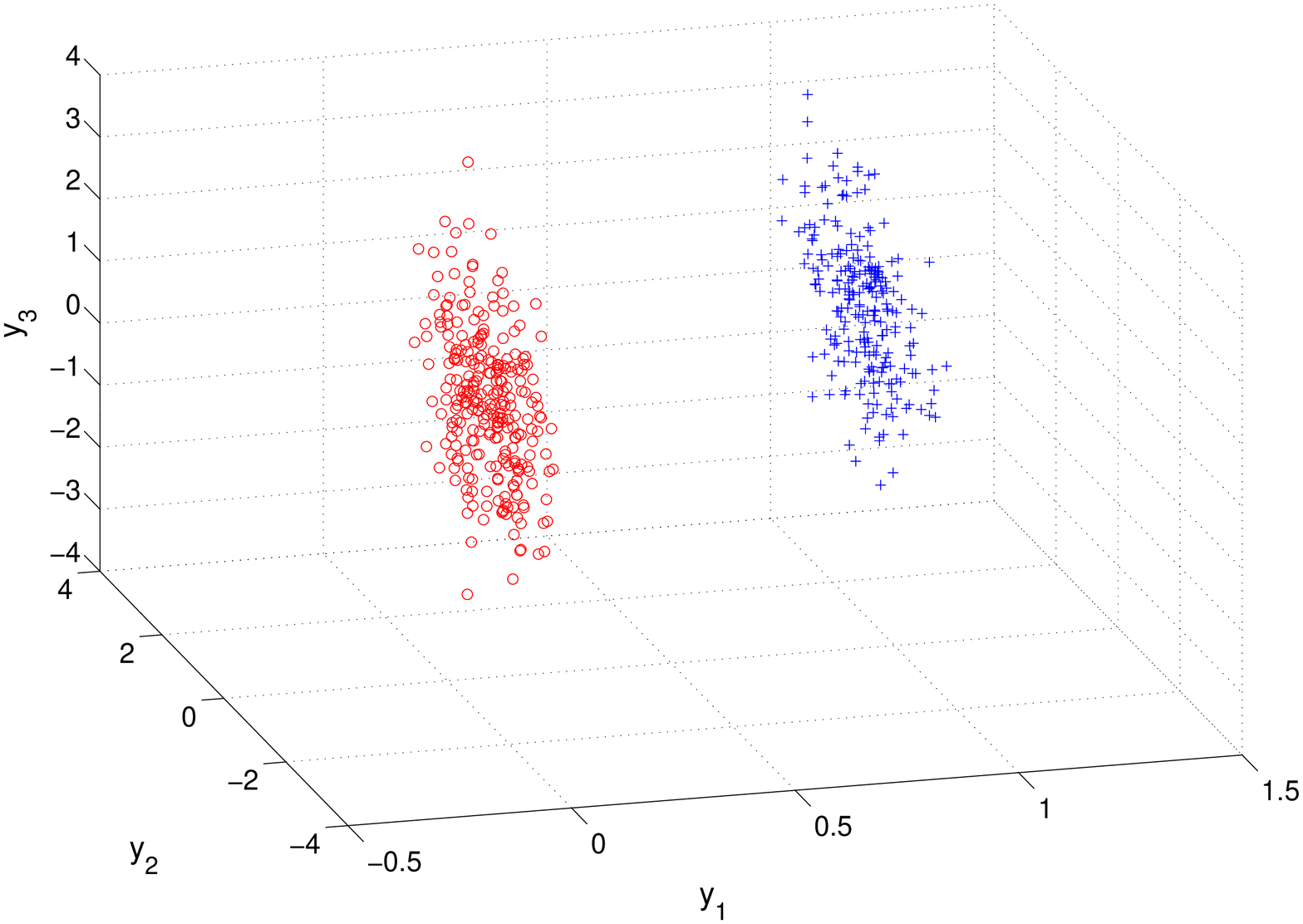}
       \label{fig:proj_d_3_nonzero}}}
\vspace{-0.2cm}
   \caption{Spatial representation of realizations of noiseless projected source signals from classes $1$ (in red circles) and $2$ (in blue crosses) for nonzero-mean classes: (a) Random Measurements, $M=1$; (b) Random Measurements, $M=2$; (c) Random Measurements, $M=3$; (d) Designed Measurements, $M=1$; (e) Designed Measurements, $M=2$; (f) Designed Measurements, $M=3$.}
   \label{fig:Fig_4}
 \end{figure*}

Figures \ref{fig:pe_comp_upper} and \ref{fig:pe_comp_real}  compare the performance  of the random measurements to the designed measurements for the zero-mean two-class classification problem. One observes that $M=3$ random measurements are necessary to eliminate the error floor in accordance with Theorem \ref{theorem2}; however, a single $M=1$ designed measurement is sufficient to drive the misclassification probability to zero in accordance with Theorem \ref{theorem4} due to the ability to focus on unique features exhibited by the classes. One also observes that it is possible to increase the diversity-order to $d_{max} =\frac{1}{4}NO_{Dim} = \frac{1}{2}$ by increasing the number of designed measurements from $M=1$ to $M=2$ and one additional measurement does not result in further increases in the diversity-order. Figures \ref{fig:pe_comp_nonzero_upper} and \ref{fig:pe_comp_nonzero_real} instead compare the performance of random to designed measurements for the nonzero-mean two-class classification problem. In line with Theorems \ref{theo:nonzero} and \ref{theorem5}, $M=3$ random measurements are necessary for the misclassification probability to decay exponentially as $\sigma^2 \to 0$ because with $M \leq r_{\mathbf{\Sigma}_{12}} = 2$ condition \eqref{eq:im} can not be satisfied, but a single designed measurement is sufficient for the purpose. Overall, this behavior is also corroborated by the spatial representation of noiseless realizations of the projected signals portrayed in Figures \ref{fig:Fig_3} and \ref{fig:Fig_4}: one can clearly see that fewer measurements are required in the designed case in relation to the random one to perfectly separate the classes.

For the multiple-class classification problem,  we consider $L=3$, $\B{\mu}_1 =\B{\mu}_2 =\B{\mu}_3 = 0$, ${\bf \Sigma}_1 = \mathrm{diag}\left(1,0,0\right)$, ${\bf \Sigma}_2 = \mathrm{diag}\left(1,1,0\right)$ and ${\bf \Sigma}_3 = \mathrm{diag}\left(0,1,1\right)$.

We also construct a measurement matrix $\mathbf{\Phi}$ by taking the first $M$ rows from the matrix
\begin{equation}
\mathbf{\Phi}_0 =\left[\begin{array}{ccc}
0&1&0\\ 1&0&0\\ 0&0&1
\end{array} \right],
\end{equation}
according to the number of measurements $M$. Note that this construction of $\B{\Phi}$ follows the approach in Algorithm~1 for the measurement budget $M=3$, as each row represents a measurement that achieves diversity-order $1/4$ for a given pair of classes. The advantage of using designed measurements in \emph{lieu} of random ones is also apparent for multiple-class problems. In the designed case one needs only two measurements to drive the misclassification probability to zero as the noise tends to zero with maximal diversity-order. In contrast, with random measurements, one needs at least three random measurements to be able to eliminate the error floor.

Finally, we also notice that the upper bound to the misclassification probability captures well the behavior of the error probability: in particular, it captures well the presence of absence of an error floor, the true diversity-order, increases in diversity-order and increases in the measurement gain. In any case, the diversity-order associated with the upper bound to the misclassification probability always lower bounds the diversity-order associated with the true misclassification probability, thereby establishing performance assurances.

 \begin{figure*}[!tbp]
   \centerline{\subfigure[]{\includegraphics[width=3.2in, height=2.5in]{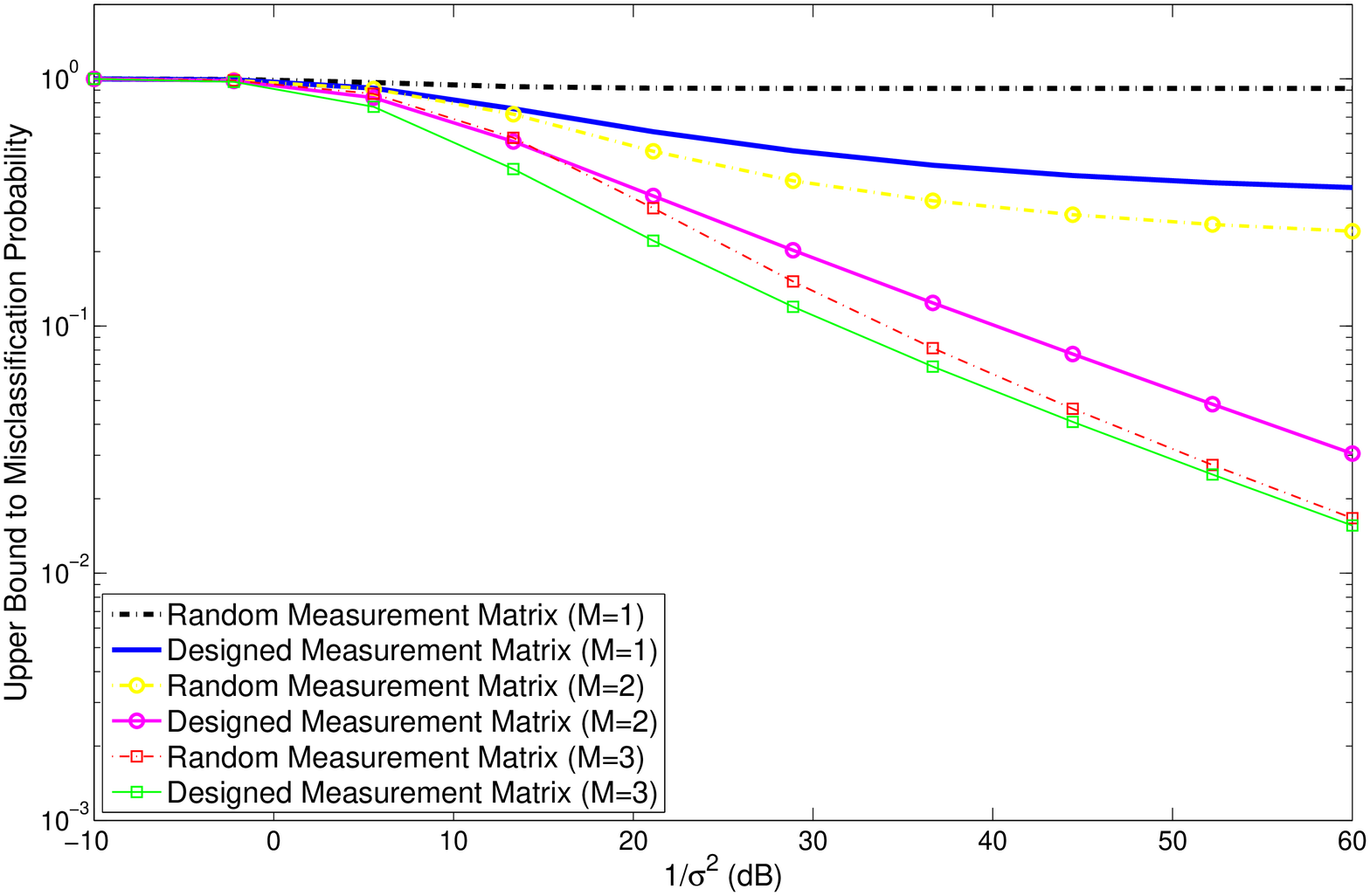}
       \label{fig:pe_comp_multi_upper}}
     \hfil
     \subfigure[]{\includegraphics[width=3.2in, height=2.5in]{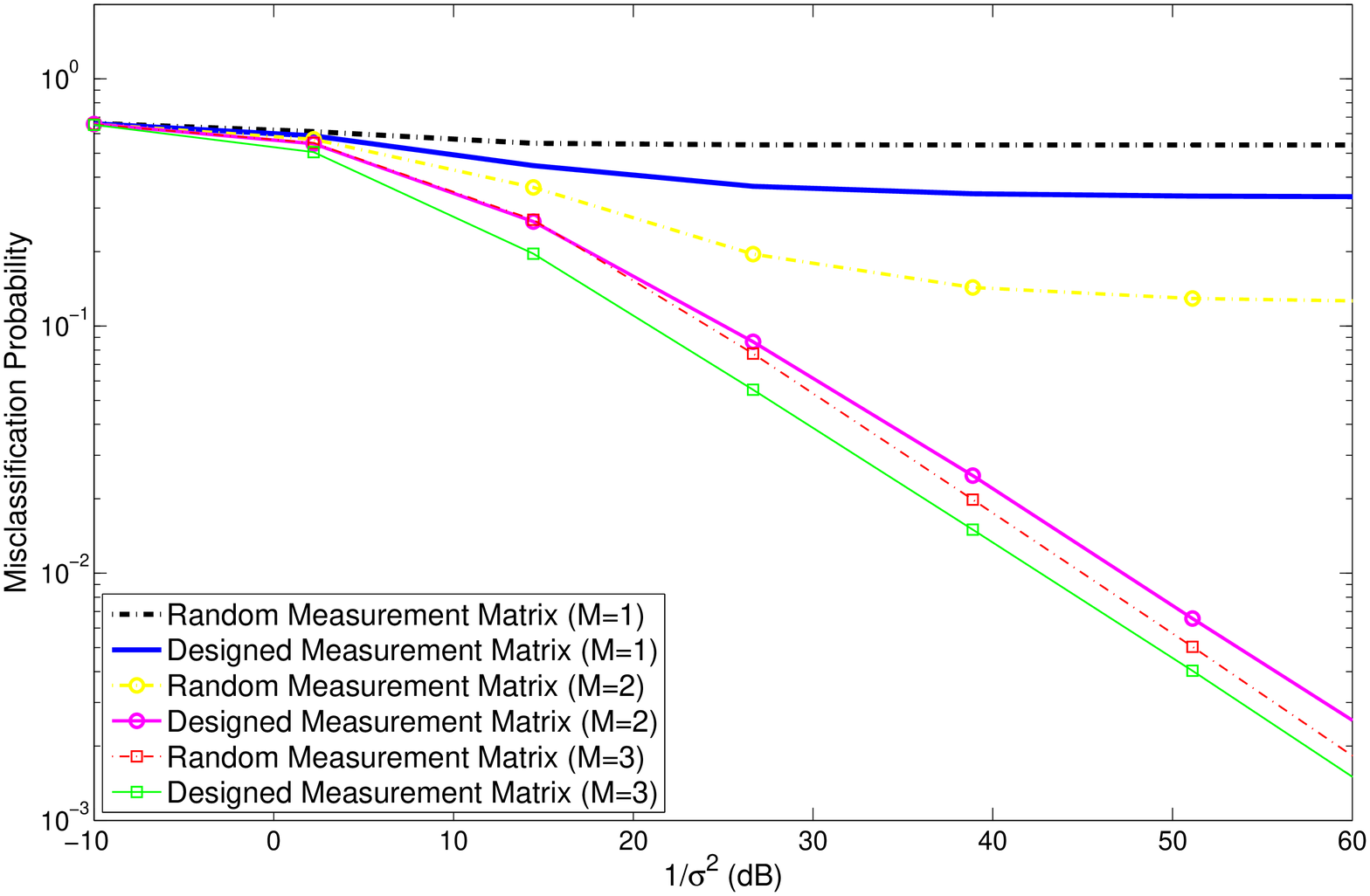}
       \label{fig:pe_comp_multi_real}}}
\vspace{-0.20cm}
   \caption{Upper bound to the probability of misclassification (a) and true probability of misclassification (b) vs $1/\sigma^2$ (in dB) for random and designed measurements (multiple classes).}
   \label{fig:pe_comp_multi}
\vspace{-0.50cm}
 \end{figure*}

\section{Concluding Remarks}
\label{conclusions} 
 
This paper studies fundamental limits in the compressive classification of a mixture of Gaussians by using performance characterizations that are the duals of performance characterizations in multiple-antenna communications systems. In particular, by considering the diversity-order and measurement gain associated with a Bhattacharrya based upper bound to the misclassification probability of the compressive classification problem, which act as the counterparts to the diversity-order and coding gain associated with upper bounds to the error probability of a multiple-antenna communications problem, it has been possible to provide more refined performance characterizations that capture well not only the presence or absence of misclassification error floors (a phase transition) but also increases in the diversity-order and increases in the measurement gain associated with the true misclassification probability of our compressive classification problem.

The proposed characterizations have been used to study the performance of two-class classification problems with zero-mean Gaussians, two-class classification problems with nonzero-mean Gaussians, and multiple-class Gaussian classification problems, both in the presence of random Gaussian i.i.d. measurements and (diversity-order) optimized measurements. One of the hallmarks of the proposed characterizations is the ability to link the concepts of diversity gain and measurement gain with certain fundamental geometrical quantities associated with the measurement and the source models. For example, it has been shown that the ultimate diversity-order in a two-class compressive classification problem with zero-mean Gaussians and, subject to some mild conditions, with nonzero-mean Gaussians, is dictated by the so-called number of non-overlapping dimensions, a quantity that can be interpreted as the number of unique features associated with two-class classification problems. One then understands that measurement is a means to probe such unique features and that designed measurements  provide a quicker route to probe such features in relation to the standard random ones.

Finally, it is also relevant to remark that one possible ramification of the asymptotic performance characterizations and results concerns dictionary learning and design for compressive classification problems. For example, if we are allowed to jointly optimize $\B {\Phi}$, $\B{\Sigma}_1$ and $\B{\Sigma}_2$ in a two-class problem with zero-mean Gaussians, one can easily show that the diversity-maximizing design is such that $r_{12} = \mathrm{rank} \left(\B{\Phi} \left(\B{\Sigma}_1+\B{\Sigma}_2\right) \B{\Phi}^T\right) = N$ and $r_1 + r_2 = \mathrm{rank}\left(\B{\Phi} \B{\Sigma}_1 \B{\Phi}^T\right) + \mathrm{rank}\left(\B{\Phi} \B{\Sigma}_2 \B{\Phi}^T\right)  = r_{12} = N$. This design procedure is in fact reminiscent of recent state-of-the-art methods associated with learning transformations for subspace clustering and classification, that seek to construct a linear transformation on subspaces using matrix rank via its convex surrogate nuclear norm: concretely, the goal is to learn a linear transformation that enforces a low-rank structure for data from the same subspace and that also enforces a high-rank structure for data from different subspaces~\cite{Sap13}.

\appendices

\section{Proof of Theorem \ref{theorem1}}
\label{app_A}

Consider ~the ~eigenvalue ~decomposition ~of ~the ~following matrices ${\bf \Phi}{\bf \Sigma}_1{\bf \Phi}^T = {\bf U}_1{\bf \Lambda}_1{\bf U}_1^{T}$, ${\bf \Phi}{\bf \Sigma}_2{\bf \Phi}^T = {\bf U}_2{\bf \Lambda}_2{\bf U}_2^{T}$, ${\bf \Phi}\left({\bf \Sigma}_1+{\bf \Sigma}_2\right){\bf \Phi}^T = {\bf U}_{12}{\bf \Lambda}_{12}{\bf U}_{12}^{T}$, where $\mathbf{U}_1, \mathbf{U}_2, \mathbf{U}_{12} \in \mathbb{R}^{M \times M}$ are orthogonal matrices and  ${\bf \Lambda}_1$, ${\bf \Lambda}_2$ and ${\bf \Lambda}_{12}$ are positive semidefinite diagonal matrices such that ${\bf \Lambda}_1 = \mathrm{diag}\left(\lambda_{1_1},\cdots,\lambda_{1_{r_1}},0,\cdots,0\right)$, ${\bf \Lambda}_2 =$ $\mathrm{diag}\left(\lambda_{2_1},\cdots,\lambda_{2_{r_2}},0,\cdots,0\right)$, ${\bf \Lambda}_{12} = \mathrm{diag}\left(\lambda_{{12}_1},\cdots,\lambda_{{12}_{r_{12}}},0,\cdots,0\right)$; and $r_1 = \mathrm{rank} \left({\bf \Phi}{\bf \Sigma}_1{\bf \Phi}^T\right)$, $r_2 = \mathrm{rank} \left({\bf \Phi}{\bf \Sigma}_2{\bf \Phi}^T\right)$ and $r_{12} = \mathrm{rank} \left({\bf \Phi}\left({\bf \Sigma}_1+{\bf \Sigma}_2\right){\bf \Phi}^T\right)$.

Therefore, we can re-express the upper bound to the misclassification error probability in \eqref{P_err_Bhat} as follows:
\begin{IEEEeqnarray}{rCl}
\label{app_a_2} 
{P}_{err}^{UB} & = & \sqrt{P_{1}P_{2}}~ e^{\left[-{ \frac{1}{2}\log\frac{\mathrm{det} \left(\frac{{\bf \Phi}\left({\bf \Sigma}_1+{\bf \Sigma}_2\right){\bf \Phi}^T + 2 \sigma^2 \mathbf{I}}{2}\right)}{\sqrt{\mathrm{det} \left({\bf \Phi}{\bf \Sigma}_1{\bf \Phi}^T + \sigma^2 \mathbf{I}\right)  \mathrm{det} \left({\bf \Phi}{\bf \Sigma}_2{\bf \Phi}^T + \sigma^2 \mathbf{I}\right)}} }\right]} \nonumber \\ 
& = & \sqrt{P_{1}P_{2}}~ e^{\left[-{ \frac{1}{2}\log \left[2^{-r_{12}} {\left(\sigma^2\right)}^{\frac{r_1 + r_2}{2} - r_{12}} \frac{   \prod_{i=1}^{r_{12}}\left(\lambda_{{12}_i} + 2\sigma^2\right)    }{\sqrt{  \prod_{i=1}^{r_{1}}\left(\lambda_{{1}_i} + \sigma^2\right)       \prod_{i=1}^{r_{2}}\left(\lambda_{{2}_i} + \sigma^2\right)    }}\right] } \right]}. 
\end{IEEEeqnarray}
The asymptotic characterization of the behavior of the upper bound to the probability of misclassification follows immediately from~\eqref{diversity} and~\eqref{gm} together with~\eqref{app_a_2}. In particular,  
\begin{itemize}
  \item	If $\frac{r_1+r_2}{2} = r_{12}$ then, ~$\displaystyle \lim_{\sigma^2 \to 0}~ {P}_{err}^{UB} = \sqrt{P_{1}P_{2}}~\left[  2^{-r_{12}}   \frac{  v_{12}  }{   \sqrt{v_1 v_2} }  \right]^{-\frac{1}{2}} \neq 0$, \\
\end{itemize}

\begin{itemize}
  \item If $\frac{r_1+r_2}{2} < r_{12}$ then, ~$\displaystyle \lim_{\sigma^2 \to 0}~ {P}_{err}^{UB} =  0$, and by using~\eqref{app_a_2} in~\eqref{diversity} and in~\eqref{gm}:
\end{itemize}
\begin{equation}
	d = -\frac{1}{2}\left(\frac{r_1+r_2}{2}- r_{12}\right)
\end{equation}
and
\begin{equation}
{g_m} = 
{ \left[2^{\frac{r_{12}}{2}}\sqrt{P_{1}P_{2}}\left[\frac{v_{12}}{\sqrt{v_1 v_2}}\right]^{-\frac{1}{2}}\right]^{-\frac{1}{d}}.
}
\end{equation}

\section{Proof of Theorem \ref{theorem2}}
\label{app_B}
Consider once again the eigenvalue decompositions in Appendix~\ref{app_A}. Consider also that  $N \geq r_{{\bf \Sigma}_{12}} \geq \max\left(r_{{\bf \Sigma}_1},r_{{\bf \Sigma}_2}\right)$ and, with  probability 1, $r_1 = \min\left(M,r_{{\bf \Sigma}_1}\right)$, $r_2= \min\left(M,r_{{\bf \Sigma}_2}\right)$ and 
$r_{12} = \min\left(M,r_{{\bf \Sigma}_{12}}\right)$. In addition, assume, without any loss of generality, that $r_{{\bf \Sigma}_1} \leq r_{{\bf \Sigma}_2}$. Therefore, we can re-express the upper bound to the misclassification error probability in (\ref{P_err_Bhat}) as follows:
\begin{IEEEeqnarray}{rCl}
\label{app_b_1} 
{P}_{err}^{UB} & = & \sqrt{P_{1}P_{2}}~ e^{\left[-{ \frac{1}{2}\log\frac{\mathrm{det} \left(\frac{{\bf \Phi}\left({\bf \Sigma}_1+{\bf \Sigma}_2\right){\bf \Phi}^T + 2 \sigma^2 \mathbf{I}}{2}\right)}{\sqrt{\mathrm{det} \left({\bf \Phi}{\bf \Sigma}_1{\bf \Phi}^T + \sigma^2 \mathbf{I}\right)  \mathrm{det} \left({\bf \Phi}{\bf \Sigma}_2{\bf \Phi}^T + \sigma^2 \mathbf{I}\right)}} }\right]} \nonumber \\ 
& = & \sqrt{P_{1}P_{2}}~ e^{\left[-{ \frac{1}{2}\log \left[2^{-\min\left(M,r_{{\bf \Sigma}_{12}}\right)} {\left(\sigma^2\right)}^{\frac{\min\left(M,r_{{\bf \Sigma}_1}\right) + \min\left(M,r_{{\bf \Sigma}_2}\right)}{2} - \min\left(M,r_{{\bf \Sigma}_{12}}\right)} \frac{   \prod_{i=1}^{\min\left(M,r_{{\bf \Sigma}_{12}}\right)}\left(\lambda_{{12}_i} + 2\sigma^2\right)    }{\sqrt{  \prod_{i=1}^{r_{1}}\left(\lambda_{{1}_i} + \sigma^2\right)       \prod_{i=1}^{r_{2}}\left(\lambda_{{2}_i} + \sigma^2\right)    }}\right] } \right]}. \IEEEeqnarraynumspace
\end{IEEEeqnarray}
The asymptotic characterization of the behavior of the upper bound to the probability of misclassification follows, once again, immediately from~\eqref{diversity} and~\eqref{gm} together with~\eqref{app_b_1}. In particular, 
\begin{itemize}
  \item If $\frac{r_{{\bf \Sigma}_1}+r_{{\bf \Sigma}_2}}{2} = r_{{\bf \Sigma}_{12}}$ then, ~$\displaystyle \lim_{\sigma^2 \to 0}~ {P}_{err}^{UB} = \sqrt{P_{1}P_{2}}~\left[  2^{-\min\left(M,r_{{\bf \Sigma}_{12}}\right)}   \frac{  v_{12}  }{   \sqrt{v_1 v_2} }  \right]^{-\frac{1}{2}} \neq 0$,  \\
\end{itemize}

\begin{itemize}
  \item If $\frac{r_{{\bf \Sigma}_1}+r_{{\bf \Sigma}_2}}{2} < r_{{\bf \Sigma}_{12}}$ and $M \leq r_{{\bf \Sigma}_1} \leq r_{{\bf \Sigma}_2} \leq r_{{\bf \Sigma}_{12}}$ then, ~$\displaystyle \lim_{\sigma^2 \to 0}~ {P}_{err}^{UB} =  \left[  2^{-M}   \frac{  v_{12}  }{   \sqrt{v_1 v_2} }  \right]^{-\frac{1}{2}} \neq 0$, \\
\end{itemize}

\begin{itemize} 
\item otherwise, ~$\displaystyle \lim_{\sigma^2 \to 0}~ {P}_{err}^{UB} =  0$, and, by using in~\eqref{app_b_1} in ~\eqref{gm} and in~\eqref{diversity}, we can write the measurement gain as:

\begin{equation}
{g_m} = 
{ \left[2^{\frac{\min\left(M,r_{\mathbf{\Sigma}_{12}}\right)}{2}}\sqrt{P_{1}P_{2}}\left[\frac{v_{12}}{\sqrt{v_1 v_2}}\right]^{-\frac{1}{2}}\right]^{-\frac{1}{d}}
}
\end{equation}
and the diversity-order as:

\begin{equation}
d = \threecases{-\frac{1}{2}\left(\frac{r_{{\bf \Sigma}_1} - M}{2}\right)}{if $r_{{\bf \Sigma}_1} < M \leq r_{{\bf \Sigma}_2} \leq r_{{\bf \Sigma}_{12}}$}{-\frac{1}{2}\left(\frac{r_{{\bf \Sigma}_1}+r_{{\bf \Sigma}_2}}{2}- M\right)}{if $r_{{\bf \Sigma}_1} \leq r_{{\bf \Sigma}_2} < M < r_{{\bf \Sigma}_{12}}$}{-\frac{1}{2}\left(\frac{r_{{\bf \Sigma}_1}+r_{{\bf \Sigma}_2}}{2}- r_{{\bf \Sigma}_{12}}\right)}{if $r_{{\bf \Sigma}_1} \leq r_{{\bf \Sigma}_2} \leq r_{{\bf \Sigma}_{12}} \leq M$}.
\end{equation}

\end{itemize}

\section{Proof of Theorem \ref{theo:nonzero}}
\label{app_C}
Consider the upper bound to the classification error probability in (\ref{P_err_Bhat}) and (\ref{exp_Bhat}). We write the exponent $K_{12}$ as follows:
\begin{equation}
K_{12} = T_1 + T_2
\end{equation}
where
\begin{equation}
T_1 = \frac{1}{8} {\left[{\bf \Phi}\left(\boldsymbol{\mu}_1 - \boldsymbol{\mu}_2\right)\right]^T \left[\frac{{\bf \Phi}\left({\bf \Sigma}_1+{\bf \Sigma}_2\right){\bf \Phi}^T + 2 \sigma^2 \mathbf{I}}{2}\right]^{-1} \left[{\bf \Phi}\left(\boldsymbol{\mu}_1 - \boldsymbol{\mu}_2\right)\right]  }
\end{equation}
and
\begin{equation}
T_2 = { \frac{1}{2}\log\frac{\mathrm{det} \left(\frac{{\bf \Phi}\left({\bf \Sigma}_1+{\bf \Sigma}_2\right){\bf \Phi}^T + 2 \sigma^2 \mathbf{I}}{2}\right)}{\sqrt{\mathrm{det} \left({\bf \Phi}{\bf \Sigma}_1{\bf \Phi}^T + \sigma^2 \mathbf{I}\right)  \mathrm{det} \left({\bf \Phi}{\bf \Sigma}_2{\bf \Phi}^T + \sigma^2 \mathbf{I}\right)}} }.
\end{equation}

Let us now define $\mathbf{M}_{12}=\mathbf{\Phi}(\B{\mu}_1 - \B{\mu}_2)  (\B{\mu}_1 - \B{\mu}_2)^T \mathbf{\Phi}^T $. Then, by recalling the eigenvalue decomposition of the matrix ${\bf \Phi}\left({\bf \Sigma}_1 + {\bf \Sigma}_2\right){\bf \Phi}^T = {\bf U}_{12}{\bf \Lambda}_{12}{\bf U}_{12}^{T}$ in Appendix~\ref{app_A}, we can also express $T_1$ as:
\begin{IEEEeqnarray}{rCl}
T_1 & = & \frac{1}{4} \tr{  \mathbf{M }_{12}  \left( \mathbf{\Phi} (\mathbf{\Sigma}_1 + \mathbf{\Sigma}_2) \mathbf{\Phi}^T + 2 \sigma^2 \mathbf{I}\right)^{-1}}\\
& = & \frac{1}{4} \tr{\mathbf{M}_{12}  \mathbf{U}_{12}  \left(\mathbf{\Lambda}_{12} + 2 \sigma^2 \mathbf{I}\right)^{-1} \mathbf{U}_{12}^T  }\\
 & = & \frac{1}{4} \sum_{i=1}^{r_{12}}  \frac{1}{\lambda_{12_i} + 2 \sigma^2} \mathbf{u}_{12,i}^T \mathbf{M}_{12} \mathbf{u}_{12,i} +  \frac{1}{ 8 \sigma^2} \sum_{i=r_{12}+1}^M  \mathbf{u}_{12,i}^T \mathbf{M}_{12} \mathbf{u}_{12,i},
 \label{eq:T1final}
\end{IEEEeqnarray}
where the vector $ \mathbf{u}_{12,i}$ corresponds to the $i$-th column of the matrix $\mathbf{U}_{12}$. 

Therefore, we can also re-write the exponent $K_{12}$ as follows:
\begin{equation}
K_{12} = \frac{1}{4} \sum_{i=1}^{r_{12}}  \frac{1}{\lambda_{12_i} + 2 \sigma^2} \mathbf{u}_{12,i}^T \mathbf{M}_{12} \mathbf{u}_{12,i} +  \frac{1}{ 8 \sigma^2} \sum_{i=r_{12}+1}^M  \mathbf{u}_{12,i}^T \mathbf{M}_{12} \mathbf{u}_{12,i} + T_2,
\label{K3terms}
\end{equation}
We are now able to extend the characterization of the asymptotic behavior of the Batthacharyya upper bound from the case of zero-mean to nonzero-mean classes by analyzing further \eqref{K3terms}. In particular,
\begin{itemize}
\item If $\boldsymbol{\mu}_1 = \boldsymbol{\mu}_2$ then the asymptotic behavior of the upper bound to the misclassification probability is identical in nonzero-mean and zero-mean cases;

\item If $\boldsymbol{\mu}_1 \neq \boldsymbol{\mu}_2$ and $M \leq r_{\B{\Sigma}_{12}} $, then the matrix ${\bf \Phi}\left({\bf \Sigma}_1 + {\bf \Sigma}_2\right){\bf \Phi}^T$ is full rank (i.e. ${r_{12}}  = M$) and, therefore, condition~\eqref{eq:im} is not verified. In such case the second term in (\ref{K3terms}) is equal to zero and the diversity-order associated with nonzero-mean classes corresponds to that for zero-mean classes unveiled in Theorem~\ref{theorem2}. In contrast, the measurement gain for nonzero-mean classes is higher than that for zero-mean classes in view of the first term in (\ref{K3terms}). In fact, it is immediate to express the measurement gain for nonzero-mean classes $g_m^{\text{NZM}}$ in terms of the measurement gain for zero-mean classes $g_m^{\text{ZM}}$ in (\ref{gm2}) as $g_m^{\text{NZM}} = a~\cdot~g_m^{\text{ZM}}$ where
\begin{equation}
a = \exp \left(  \frac{1}{4d} \sum_{i=1}^{M}  \frac{1}{\lambda_{12_i} } \mathbf{u}_{12,i}^T \mathbf{M}_{12} \mathbf{u}_{12,i} \right)>1.
\end{equation}
\item If $\boldsymbol{\mu}_1 \neq \boldsymbol{\mu}_2$ and $M > r_{\B{\Sigma}_{12}} $, then the second term in (\ref{K3terms}) is given by
\begin{equation}
\frac{1}{8\sigma^2}  \sum_{i=r_{\mathbf{\Sigma}_{12}}+1}^M  \mathbf{u}_{12,i}^T \mathbf{M}_{12} \mathbf{u}_{12,i},
\label{K2term}
\end{equation}
where the vectors $\mathbf{u}_{12,r_{\mathbf{\Sigma}_{12}}+1},\ldots, \mathbf{u}_{12,M}$ are the eigenvectors of the matrix $\mathbf{\Phi} (\mathbf{\Sigma}_1 + \mathbf{\Sigma}_2) \mathbf{\Phi}^T$ corresponding to the zero eigenvalues, which form an orthonormal basis of the null space of that matrix. In this case, since the matrices $\mathbf{\Phi} (\mathbf{\Sigma}_1 + \mathbf{\Sigma}_2) \mathbf{\Phi}^T$ and $\mathbf{M}_{12}$ are positive semi-definite, by using the fundamental theorem of algebra, we can conclude that (\ref{K2term}) is equal to zero if and only if
\begin{equation}
 \mathrm{Null}(\mathbf{\Phi} (\mathbf{\Sigma}_1 + \mathbf{\Sigma}_2) \mathbf{\Phi}^T) \subseteq \mathrm{Null}(\mathbf{M}_{12}) \Leftrightarrow   \mathrm{im}(\mathbf{M}_{12}) \subseteq \mathrm{im}(\mathbf{\Phi} (\mathbf{\Sigma}_1 + \mathbf{\Sigma}_2) \mathbf{\Phi}^T) \Leftrightarrow \mathbf{\Phi} (\B \mu_1- \B \mu_2) \in \mathrm{im}(\mathbf{\Phi} (\mathbf{\Sigma}_1 + \mathbf{\Sigma}_2) \mathbf{\Phi}^T
 \label{nullim}
\end{equation}

Therefore, if we do not satisfy \eqref{nullim} then \eqref{K2term} is strictly greater than zero and the upper bound to the misclassification probability decays exponentially with $1/\sigma^2$ as $\sigma^2 \to 0$.

\end{itemize}

\section{Proof of Theorem \ref{theorem4}}
\label{app_D}
Recall that the diversity-order is given by:
\begin{equation}
\label{diver_design}
	d = -\frac{1}{2}\left(\frac{r_1+r_2}{2}- r_{12}\right) = \frac{1}{4}\left(2r_{12} - r_1 - r_2\right) .
\end{equation}
We establish the designs by pursuing a two-step approach where we first determine an upper bound to the maximum diversity-order and we then determine a design that achieves such a maximum diversity-order. We consider the cases where $M \geq NO_{Dim}$ and $M < NO_{Dim}$ separately.

\subsection{Case Where $M \geq NO_{Dim}$}

The maximum diversity-order that we can achieve with any measurement matrix when $M \geq NO_{Dim}$ is given by:
\begin{equation}
d \leq \frac{1}{4}  NO_{Dim}.
\label{max_div_ord}
\end{equation}
This upper bound can be proven by showing that:
\begin{equation}
2r_{12} - r_1 - r_2 \leq 2r_{\B{\Sigma}_{12}} - r_{\B{\Sigma}_{1}} - r_{\B{\Sigma}_{2}} =  NO_{Dim},
\label{r_rsigma}
\end{equation}
or, instead,
\begin{equation}
r_{\B{\Sigma}_{12}} - r_{12} \geq r_{\B{\Sigma}_{1}} - r_1 \wedge r_{\B{\Sigma}_{12}} - r_{12} \geq r_{\B{\Sigma}_{2}}  - r_2,
\label{nes_suf_cond}
\end{equation}
since (\ref{nes_suf_cond}) implies (\ref{r_rsigma}). Consider the generalized eigenvalue decomposition of the positive semidefinite matrices $\mathbf{\Sigma}_1$ and $\mathbf{\Sigma}_2$ given by~\cite[Theorem 8.7.1]{Golub96}, namely,  
${\B{\Sigma}_{1}} = \mathbf{X}^{-T}\mathbf{D}_1\mathbf{X}^{-1} = \mathbf{X}^{-T}\ \mathrm{diag}\left(d_{1_1}, \ldots, d_{1_N}\right)\mathbf{X}^{-1}$ with $d_{1_i} \geq 0, i = 1, \ldots, N$ and 
${\B{\Sigma}_{2}} = \mathbf{X}^{-T}\mathbf{D}_2\mathbf{X}^{-1} = \mathbf{X}^{-T}\ \mathrm{diag}\left(d_{2_1}, \ldots, d_{2_N}\right)\mathbf{X}^{-1}$ with $d_{2_i} \geq 0, i = 1, \ldots, N$, where $\mathbf{X}$ is a non-singular matrix. 

Now, we can write $r_{12} = \mathrm{rank} \left( \B{\Phi}\mathbf{X}^{-T}\left(\mathbf{D}_1 + \mathbf{D}_2\right)\mathbf{X}^{-1}\B{\Phi}^{-T}\right) = \mathrm{rank} \left( \B{\tilde{\Phi}}\left(\mathbf{D}_1 + \mathbf{D}_2\right)^{\frac{1}{2}}\right)$ and likewise, $r_{1} = \mathrm{rank} \left( \B{\Phi}\mathbf{X}^{-T}\left(\mathbf{D}_1 \right)\mathbf{X}^{-1}\B{\Phi}^{-T}\right) = \mathrm{rank} \left( \B{\tilde{\Phi}}\left(\mathbf{D}_1 \right)^{\frac{1}{2}}\right)$ and $r_{2} = \mathrm{rank} \left( \B{\Phi}\mathbf{X}^{-T}\left(\mathbf{D}_2 \right)\mathbf{X}^{-1}\B{\Phi}^{-T}\right) = \mathrm{rank} \left( \B{\tilde{\Phi}}\left(\mathbf{D}_2 \right)^{\frac{1}{2}}\right)$, where $\B{\tilde{\Phi}} = \B{\Phi}\mathbf{X}^{-T}$.

On the other hand, the ranks of the input covariance matrices can be expressed as $r_{\B{\Sigma}_{12}} = \mathrm{rank} \left(\mathbf{X}^{-T}\left(\mathbf{D}_1 + \mathbf{D}_2\right)\mathbf{X}^{-1}\right) = \mathrm{rank} \left( \mathbf{X}^{-T}\left(\mathbf{D}_1 + \mathbf{D}_2\right)^{\frac{1}{2}}\right) =  \mathrm{rank} \left( \left(\mathbf{D}_1 + \mathbf{D}_2\right)^{\frac{1}{2}}\right)$ and $r_{\B{\Sigma}_{1}} = \mathrm{rank} \left( \left(\mathbf{D}_1 \right)^{\frac{1}{2}}\right)$
and $r_{\B{\Sigma}_{2}} = \mathrm{rank} \left( \left(\mathbf{D}_2 \right)^{\frac{1}{2}}\right)$.

Let us now define the cardinalities of the following sets: $k_c = \left|\left\{i:d_{1_i} > 0 \wedge d_{2_i} > 0\right\}\right|$, $k_1 =  \left|\left\{i: d_{1_i} > 0 \right\}\right|$ and $k_2 =  \left|\left\{i: d_{2_i} > 0\right\}\right|$. Then, it becomes evident that, $r_{\B{\Sigma}_{12}} -  r_{\B{\Sigma}_{1}} = k_1 + k_2 - k_c - k_1 = k_2 - k_c$ and $r_{\B{\Sigma}_{12}} -  r_{\B{\Sigma}_{2}} = k_1 + k_2 - k_c - k_2 = k_1 - k_c$, and, in view of the possible dependence between columns of $\B{\tilde{\Phi}}$, $r_{12} -  r_{1} \leq k_2 - k_c$ and $r_{12} -  r_{2} \leq k_1 - k_c$, thus concluding the proof for the upper bound on the diversity-order.

This upper bound -- in view of (\ref{r_rsigma}) -- can be achieved by a projections matrix design that satisfies
\begin{equation}
\label{cond_desig_zero}
NO_{Dim} = 2r_{12} - r_1 - r_2  = 2r_{\B{\Sigma}_{12}} - r_{\B{\Sigma}_{1}} - r_{\B{\Sigma}_{2}} = {n_{\B{\Sigma}_{1}}} + {n_{\B{\Sigma}_{2}}},
\end{equation}
where we have used the fact that $r_{\B{\Sigma}_{1}} = N - \mathrm{dim}\left(\mathrm{Null}\left({\B{\Sigma}_{1}}\right)\right)= N - n_{12}-{n_{\B{\Sigma}_{1}}}$, $r_{\B{\Sigma}_{2}} = N - \mathrm{dim}\left(\mathrm{Null}\left({\B{\Sigma}_{2}}\right)\right)= N - n_{12}-{n_{\B{\Sigma}_{2}}}$ and $r_{\B{\Sigma}_{12}} = N - \mathrm{dim}\left(\mathrm{Null}\left(\B{\Sigma}_{1}\right)  \bigcap \mathrm{Null}\left(\B{\Sigma}_{2}\right)\right)= N - n_{12}$, where $n_{12}$, $n_{12}+n_{\B{\Sigma}_{1}}$ and  $n_{12}+n_{\B{\Sigma}_{2}}$ are the dimensions of the sub-spaces $\mathrm{Null}\left(\B{\Sigma}_{1}\right)  \bigcap \mathrm{Null}\left(\B{\Sigma}_{2}\right)$, $\mathrm{Null}\left(\B{\Sigma}_{1}\right)$ and $\mathrm{Null}\left(\B{\Sigma}_{2}\right)$, respectively. Note that, in order to guarantee \eqref{cond_desig_zero}, the two conditions in \eqref{nes_suf_cond} have to hold with equality, thus implying that $r_2 \geq 2r_{12} - r_1 = r_{\B{\Sigma}_{12}} - r_{\B{\Sigma}_{1}}$ and $r_1 \geq 2r_{12} - r_2 = r_{\B{\Sigma}_{12}} - r_{\B{\Sigma}_{2}}$.

A possible measurement matrix construction that achieves the maximum diversity-order is
\begin{equation}
\B{\Phi} = \left[\B{v}_1, \B{v}_2,	\ldots, \B{v}_{n_{\B{\Sigma}_{1}}},\B{w}_1,	\B{w}_2,\ldots,\B{w}_{n_{\B{\Sigma}_{2}}}\right]^T,
\label{phi_opt}
\end{equation}
where the set of vectors $\left[\B{u}_1,\ldots,\B{u}_{n_{12}}\right],~\left[\B{u}_1,\ldots,\B{u}_{n_{12}},\B{v}_1,\ldots,\B{v}_{n_{\B{\Sigma}_{1}}}\right]$, $\left[\B{u}_1,\ldots,\B{u}_{n_{12}},\B{w}_1,\ldots,\B{w}_{n_{\B{\Sigma}_{2}}}\right]$,  $\B{u}_i, \B{v}_i, \B{w}_i  \in \mathbb{R}^N$, constitute an orthonormal basis of the linear spaces $\mathrm{Null}\left(\B{\Sigma}_{1}\right)  \bigcap \mathrm{Null}\left(\B{\Sigma}_{2}\right)$, $\mathrm{Null}\left(\B{\Sigma}_{1}\right)$ and $\mathrm{Null}\left(\B{\Sigma}_{2}\right)$, respectively. This can be verified by writing
\begin{equation}
\mathbf{\Phi} \mathbf{\Sigma}_1 \mathbf{\Phi}^T=\left[
\begin{array}{c|c}
  \B{0} & \B{0} \\
  \hline
  \B{0} & \mathbf{Q}
 \end{array}\right] \qv
\mathbf{Q} = \begin{bmatrix}
	\B{w}_1   & \B{w}_2	& \cdots & \B{w}_{n_{\B{\Sigma}_{2}}} \end{bmatrix}^T  \B{\Sigma}_{1} \begin{bmatrix}
	\B{w}_1   & \B{w}_2	& \cdots & \B{w}_{n_{\B{\Sigma}_{2}}} \end{bmatrix}
\end{equation}
where $r_1=\mathrm{rank}\left(\mathbf{\Phi} \mathbf{\Sigma}_1 \mathbf{\Phi}^T\right) = \mathrm{rank}\left(\mathbf{Q} \right)$.

Now, note that the matrix $\mathbf{Q}$ is the Gram matrix of the set of vectors $\B{q}_i= \B{\Sigma}_{1}^{\frac{1}{2}}\B{w}_i$, $i=1, \ldots, {n_{\B{\Sigma}_{2}}}$, and, therefore, $r_1=\mathrm{rank}\left(\mathbf{Q} \right) = {n_{\B{\Sigma}_{2}}}$ if and only if the vectors $\B{q}_i$, $i=1, \ldots, {n_{\B{\Sigma}_{2}}}$, are linearly independent.

Assume by contradiction that the vectors $\B{q}_i$ are linearly dependent. Then, there exists a set of $n_{\mathbf{\Sigma}_2}$ scalars $\alpha_i$ (with $\alpha_i \neq 0$ for at least one index $i$) such that $\B{\Sigma}_{1}^{\frac{1}{2}} \sum_i \alpha_i \B{w}_i = \mathbf{0}$. It is known that $\sum_i \alpha_i \B{w}_i \neq \mathbf{0}$ because $\B{w}_i$ are linearly independent by construction. Therefore, the linearly dependence among the vectors $\B{q}_i$ implies that
\begin{equation}
\sum_i \alpha_i \B{w}_i \in \mathrm{Null}\left(\B{\Sigma}_{1}^{\frac{1}{2}}\right)\  \mbox{or}  \  \sum_i \alpha_i \B{w}_i \in \mathrm{Null}\left(\B{\Sigma}_{1}\right)
\end{equation}
which is false since, by construction, $\sum_i \alpha_i \B{w}_i \in \mathrm{Null}\left(\B{\Sigma}_{2}\right)$ and $\sum_i \alpha_i \B{w}_i \notin \mathrm{Null}\left(\B{\Sigma}_{1}\right)  \bigcap \mathrm{Null}\left(\B{\Sigma}_{2}\right)$. Therefore, we can establish that $r_1 =\mathrm{rank}\left(\mathbf{\Phi} \mathbf{\Sigma}_1 \mathbf{\Phi}^T\right) = \mathrm{rank}\left(\mathbf{Q} \right) = {n_{\B{\Sigma}_{2}}}$. We can similarly establish that $r_2 =\mathrm{rank}\left(\mathbf{\Phi} \mathbf{\Sigma}_2 \mathbf{\Phi}^T\right) =  {n_{\B{\Sigma}_{1}}}$ and $r_{12} = \mathrm{rank}\left(\mathbf{\Phi} (\mathbf{\Sigma}_i + \mathbf{\Sigma}_j) \mathbf{\Phi}^T\right) = {n_{\B{\Sigma}_{1}}} + {n_{\B{\Sigma}_{2}}}$, that is, this matrix construction, which satisfies the condition in (\ref{cond_desig_zero}), achieves the maximum diversity-order in (\ref{max_div_ord}).

\subsection{Case Where $M < NO_{Dim}$}

The maximum diversity-order that we can achieve with any measurement matrix when $M < NO_{Dim}$ is now given by:
\begin{equation}
d \leq \frac{1}{4} M.
\end{equation}
This upper bound follows from the solution to the following integer-valued optimization problem\footnote{Note that this problem represents a relaxation of the actual diversity-order maximization problem, as it incorporates only some of the constraints dictated by the geometrical description of the scenario. For example, it does not take into account the actual value of some parameters of the input description as $r_{\mathbf{\Sigma}_1},r_{\mathbf{\Sigma}_2}$ and $r_{\mathbf{\Sigma}_{12}}$.
}:
\begin{equation}
\label{opt_1}
\max_{r_{1}, r_{2}, r_{12}}  -\frac{1}{2}\left( \frac{r_{1}+r_{2}}{2} - r_{12} \right)
\end{equation}
subject to: $r_{1}+r_{2} \geq r_{12}$, $r_{1} \leq M$, $r_{2} \leq M$, $r_{12} \leq M$ and $r_{1}, r_{2}, r_{12} \in \mathbb{Z}_0^+$.

The solution, which can be obtained by considering a linear programming relaxation along with a Branch and Bound approach~\cite{Schrijver98}, is given by\footnote{Note that the solution of the optimization problem is not unique. Nevertheless, the maximum value achieved by the objective function is indeed unique.}:
\begin{equation}
r_{1}+ r_{2}=r_{12}
\qv
r_{12} = M
\end{equation}
and\begin{equation}
d =-\frac{1}{2}\left( \frac{r_{1}+r_{2}}{2} - r_{12} \right) =  \frac{1}{4}M.
\label{up_div}
\end{equation}

A possible measurement matrix construction that achieves such maximum diversity-order in (\ref{up_div}) is obtained by picking arbitrarily only $M$ among its $n_{\mathbf{\Sigma}_1} + n_{\mathbf{\Sigma}_2}$ row vectors of the matrix $\mathbf{\Phi}$ in (\ref{phi_opt}). In particular, we take $M_1$ rows from the set $\left[ \mathbf{v}_1,\ldots, \mathbf{v}_{n_{\mathbf{\Sigma}_1}}\right]$ and $M_2$ rows from the set $\left[ \mathbf{w}_1,\ldots, \mathbf{w}_{n_{\mathbf{\Sigma}_2}}\right]$, where $M_1 + M_2=M$, which is always possible as $M < NO_{Dim}=n_{\mathbf{\Sigma}_1}+n_{\mathbf{\Sigma}_2}$. Then, by following steps similar to the previous ones, it is possible to show that $r_1 = \mathrm{rank}\left(\mathbf{\Phi} \mathbf{\Sigma}_1 \mathbf{\Phi}^T\right) =  M_2$, $r_2 = \mathrm{rank}\left(\mathbf{\Phi} \mathbf{\Sigma}_2 \mathbf{\Phi}^T\right) =  M_1$ and $r_{12} = \mathrm{rank}\left(\mathbf{\Phi} (\mathbf{\Sigma}_i + \mathbf{\Sigma}_j) \mathbf{\Phi}^T\right) = M_1+M_2=M$, that is, this matrix construction achieves the maximum diversity-order in (\ref{up_div}).

\section{Proof of Theorem \ref{theorem5}}
\label{app_E}

As presented in Appendix \ref{app_C}, the upper bound to the misclassification probability decays exponentially with $1/\sigma^2$, as $\sigma^2 \to 0$ (achieving a diversity-order equal to infinity) if
\begin{equation}
 \sum_{i=r_{12}+1}^M  \mathbf{u}_{12,i}^T \mathbf{\Phi}(\B{\mu}_1 - \B{\mu}_2)  (\B{\mu}_1 - \B{\mu}_2)^T \mathbf{\Phi}^T \mathbf{u}_{12,i} > 0.
\end{equation}
However, we underline that, in view of the fact that the condition $M \leq r_{\B{\Sigma}_{12}}$ no longer implies $r_{12} = M$, the number of measurements required to achieve infinite diversity-order with a optimized $\mathbf{\Phi}$ can be lower than those with a random $\mathbf{\Phi}$.

Assume that
\begin{equation}
  (\B{\mu}_1 - \B{\mu}_2) \notin  \mathrm{im}\left(\mathbf{\Sigma}_1 + \mathbf{\Sigma}_2\right).
  \label{cond_nonzero}
\end{equation}
We can show that that it is possible to achieve infinite diversity-order with the measurement kernel
\begin{equation}
\B{\Phi} = \begin{bmatrix}
	\B{\phi}^T     
     \end{bmatrix}
\end{equation}
where $\B{\phi} \in \mathrm{Null} (\mathbf{\Sigma}_1 + \mathbf{\Sigma}_2)$ -- note that there exists such a $\B{\phi}$ because $\mathrm{Null} (\mathbf{\Sigma}_1 + \mathbf{\Sigma}_2)$ is the orthogonal complement of $\mathrm{im}\left(\mathbf{\Sigma}_1 + \mathbf{\Sigma}_2\right)$ and hence it does not contain only the zero vector. In fact, let us consider the standard decomposition~\cite{Horn}
\begin{equation}
(\B{\mu}_1- \B{\mu}_2) = {\B \mu}_{im} + {\B \mu}_{Null},
\end{equation}
where $\B \mu_{im} \in \mathrm{im}\left(\mathbf{\Sigma}_1 + \mathbf{\Sigma}_2\right)$, $\B \mu_{Null} \in \mathrm{Null}\left(\mathbf{\Sigma}_1 + \mathbf{\Sigma}_2\right)$ and, given (\ref{cond_nonzero}), $\B \mu_{Null} \neq \mathbf{0}$. Then, $\B{\Phi}(\B{\mu}_1- \B{\mu}_2) = 0 + \B{\phi}^T{\B \mu}_{Null} > 0$ and, therefore, $\sum_{i=r_{12}+1}^M  \mathbf{u}_{12,i}^T \mathbf{\Phi}(\B{\mu}_1 - \B{\mu}_2)  (\B{\mu}_1 - \B{\mu}_2)^T \mathbf{\Phi}^T \mathbf{u}_{12,i} = |\B{\phi}^T{\B \mu}_{Null} |^2  > 0$.

Assume now that
\begin{equation}
  (\B{\mu}_1 - \B{\mu}_2) \in  \mathrm{im}\left(\mathbf{\Sigma}_1 + \mathbf{\Sigma}_2\right).
\end{equation}
We can now show that one cannot achieve infinite diversity-order for all possible choices of the measurement kernel because $ \mathbf{\Phi} (\B{\mu}_1 - \B{\mu}_2)\in  \mathrm{im}\left(\mathbf{\Phi} (\mathbf{\Sigma}_1 + \mathbf{\Sigma}_2) \mathbf{\Phi}^T\right)$ and hence $ \sum_{i=r_{12}+1}^M  \mathbf{u}_{12,i}^T \mathbf{\Phi}(\B{\mu}_1 - \B{\mu}_2)  (\B{\mu}_1 - \B{\mu}_2)^T \mathbf{\Phi}^T  \mathbf{u}_{12,i} = 0$. In fact, in view of this last result, the maximum diversity-order and the measurement kernel that achieves such a diversity-order are now given by Theorem~\ref{theorem4}.

\bibliographystyle{IEEEtran}
\bibliography{IEEEabrv,cl}

\end{document}